\documentclass[a4paper]{article}
\usepackage{amsmath}
\usepackage{graphicx}
\usepackage{geometry}
\usepackage{floatrow}
\usepackage{float}
\usepackage{layout}
\usepackage{amssymb} 
\usepackage{multirow}
\usepackage{caption}
\usepackage{subcaption} 
\usepackage{authblk}
\usepackage{hyperref}
\usepackage{indentfirst}
\usepackage{physics}
\usepackage[toc,page]{appendix}
\usepackage{tikz}
\usepackage{cite}
\usetikzlibrary{calc}
\usepackage{relsize}
\tikzset{fontscale/.style = {font=\relsize{#1}}}
\usetikzlibrary{arrows}
\usetikzlibrary{shapes,arrows,shapes.multipart}
\usepackage{circuitikz}
\usepackage{enumitem}
\usepackage{amsmath, amssymb, amsthm, mathrsfs}

\newtheorem{theorem}{Theorem}
\newtheorem{proposition}{Proposition}
\newtheorem{assumption}{Assumption}

\begin{document}

\newcommand{\beq}{\begin{equation}}
\newcommand{\eeq}{\end{equation}}
\newcommand{\bea}{\begin{eqnarray}}
\newcommand{\eea}{\end{eqnarray}}

\newcommand{\blue}[1]{\textcolor{blue}{#1}}
\newcommand{\red}[1]{\textcolor{red}{#1}}
\newcommand{\purple}[1]{\textcolor{purple}{#1}}

\title{\textbf{\huge{Conformal Cauchy Slice Holography:\\ An Alternative Phase Space For Gravity}}}
\author{\large{Rifath Khan\footnote{rifathkhantheo@gmail.com} }}
\affil{SITP, Stanford University} 
\date{\today}

\maketitle

\maketitle

\begin{abstract}

The phase space of gravitational theories in asymptotically Anti-de Sitter (AAdS) spacetimes consists of geometries, matter configurations, and their conjugate momenta on a Cauchy surface, subject to the Hamiltonian, momentum, and matter-gauge constraints. When a unique maximal volume slice exists in all classical solutions of the bulk equations of motion, and the matter fields satisfy certain conditions, we show that this phase space is physically equivalent to an alternative phase space in which the Hamiltonian constraint is replaced by the real Weyl-anomaly constraint, while the momentum and matter-gauge constraints remain unchanged. A necessary requirement for a functional of the metric and matter configurations to qualify as a valid quantum gravity state is that it satisfies the operator gauge constraints. Partition functions of certain conformal field theories with imaginary central charge, defined on bulk Cauchy slices, satisfy these operator gauge constraints and therefore provide candidate quantum gravity states in the alternative phase space formulation.

\end{abstract}

\tableofcontents

\section{Introduction}

The AdS/CFT correspondence provides us with a non-perturbative definition of a theory of quantum gravity in a $(d+1)$-dimensional asymptotically Anti-de Sitter (AAdS) spacetime $\mathcal{M}$ in terms of a dual $d$-dimensional conformal field theory (CFT) living on the asymptotic timelike boundary $\partial\mathcal{M}$ \cite{Maldacena:1997re,Witten:1998qj}. The UV completeness of the CFT implies that its dual bulk quantum gravity theory is UV complete \cite{kaplanAdSCFT}. Moreover, because no explicit background is chosen in the bulk, the theory is background independent in principle \cite{Hubeny:2014bla}. 
However, the description is phrased in the language of a dual $d$-dimensional CFT, and one needs to translate it into the language of a $(d+1)$-dimensional bulk gravity theory. To achieve this, one must construct a map from the boundary CFT Hilbert space to the Hilbert space of the bulk quantum gravity theory. A necessary precursor to this construction is a thorough understanding of the classical phase space of the bulk gravity theory and its constraints.

In the classical theory of gravity coupled to arbitrary matter in AAdS spacetimes, the covariant phase space is the set of all spacetime field configurations satisfying the bulk equations of motion and boundary conditions within the bulk domain of dependence of a given boundary Cauchy surface $\partial\Sigma$, modulo gauge redundancies \cite{Wald:1993nt,Harlow:2019yfa}. Here, $\partial\Sigma$ is a $(d-1)$-dimensional Cauchy surface of the boundary $\partial\mathcal{M}$, and the bulk domain of dependence (henceforth denoted by $\Omega$) of some $\partial\Sigma$ is defined as the domain of dependence of any bulk Cauchy surface $\Sigma$ whose boundary is $\partial\Sigma$. In the ADM formalism, this covariant phase space can be equivalently represented by what is known as the ADM phase space: the space of initial‑value data on $\Sigma$, namely $(g_{ab},\Pi^{ab},\Phi^i,\Pi_{\Phi^i})$, where $g_{ab}$ and $\Pi^{ab}$ are the spatial metric and its conjugate momentum, and $\Phi^i$ and $\Pi_{\Phi^i}$ are the matter fields and their conjugate momenta \cite{Arnowitt:1962hi}. To map it back to the full covariant phase space, one solves the bulk equations of motion for these initial datas. The ADM phase space is subject to the first‑class constraints: the Hamiltonian constraint $\mathcal{H}$ (which generates temporal diffeomorphisms), the momentum constraints $D_a$ (which generate spatial diffeomorphisms and are also known as spatial diffeomorphism constraints), and the matter‑gauge constraints $G^A$ (which generate matter‑gauge transformations). The explicit form of each constraint depends on the specifics of the theory.

Quantum gravity states $\Psi_{\text{WDW}}[g_{ab},\Phi^i]$ are wave functionals of the metric and matter configurations on $\Sigma$, which are annihilated by the constraints (now imposed as operator constraints): $\widehat{\mathcal{H}} \Psi_{\text{WDW}} = \widehat{D}_a \Psi_{\text{WDW}} = \widehat{G}^A \Psi_{\text{WDW}} = 0$,
and are known as Wheeler-DeWitt (WDW) states \cite{DeWitt:1967yk,DeWitt:1967ub,DeWitt:1967uc}. In this paper, we use the hat symbol $(\, \widehat{\ }\, )$ to denote operators, following the standard convention in quantum mechanics, in order to distinguish them from their classical counterparts. These constraints ensure the invariance of the wave functional under temporal diffeomorphisms, spatial diffeomorphisms, and matter-gauge transformations, respectively. The condition that the Hamiltonian constraint annihilates the wave functional is also known as the Wheeler-DeWitt equation: \begin{equation}\label{ham}
\widehat{\mathcal{H}}\Psi[g,\Phi^i]:=\Bigg\{\frac{2\kappa}{\sqrt{g}}\!\Big(\widehat{\Pi}_{ab}\widehat{\Pi}^{ab}-\frac{1}{d-1}\widehat{\Pi}^2\Big)\!-\frac{\sqrt{g}}{2\kappa}(R-2\Lambda) + \widehat{\mathcal{H}}_{\text{matter}}\Bigg\}\Psi[g,\Phi^i]=0,
\end{equation}
where $\kappa=8\pi G_N$, $\mathcal{H}_{\text{matter}}$ is the matter Hamiltonian, $\widehat{\mathcal{H}}_{\text{matter}}$ is its corresponding quantum operator, and the momentum operator is given by $\widehat{\Pi}^{ab} = -i \frac{\delta}{\delta g_{ab}}$. This equation has historically been notoriously difficult to solve, as it is a \emph{second-order functional differential equation}. In addition, there are problems related to operator ordering ambiguities, as well as issues in defining the second-order functional differential operator $\frac{\delta}{\delta g_{ab}(x)}\frac{\delta}{\delta g_{cd}(x)}$ at coincident points. As a result, solving this equation has generally been limited to simplified settings such as the minisuperspace approximation or lower-dimensional models like JT gravity.


The WDW equation also appears in the context of finite cutoff holography in a different form: the radial WDW equation \cite{Freidel:2008sh}. One of the central statements of the AdS/CFT correspondence is that the bulk gravitational path integral in asymptotically AdS spacetimes, with Dirichlet boundary conditions imposed at the asymptotic boundary, is equal to the partition function of the boundary CFT. This is the case when the boundary is at the asymptotic infinity. However, if the timelike boundary is instead placed at a finite radial distance, the gravitational path integral evaluated with Dirichlet boundary conditions on this finite boundary satisfies the radial WDW equation and no longer equals the CFT partition function. Rather, it corresponds to the partition function of the so-called $T^2$ theory \cite{McGough:2016lol,Hartman:2018tkw,Zamolodchikov:2004ce,Taylor:2018xcy,Lewkowycz:2019xse}. This theory is obtained by deforming the CFT with an irrelevant operator that is quadratic in the stress tensor—hence the name $T^2$—followed by the inclusion of appropriate holographic counterterms. The partition function of this $T^2$ theory satisfies the radial WDW equation.


In Cauchy slice holography \cite{CSH}, we employed the $T^2$ deformation to construct quantum gravity states. We began by Wick‑rotating the holographic CFT—originally defined on the asymptotic timelike boundary $\partial\mathcal{M}$—and placing it on a (hypothetical) bulk Cauchy slice $\Sigma$. Its partition function then becomes a functional of the metric $g_{ab}$ and matter fields $\Phi^i$ on $\Sigma$, which act as background sources to the CFT.\footnote{The matter source $\Phi^i$ is often mistaken for the CFT’s dynamical fields. In fact, $\Phi^i$ is not part of the CFT field content but a background source, analogous to the metric.} Since $\Sigma$ is an open manifold with boundary $\partial \Sigma$, one must supply boundary conditions to define the partition function uniquely. Imposing the CFT state as the boundary condition yields $Z_{\text{CFT}}[g_{ab},\Phi^i;\psi_{\text{CFT}}]$. This partition function automatically satisfies the momentum constraints, $\widehat{D}^a Z_{\text{CFT}} = 0$, as do all QFT partition functions due to covariance, and also satisfies the matter gauge constraints, $\widehat{G}^A Z_{\text{CFT}} = 0$, which follow from the duality.\footnote{The gravitational path integral with Dirichlet boundary conditions on the timelike boundary is invariant under matter‑gauge transformations—whether the boundary lies at finite distance or at infinity. By duality, the CFT partition function shares this invariance and thus satisfies $\widehat{G}^A Z_{\text{CFT}} = 0$.} However, it cannot serve as a WDW state because it does not satisfy the Hamiltonian constraint. Instead, it obeys the Weyl‑anomaly constraint (also known as the conformal Ward identity)
\begin{equation} \label{imaginary Weyl-anomaly constraint equation}
\left(\widehat{\mathcal{W}}(x)-i\mathcal{A}(x)\right) Z^{(\epsilon)}_{\text{CFT}} = 0, 
\end{equation} 
where $\widehat{\mathcal{W}}(x)$ is the Weyl operator (generating local Weyl transformations) and $\mathcal{A}(x)$ is the Weyl (conformal) anomaly in the CFT.\footnote{For example, in $d=2$ with only a background metric (and no matter sources), one has $\widehat{\mathcal{W}}=2\widehat{\Pi}=2 g_{ab}\widehat{\Pi}^{ab}$ and the Weyl anomaly $\mathcal{A}=-\frac{c}{24 \pi}\sqrt{g}R$, where $R$ is the Ricci scalar of the background metric $g_{ab}$ and $c$ the central charge. In its more familiar form, this equation reads $\langle \widehat{T} \rangle = -\frac{c}{24\pi} R$. For a detailed explanation, see Section~\ref{CFT Partition Functions as Quantum Gravity States in the Alternative Phase Space}.} The regulator $\epsilon$ regulates logarithmic divergences arising from the anomaly, which cannot be removed by counterterms. Moreover, the anomaly $\mathcal{A}$ of the holographic CFT is directly related to the bulk Hamiltonian constraint $\mathcal{H}$. We then deformed this CFT by the $T^2$ operator to obtain $Z_{T^2}$:
\begin{align} 
Z_{T^2}[g,\Phi^i;\psi_{\text{CFT}}] = e^{\widehat{CT}(\mu)} \left( \text{P} \exp \int_{\epsilon}^{\mu}\frac{d\lambda}{\lambda} \widehat{O}(\lambda) \right) Z_\text{CFT}^{(\epsilon)}[g,\Phi^i;\psi_{\text{CFT}}], 
\end{align} 
where $\widehat{O}(\lambda)$ is the $T^2$ deformation operator, $\mu$ is the deformation parameter, and $\widehat{CT}$ are the holographic counterterms. The precise form of the $T^2$ operator and of the counterterms depends on the Hamiltonian constraint under consideration.\footnote{The term “$T^2$ deformation” is used broadly: the precise operator varies with the Hamiltonian constraint. Its derivation for a general Hamiltonian constraint appears in detail in \cite{CSH}.} The deformed partition function then satisfies the WDW equation $\widehat{\mathcal{H}} Z_{T^2} = 0$. Because this deformation is covariant and matter‑gauge invariant, $Z_{T^2}$ is annihilated by all the bulk constraints and thus defines a valid WDW state:
\begin{equation} 
\Psi_{\text{WDW}}[g,\Phi^i] = Z_{T^2}[g,\Phi^i;\psi_{\text{CFT}}]. 
\end{equation} 
For each boundary CFT state $\psi_{\text{CFT}}$, we obtain a corresponding bulk WDW state. This correspondence provides a natural dictionary from the boundary CFT Hilbert space to the bulk WDW states, as proposed in \cite{CSH}.


The $T^2$ theory offers key advantages for understanding the WDW equation. First, the previously ill‑defined second‑order functional derivative in the WDW equation can now be made well‑defined, at least in the large‑$N$ limit.\footnote{See \cite{CSH} for full details; here, $N$ denotes the rank of the CFT gauge group in the AdS/CFT correspondence.} Second, all quantum‑gravitational information encoded by the WDW equation becomes accessible via derivatives of the WDW wave function with respect to its arguments—these are precisely the correlation functions of the $T^2$ theory. The $T^2$ deformation is an irrelevant deformation, and it satisfies the renormalization group (RG) equations. Consequently, the $T^2$-deformed theory should be understood as an effective quantum field theory in the Wilsonian RG sense. Consequently, its correlation functions lack infinite resolution, and the theory requires a UV completion (just as a QFT) to capture UV‑sensitive aspects of quantum gravity.


The need to address the WDW equation arose because we did not fix the Hamiltonian constraint at the classical level (\emph{before} quantizing the theory), primarily due to lack of a suitable method. However, a much better understanding now exists via the conformal decomposition \cite{Isenberg_1995,Andersson_Chrusciel,Sakovich:2009nb,Witten:2022xxp}. In this decomposition, the metric $g_{ab}$ is split into its conformal part $\gamma_{ab}$ and a conformal factor $\phi$, such that $g_{ab} = \phi^{4/(d-2)} \gamma_{ab}$.\footnote{One can think of $\gamma_{ab}$ as representing a conformal class of metrics, i.e., the set of all metrics that are related to $g_{ab}$ by a Weyl transformation. Equivalently, as is often necessary in practice, one can also think of $\gamma_{ab}$ as a chosen representative of the conformal class. One often selects a representative element for each conformal class by imposing a condition on $\sqrt{\det(\gamma)}$. A common choice is to fix it to a constant, such as $\sqrt{\det(\gamma)} = 1$, but this is coordinate dependent and thus breaks manifest coordinate invariance. To preserve coordinate invariance, a better approach is to fix $\sqrt{\det(\gamma)}$ by equating it to a scalar density of weight one. However, this is also unsatisfactory, since the theory does not naturally contain such scalar densities; introducing one would require fictitious auxiliary structures and render the formalism dependent on them. Instead, we will impose a fully covariant condition by requiring the Ricci scalar of the conformal metric $\gamma_{ab}$ to be constant and equal to $2\Lambda$ everywhere on $\Sigma$. We refer to this choice as the covariant conformal decomposition and will explain it in detail later in the paper.} Similarly, the conjugate momentum $\Pi^{ab}$ is decomposed into its trace part $\Pi = g_{ab} \Pi^{ab}$ and its traceless part $\Pi^{ab}_{\text{traceless}} = \phi^{4/(d-2)}(\Pi^{ab} - \frac{1}{d} g^{ab} \Pi)$. The variables $\frac{4}{(d-2)\phi}\Pi$ and $\Pi^{ab}_{\text{traceless}}$ are canonically conjugate to $\phi$ and $\gamma_{ab}$, respectively.\footnote{In the covariant conformal decomposition, we will also decompose $\Pi^{ab}$ differently from the standard one given above: it will be chosen so that the resulting variables naturally pair canonically with the conformal factor and the conformal metric. We will return to this refined decomposition later; for now, the reader may simply keep the standard conformal decomposition in mind.} This decomposition is a canonical transformation on the phase space and coordinatises it by $(\phi,\Pi,\gamma_{ab},\Pi^{ab}_{\text{traceless}},\Phi^i,\Pi_{\Phi^i})$. The Hamiltonian constraint then becomes the Lichnerowicz equation for $\phi$, which is best understood on constant mean curvature (CMC) slices---Cauchy surfaces on which the trace of the extrinsic curvature, $K$, is constant \cite{Isenberg_1995,Andersson_Chrusciel}.\footnote{See also \cite{Isenberg:1996kz} for an analysis on non-constant mean curvature slices.} A maximal volume slice is a CMC slice on which $K = 0$ everywhere. For a negative cosmological constant and matter fields satisfying certain conditions, it has been proven that a unique solution to the Lichnerowicz equation exists on such a slice \cite{Sakovich:2009nb}.


For every point in the covariant phase space (i.e., every classical saddle in $\Omega$), a maximal volume slice exists in $\Omega$ if one assumes a plausible form of cosmic censorship, and it is unique provided the strict generic strong energy condition holds (i.e., that gravity acts as an attractive force)\cite{Couch:2018phr}.\footnote{See also \cite{Chrusciel:2022cjz} for discussions on existence.} When a unique maximal volume slice exists in $\Omega$ for all classical solutions, the condition $K=0$ can serve as a valid gauge‑fixing condition for the temporal gauge transformations generated by the Hamiltonian constraint. Hence, on the constraint surface $\mathcal{H}=0$ in phase space, after imposing $K=0$ to fix temporal gauge transformations and determining the conformal factor $\phi$ uniquely via the Lichnerowicz equation, one obtains a reduced phase space parameterized by $(\gamma_{ab},\Pi^{ab}_{\text{traceless}},\Phi^i,\Pi_{\Phi^i})$, with only the residual constraints $D_a$ and $G^A$ generating spatial diffeomorphisms and matter‑gauge transformations, respectively \cite{Witten:2022xxp}.


With this reduced phase space, after quantization, quantum gravity states $\Psi_{\text{reduced}}[\gamma_{ab},\Phi^i]$ become wavefunctionals of the conformal part of the metric $\gamma_{ab}$ (or, more precisely, its conformal classes $[\gamma_{ab}]$) and of the matter fields $\Phi^i$, which are annihilated by the spatial diffeomorphism and matter‑gauge constraints $\widehat{D}^a \Psi_{\text{reduced}} = \widehat{G}^A \Psi_{\text{reduced}}=0$. Alternatively, one may choose to deal with these residual constraints using BRST techniques. This involves trading spatial diffeomorphism and matter‑gauge invariance for BRST invariance and introducing ghost fields corresponding to these gauge symmetries. Then, quantum gravity states $\Psi_{\text{BRST}}$ become those wavefunctionals of conformal classes of metrics, matter configurations, and ghost fields on the maximal volume slice that belongs to the BRST cohomology. To obtain BRST‑closed states, one solves the functional differential equation $\widehat{Q} \Psi_{\text{BRST}}=0$, where $\widehat{Q}$ is the BRST charge operator—generally a complicated task. One approach is to prepare such states by the BRST gravitational path integral with conformal boundary conditions: one fixes $K$ and $\gamma_{ab}$ on the Cauchy slice and integrates over the spacetime metric to the past of the slice, including appropriate gauge‑fixing and ghost terms in the action. Since the slice is maximal volume, $K=0$. Performing this path integral exactly remains an open problem, but it can be done perturbatively by expanding the metric around a classical saddle, integrating over the perturbations, and imposing a gauge such as the harmonic gauge. Witten demonstrates this procedure in \cite{Witten:2022xxp}, constructing a perturbative Hilbert space for quantum gravity valid to all orders in the metric expansion. Consequently, the BRST‑closure condition holds order by order: when $\widehat{Q}\Psi_{\text{BRST}}=0$ is expanded in powers of the metric perturbation, the state obtained by truncating the BRST path integral at a given order satisfies the corresponding truncated BRST closure equation. Witten also provides a comprehensive discussion of phase‑space reduction, the existence and uniqueness of maximal volume slices, and the Lichnerowicz equation in \cite{Witten:2022xxp}.


In this paper, we show that, when a unique maximal volume slice exists in $\Omega$ for every classical saddle, and the matter fields satisfy certain conditions, there exists an alternative phase space for gravity and matter fields that is physically equivalent to the ADM phase space. This alternative phase space consists of the set of geometries, matter fields on a Cauchy slice, and their conjugate momenta $(\mathfrak{g}_{ab},\mathsf{\Pi}^{ab},\Phi^i,\Pi_{\Phi^i})$, subject to the spatial diffeomorphism constraints $D_a$, matter gauge constraints $G^A$, and the real Weyl‑anomaly constraint instead of the Hamiltonian constraint. We will explain what this real Weyl‑anomaly constraint is later. Briefly, one obtains this alternative phase space by reinstating the conformal factor and the trace of the extrinsic curvature into the reduced phase space and imposing an additional gauge constraint—the real Weyl‑anomaly constraint—to eliminate these variables. One can extend any phase space in various ways by introducing additional pure-gauge degrees of freedom, along with suitable gauge constraints that eliminate them. As we will see, the particular choice of extension we consider here is relevant for constructing candidate quantum gravity states.  

Upon quantization, these constraints become operator gauge constraints. A necessary requirement for a functional $\Psi_{\text{QG}}[\mathfrak{g}_{ab},\Phi^i]$ of the metric and matter configurations to qualify as a valid quantum gravity state is that it satisfies these operator gauge constraints. We will see that certain CFT partition functions with imaginary central charge satisfy these constraints, and thus provide candidate quantum gravity states in the alternative phase space formulation. We emphasize the term “candidate” because satisfying the operator constraint equations is merely a necessary condition; additional requirements must be fulfilled for a wavefunctional to qualify as a valid quantum gravity state. Establishing this fully requires further work. We will discuss some of these additional conditions later.

Demonstrating the physical equivalence of this alternative phase space with the original phase space and constructing candidate quantum gravity states non‑perturbatively constitute the key novel contributions of this paper.

\subsection*{Plan of the paper}

In Section \ref{Statement of Main Result}, we state the main theorems precisely and provide a sketch of the argument. Section \ref{Phase Space Reduction} reviews preliminary concepts and explains the phase‑space reduction in a self‑consistent style. In Section \ref{Phase Space Enlargement}, we construct the alternative phase space, demonstrate its physical equivalence to the ADM phase space, and explain how to translate between them. Section \ref{Quantum Gravity Wave Function} describes the construction of candidate quantum gravity states in the new phase space. Finally, in Section~\ref{Discussion}, we discuss outstanding issues, open questions, and future work.

\section{Statement of Main Results}\label{Statement of Main Result}

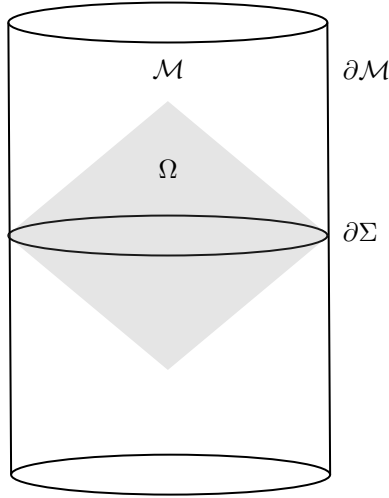
\begin{figure}[ht]
    \centering
\tikzset{every picture/.style={line width=0.75pt}} 
\begin{tikzpicture}[x=0.75pt,y=0.75pt,yscale=-1,xscale=1]

\draw    (242.5,22.54) -- (243.88,249.96) ;
\draw    (401.73,22.54) -- (403.12,249.96) ;
\draw   (242.5,129.69) .. controls (242.5,124.14) and (278.15,119.64) .. (322.12,119.64) .. controls (366.09,119.64) and (401.73,124.14) .. (401.73,129.69) .. controls (401.73,135.24) and (366.09,139.73) .. (322.12,139.73) .. controls (278.15,139.73) and (242.5,135.24) .. (242.5,129.69) -- cycle ;
\filldraw[gray,opacity=0.2]  (322.12,62.72) -- (401.73,129.69) -- (322.12,196.65) -- (242.5,129.69) -- cycle ;
\draw   (243.88,249.96) .. controls (243.88,244.41) and (279.53,239.91) .. (323.5,239.91) .. controls (367.47,239.91) and (403.12,244.41) .. (403.12,249.96) .. controls (403.12,255.5) and (367.47,260) .. (323.5,260) .. controls (279.53,260) and (243.88,255.5) .. (243.88,249.96) -- cycle ;
\draw   (242.5,22.54) .. controls (242.5,17) and (278.15,12.5) .. (322.12,12.5) .. controls (366.09,12.5) and (401.73,17) .. (401.73,22.54) .. controls (401.73,28.09) and (366.09,32.59) .. (322.12,32.59) .. controls (278.15,32.59) and (242.5,28.09) .. (242.5,22.54) -- cycle ;

\draw (408,121.86) node [anchor=north west][inner sep=0.75pt]   [align=left] {$\partial\Sigma$};
\draw (408,40) node [anchor=north west][inner sep=0.75pt]   [align=left] {$\partial\mathcal{M}$};
\draw (313,40) node [anchor=north west][inner sep=0.75pt]   [align=left] {$\mathcal{M}$};
\draw (316,90) node [anchor=north west][inner sep=0.75pt]   [align=left] {$\Omega$};

\end{tikzpicture}
    \caption{\small The $(d+1)$-dimensional bulk spacetime $\mathcal{M}$ has a timelike boundary $\partial\mathcal{M}$. $\partial\Sigma$ is a $(d-1)$-dimensional Cauchy surface of $\partial\mathcal{M}$. The bulk domain of dependence $\Omega$ of $\partial\Sigma$ is represented by the shaded region. All bulk Cauchy surfaces anchored on $\partial\Sigma$—that is, sharing the same boundary—lie entirely within $\Omega$.}
    \label{AdStincan1}
\end{figure}

Let $\mathcal{M}$ be a $(d+1)$-dimensional AAdS spacetime with a $d$-dimensional asymptotic timelike boundary $\partial\mathcal{M}$, and let $\partial\Sigma$ be a $(d-1)$-dimensional Cauchy surface of $\partial\mathcal{M}$. We now briefly recall prior definitions: the bulk domain of dependence $\Omega$ of $\partial\Sigma$ is the domain of dependence of any bulk Cauchy surface $\Sigma$ whose boundary is $\partial\Sigma$ (see Figure \ref{AdStincan1}); the covariant phase space is the set of all solutions to the bulk equations of motion (i.e., classical saddles) in $\Omega$, modulo gauge redundancies; equivalently, in the ADM formalism, the phase space is the set of all initial-value data $(g_{ab},\Pi^{ab},\Phi^i,\Pi_{\Phi^i})$ on $\Sigma$, subject to the constraints $\mathcal{H}$, $D_a$, and $G^A$; and a maximal volume slice $\Sigma_{K=0}$ is a bulk Cauchy surface on which the trace of the extrinsic curvature vanishes everywhere, i.e., $K=0$.

The matter Hamiltonian $\mathcal{H}_{\text{matter}}$ depends on the fields $\Phi^i$, their conjugate momenta $\Pi_{\Phi^i}$, and the spatial metric $g_{ab}$. Under a Weyl transformation of the form $g_{ab} = \phi^\alpha \gamma_{ab}$, with $\alpha = \frac{4}{d-2}$, $\mathcal{H}_{\text{matter}}$ may, for generic matter fields, acquire dependence on derivatives of the conformal factor $\phi$, particularly through the Weyl transformation of covariant derivatives acting on matter fields. To simplify the analysis, we restrict attention to matter theories for which $\mathcal{H}_{\text{matter}}$ depends on $\phi$ only algebraically. In such cases, $\frac{2\kappa}{\sqrt{g}} \mathcal{H}_{\text{matter}}$ can be expressed as a power series in $\phi$:
\begin{align}
\frac{2\kappa}{\sqrt{g}} \mathcal{H}_{\text{matter}} = A_i \phi^{n_i} + B \phi^{-\alpha} + C_j \phi^{m_j},
\end{align}
where $n_i < -\alpha < m_j$, and the coefficients $A_i$, $B$, and $C_j$ depend only on $\Phi^i$, $\Pi_{\Phi^i}$, and $\gamma_{ab}$.


\vspace{2mm}
We make two key assumptions:
\begin{assumption} \label{assump: MVC}
For every on-shell spacetime configuration in $\Omega$, there exists a unique maximal volume slice.
\end{assumption}

\begin{assumption} \label{assump: matter Hamiltonian}
    The matter Hamiltonian $\mathcal{H}_{\text{matter}}$ depends on the conformal factor $\phi$ algebraically, and the coefficients in its power series expansion satisfy $A_i \geq 0$, $B > 2\Lambda$, and $C_j \leq 0$, with the weaker bound $C_j < -2\Lambda$ permitted when $m_j = 0$.
\end{assumption}

\noindent
A maximal volume slice is expected to exist under a condition on singularities analogous to cosmic censorship \cite{Witten:2022xxp}. The uniqueness of the maximal volume slice was proven in Appendix A of \cite{Couch:2018phr} under the strict generic strong energy condition, which demands that the stress-energy tensor $T_{\mu\nu}$ of the matter fields satisfies
\begin{equation}
    T_{\hat{t}\hat{t}}  + \frac{1}{d-1} T > \frac{2\Lambda}{(d-1)\kappa} \quad \text{for all unit timelike vectors } \hat{t}.
\end{equation}
Examples of matter fields that satisfy the strict generic strong energy condition and also meet the requirements of Assumption~\ref{assump: matter Hamiltonian} include minimally coupled scalar fields whose total potential—including the contribution from the cosmological constant—is strictly negative, electromagnetic fields, Yang–Mills fields, and abelian $p$-form gauge fields.


\vspace{2mm}
We now state the first main theorem.
\begin{theorem} \label{theorem: phase space equivalence}
Under Assumptions \ref{assump: MVC} and \ref{assump: matter Hamiltonian}, the following two phase spaces 
\begin{itemize}
    \item the ADM phase space $(g_{ab},\Pi^{ab},\Phi^i,\Pi_{\Phi^i})$, subject to the first-class constraints $\mathcal{H}, D_a, G^A$;
    \item the alternative phase space $(\mathfrak{g}_{ab},\mathsf{\Pi}^{ab},\Phi^i,\Pi_{\Phi^i})$, subject to the first-class constraints $\mathcal{W}+\mathcal{A}, D_a, G^A$
\end{itemize}
are physically equivalent.
\end{theorem}

\noindent
Two phase spaces with first-class constraints (i.e., gauge constraints) are said to be physically equivalent if their reduced phase spaces are symplectomorphic to each other, meaning they coincide as manifolds equipped with the same symplectic form. We use distinct notation for the metric $\mathfrak{g}_{ab}$ and its conjugate momentum $\mathsf{\Pi}^{ab}$ in the alternative phase space to distinguish them from their counterparts in the ADM phase space. We choose $\mathcal{W} + \mathcal{A}$ rather than $\mathcal{W} - i\mathcal{A}$ as one of the first-class constraints in the alternative phase space, because constraint functions must be real-valued on phase space. Both the Weyl generator $\mathcal{W}$ and the conformal anomaly $\mathcal{A}$ of the holographic CFT are real functions on phase space. When there is no risk of confusion, we refer to either expression as the Weyl-anomaly constraint, based on context. However, when a clear distinction is necessary, we refer to $\mathcal{W} + \mathcal{A}$ as the real Weyl-anomaly constraint and to $\mathcal{W} - i\mathcal{A}$ as the imaginary Weyl-anomaly constraint. Although we will prove this theorem in later sections, let us first outline the crux of the argument. The reasoning is actually quite straightforward. First, the Weyl-anomaly constraint closes under the Poisson bracket with the spatial diffeomorphism and matter-gauge constraints, thereby validating it as a first-class constraint. This Poisson closure would generally fail if one were to substitute an arbitrary function in place of the conformal anomaly of the holographic CFT. Moreover, retaining this specific form of the anomaly will be crucial for establishing the second theorem later. Second, imposing this constraint at the classical level eliminates the conformal factor $\varphi$ and its conjugate momentum $\mathsf{\Pi} = \mathfrak{g}_{ab}\mathsf{\Pi}^{ab}$ from the alternative phase space, yielding the same reduced phase space that one obtains by fixing the Hamiltonian constraint in the ADM formulation.


\vspace{2mm}
Now let us state the next main theorem.
\begin{theorem} \label{theorem: QG states}
Assume that the CFT partition function $Z^{(c)}_{\text{CFT}}[\mathfrak{g}_{ab},\Phi^i;\psi_{\text{CFT}}]$ is analytic in its central charge $c$. Then, upon analytically continuing to imaginary central charge $c \to i c$, the CFT partition function $Z^{(ic)}_{\text{CFT}}[\mathfrak{g}_{ab},\Phi^i;\psi_{\text{CFT}}]$ satisfies all of the operator gauge constraints in the alternative phase space:
\begin{equation}
    (\widehat{\mathcal{W}}+\mathcal{A}) Z^{(ic)}_{\text{CFT}} = \widehat{D}_a Z^{(ic)}_{\text{CFT}} = \widehat{G}^A Z^{(ic)}_{\text{CFT}}= 0,
\end{equation}
and thus provide candidate quantum gravity states in the alternative phase space formulation:
\begin{equation}
    \Psi_{\text{QG}}[\mathfrak{g}_{ab},\Phi^i] = Z^{(ic)}_{\text{CFT}}[\mathfrak{g}_{ab},\Phi^i;\psi_{\text{CFT}}].
\end{equation}
\end{theorem}

\noindent
The first constraint equation follows from the original CFT partition function satisfying the imaginary Weyl-anomaly constraint equation \eqref{imaginary Weyl-anomaly constraint equation}: when you send $c\to i c$ the CFT partition function becomes $Z^{(c)}_{\text{CFT}} \to Z^{(ic)}_{\text{CFT}}$ and the anomaly becomes $\mathcal{A} \to i\mathcal{A}$ (because the coefficient of the anomaly is the central charge), and so the imaginary Weyl-anomaly constraint becomes the real Weyl-anomaly constraint $(\widehat{\mathcal{W}}-i\mathcal{A}) \to (\widehat{\mathcal{W}}+\mathcal{A})$.

\section{Phase Space Reduction}\label{Phase Space Reduction}

In this section, we review the key steps of phase space reduction, assembling them into a single coherent presentation. The necessary ingredients—existence and uniqueness of maximal volume slices and of solutions to the Lichnerowicz equation—were originally established in works such as \cite{Isenberg_1995, Andersson:1992yk, Andersson_Chrusciel, Sakovich:2009nb} and later reviewed in \cite{Witten:2022xxp}, where it is shown that the reduced phase space is a cotangent bundle over the space of conformal classes of metrics modulo spatial diffeomorphisms. Our goal is not merely to rehash prior proofs, but to highlight the specific viewpoint and notation we will use throughout the remainder of the paper. In particular, we follow a slightly different ordering and emphasis than, for instance, the treatment in \cite{Witten:2022xxp}. By recasting these results in the language of first-class constraints and gauge conditions—imposing $K=0$ to fix the Hamiltonian constraint—we recover the standard reduced phase space and set the stage for the phase space enlargement in Section \ref{Phase Space Enlargement}. This discussion not only makes the translation between the ADM phase space and the alternative phase space transparent, but also lays the groundwork for the developments that follow.

\subsection{The ADM Phase Space}

Consider the action for Einstein gravity coupled to arbitrary matter fields satisfying assumptions \ref{assump: MVC} and \ref{assump: matter Hamiltonian} in an asymptotically AdS$_{d+1}$ spacetime $\mathcal{M}$ with boundary~$\partial\mathcal{M}$, subject to Dirichlet boundary conditions:
\begin{equation}
S \;=\; \frac{1}{2\kappa}
\int_{\mathcal{M}} d^{\,d+1}x\,\sqrt{-\mathbf{g}}\;\bigl(R[\mathbf{g}] \;-\; 2\Lambda\bigr)
\;+\; \frac{1}{\kappa}
\int_{\partial\mathcal{M}} d^{\,d}y\,\sqrt{-g_0}\;K_{\partial\mathcal{M}}
\;+\; S_{\text{matter}}[\mathbf{\Phi}^i,\mathbf{g}]\,,
\end{equation}
where $\mathbf{g}$ is the spacetime metric on $\mathcal{M}$, $g_0$ is the induced metric on $\partial\mathcal{M}$, and $K_{\partial\mathcal{M}}$ is the trace of the extrinsic curvature of $\partial\mathcal{M}$. The collection $\{\mathbf{\Phi}^i\}$ represents matter fields on $\mathcal{M}$, and $S_{\text{matter}}[\mathbf{\Phi}^i,\mathbf{g}]$ denotes their action—including any boundary terms required for a well-posed variational principle—which we leave unspecified to maintain full generality. However, we will only consider models that do not lead to any second-class constraints. We will keep this level of generality throughout, specializing to explicit matter Lagrangians only when needed. Most of the discussion in this section is more general than matter fields satisfying assumptions \ref{assump: MVC} and \ref{assump: matter Hamiltonian}, but the phase space reduction applies to matter fields that do satisfy these assumptions.

In the Hamiltonian formalism, we start with the phase space. The key elements of any phase space are the underlying manifold, the symplectic structure, and the constraints. These determine the physics at the kinematical level, and the dynamics is then provided by the Hamiltonian, which is a function on this phase space. The underlying manifold $\mathcal{P}_{\text{ADM}}$ of the ADM phase space is coordinatized by the set of geometries and matter fields on a hypersurface $\Sigma$, along with their conjugate momenta: ($g_{ab}, \Pi^{ab}, \Phi^i, \Pi_{\Phi^i}$). The hypersurface $\Sigma$ is an asymptotically hyperbolic (AH) manifold, so the Ricci scalar of all possible metrics $g_{ab}$ approaches $2\Lambda$ at the asymptotic boundary $\partial\Sigma$. The symplectic structure is given by the symplectic potential
\begin{equation}
\theta_{\text{ADM}} = \int_{\Sigma} d^dx (\Pi^{ab} \delta g_{ab} + \Pi_{\Phi^i} \delta \Phi^i),
\end{equation}
and equivalently by the symplectic two-form \begin{equation}
\omega_{\text{ADM}} = \delta \theta_{\text{ADM}} = \int_{\Sigma} d^dx (\delta \Pi^{ab} \wedge \delta g_{ab} + \delta \Pi_{\Phi^i} \wedge \delta \Phi^i ),
\end{equation}
where there are no boundary terms due to Dirichlet boundary conditions on all fields. Here, $\delta$ and $\wedge$ denote, respectively, the exterior derivative and wedge product on $\mathcal{P}_{\text{ADM}}$. The wedge product on spacetime will not be written explicitly and must be understood from context whenever it is present. This gives the canonical relations for the Poisson brackets among the fields:
\begin{subequations}
\begin{align}
\{g_{ab}(x),\,\Pi^{cd}(y)\} 
&= \delta_{ab}^{cd}\,\delta^{(d)}(x,y), \\[6pt]
\{\Phi^i(x),\,\Pi_{\Phi^j}(y)\} 
&= \delta^i_{\;j}\,\delta^{(d)}(x,y),
\end{align}
\end{subequations}
where $\delta_{ab}^{cd} := \frac{1}{2}\bigl(\delta_{a}^{c}\,\delta_{b}^{d} + \delta_{a}^{d}\,\delta_{b}^{c}\bigr)$, and all other Poisson brackets among the fields vanish.

The constraints are the Hamiltonian constraints
\begin{align}
    \mathcal{H}(x) &= \frac{2\kappa}{\sqrt{g}}\!\Big(\Pi_{ab}\Pi^{ab}-\frac{1}{d-1}\Pi^2\Big)\!-\frac{\sqrt{g}}{2\kappa}(R-2\Lambda) + \mathcal{H}_{\text{matter}}(x) = 0,
\end{align}
the spatial diffeomorphism constraints (also known as momentum constraints or diff constraints)
\begin{align}
    D_a(x) = -2 g_{ab} \nabla_c \Pi^{bc} + g_{ab} D^b_{\text{matter}}(x)=0,
\end{align}
and the matter gauge constraints
\begin{equation}
    G^A(x)=0.
\end{equation}
The specific form of $\mathcal{H}_{\text{matter}}$, $D^a_{\text{matter}}$, and $G^A$ depends on the matter fields. We assume that $\mathcal{H}_{\text{matter}}$ is bounded from below. Here, $G^{A}(x)$ denotes a tensor density of weight $1$ (carrying any necessary spatial tensor indices), and the index $A$ runs over the internal Lie-algebra generators. In the smeared form, the constraints are
\begin{subequations}
\begin{align}
    \mathcal{H}[N] &= \int_{\Sigma} d^dx \, N(x)  \, \mathcal{H}(x),\\
    D[N^a]&= \int_{\Sigma} d^dx \, N^a(x) \, D_a(x),\\
    G[\alpha] &= \int_{\Sigma} d^dx \, \alpha_A(x) \, G^A(x),
\end{align}
\end{subequations}
where the smearing functions (or tensors) $N$, $N^a$, and $\alpha$, respectively called the lapse, shift, and matter gauge parameters, are taken to be arbitrary but with compact support—meaning they are required to vanish at the boundary.\footnote{When these parameters don't vanish at the boundary, they are no longer true gauge transformations (i.e., redundancies in the theory), but are called large-gauge transformations and are symmetries of the theory.} The matter gauge parameter $\alpha$ could be any Lie-algebra-valued tensor field depending on the theory. Invoking index notation, it would be denoted as $\alpha^{A\,a_{1}\dots a_{r}}{}_{b_{1}\dots b_{s}}(x)$. Since these are arbitrary in the interior of $\Sigma$, they imply the constraints in the local form.

Let the Poisson bracket algebra among the constraints be\footnote{Throughout, we restrict attention to matter sectors that do not disturb this ADM constraint algebra. Given a set of $n$ constraints, one can redefine the constraints by taking any set of $n$ linearly independent linear combinations of them. The resulting constraints define the same constraint surface, but the Poisson bracket algebra among them may change. The specific form in which one expresses the constraints is a choice. The form we adopt is preferred because it renders the diffeomorphism constraints $D_a$ as generators of pure spatial diffeomorphisms. Alternatively, one could define momentum constraints $\widetilde{D}_a$ as the normal-tangential components of the Einstein equations. These would commute with the matter-gauge constraints, i.e., $\{\widetilde{D}[N], G[\alpha]\} = 0$, because the Einstein equations are matter-gauge invariant. However, $\widetilde{D}_a$ would then generate a combination of spatial diffeomorphisms and matter-gauge transformations, rather than spatial diffeomorphisms alone.}
\begin{subequations} \label{PBrelations}
\begin{align} 
\{\mathcal H[N],\,\mathcal H[M]\} 
&=  D\!\bigl[g^{ab}(N\,\partial_{b}M - M\,\partial_{b}N)\bigr],\label{PBHH} \\[6pt]
\{D[N^{a}],\,\mathcal H[M]\} 
&= \mathcal H\!\bigl[\mathcal L_{N}M\bigr],\label{PBDH} \\[6pt]
\{D[N^{a}],\,D[M^{a}]\} 
&= D\!\bigl[(\mathcal L_{N}M)^{a}\bigr] ,\label{PBDD} \\[6pt]
\{D[N^{a}],\,G[\alpha]\} 
&= G\!\bigl[\mathcal L_{N}\alpha\bigr], \label{PBDG}\\[6pt]
\{G[\alpha],\,G[\beta]\} 
&= G\!\bigl[[\alpha,\beta]\bigr],\label{PBGG}\\[6pt]
\{\mathcal H[N],\,G[\alpha]\} 
&=0. \label{PBHG}
\end{align}
\end{subequations}
Here, $\mathcal{L}_{N}$ is the Lie derivative with respect to the vector field $N^a$. The Lie derivatives of a scalar field $M$ and a vector field $M^a$ with respect to a vector field $N^a$ are $\mathcal{L}_N M = N^a \partial_a M$ and $(\mathcal{L}_N M)^a = [N, M]^a = N^b \partial_b M^a - M^b \partial_b N^a$, respectively. The bracket $[\alpha, \beta]$ denotes the Lie bracket of the gauge parameters, defined appropriately based on the nature of the matter fields.

Equation \eqref{PBHH} is necessarily satisfied by any locally Lorentz-invariant theory with a Hamiltonian description (modulo redefinitions of the fields and constraints). Equations (\ref{PBDH}--\ref{PBDG}) are the consequences of the diffeomorphism constraint generating spatial diffeomorphisms. Equation \eqref{PBGG} is the standard closure relation among the matter gauge constraints. The Hamiltonian constraint arises from the normal–normal component of the Einstein equation, where the normal is to the spatial hypersurfaces. Since the matter stress-energy tensor is defined as the variation of the matter action with respect to the metric, and the matter action is invariant under its internal gauge symmetries, the stress-energy tensor—and hence the Hamiltonian constraint—is also invariant under those gauge transformations. This invariance leads directly to equation \eqref{PBHG}.

Let us now express all components of the ADM phase space $\Gamma_{\text{ADM}}$ together as
\begin{equation}
    \Gamma_{\text{ADM}} = \left(\,\mathcal{P}_{\text{ADM}}, \,\omega_{\text{ADM}}\ ;\ \mathcal{H}, \, D_a,\,G^A\,\right).
\end{equation}

A constraint surface is the submanifold of the phase space where all the constraints vanish. Let $\mathcal{C}_{\mathcal{H},D,G} \subset \mathcal{P}_{\text{ADM}}$ be the constraint surface where all the constraints—$\mathcal{H}[N]$, $D[N^a]$, and $G[\alpha]$—vanish. A constraint is said to be first-class if its Poisson brackets with all other constraints vanish weakly, meaning they vanish on the constraint surface; otherwise, the constraint is said to be second-class. The constraints $\mathcal{H}[N]$, $D[N^a]$, and $G[\alpha]$ are first-class, as their Poisson brackets close among themselves and hence vanish weakly. According to Dirac's conjecture, all first-class constraints generate gauge transformations and are also called the gauge constraints. See Section~1.2 of \cite{Henneaux:1992ig} for a discussion of Dirac’s conjecture.

The extended Hamiltonian is
\begin{equation}\label{extendedHamiltonian}
    H_{\text{ADM}} = \int_{\Sigma} d^dx \left( N \mathcal{H} + N^a D_a + \alpha_A G^A \right) + H_{\text{bdy}} = \mathcal{H}[N] + D[N^a] + G[\alpha] + H_{\text{bdy}},
\end{equation}
where the first integral term enacts pure gauge deformations of the bulk slice—specifically, \(N\,\mathcal{H}\) generates time shifts of \(\Sigma\), \(N^a\,D_a\) produces spatial diffeomorphisms along \(\Sigma\), and \(\alpha_A\,G^A\) implements internal gauge rotations—each leaving \(\partial\Sigma\) fixed as long as \(N\), \(N^a\), and \(\alpha_A\) vanish on the boundary. By contrast, \(H_{\mathrm{bdy}}\) is the genuine boundary ADM Hamiltonian (the specific form depends on the theory): it alone translates \(\partial\Sigma\) forward or backward in time.\footnote{See \cite{Hayward:1992ix} for a derivation of the boundary term $H_{\text{bdy}}$ in gravity.} In particular, setting \(N|_{\partial\Sigma} = 0\) ensures that the boundary stays fixed while the rest of \(\Sigma\) moves in time, whereas allowing a nonzero lapse at \(\partial\Sigma\) advances the entire slice—including its boundary—in time. 

Before moving on, note a subtle point: it is straightforward to interpret “time-translation” of the boundary slice—the boundary metric is fixed, so pushing $\partial\Sigma$ along the boundary time direction is unambiguous. By contrast, the statement that the bulk Hamiltonian constraint “moves” the interior slice $\Sigma$ requires more care, since the bulk spacetime metric is dynamical. In other words, what does it really mean to deform a dynamical hypersurface in a fluctuating geometry? We will now explain this by focusing on the gauge orbits of the Hamiltonian and diff constraints.

\subsection{Gauge Orbits of \texorpdfstring{$\mathcal{H} \ \text{and} \ D_a$}{}}\label{Gauge orbits of H and D}

The gauge orbits of $\mathcal{H}$ and $D_a$ can be easily understood by translating between the covariant phase space and the ADM phase space. The covariant phase space is the solution space, modulo gauge redundancies. By solution space, we refer to the set of all solutions to the bulk equations of motion and boundary conditions in $\Omega$ (the domain of dependence of a given boundary Cauchy surface $\partial\Sigma$). Given an element of the solution space, one can take any bulk Cauchy surface embedded within $\Omega$ (and thus anchored on $\partial\Sigma$), and then take the induced data on that Cauchy surface to obtain a point on $\mathcal{C}_{\mathcal{H},D,G}$. 

Via this translation, all different choices of these Cauchy surfaces yield different points on $\mathcal{C}_{\mathcal{H},D,G}$ that are related by gauge transformations. Now, given a point on $\mathcal{C}_{\mathcal{H},D,G}$—i.e., a valid initial value dataset—one can use the extended Hamiltonian in \eqref{extendedHamiltonian} with a particular choice of the lapse, shift, and matter gauge parameter to evolve the initial data and construct a solution in $\Omega$, thereby obtaining a point in the solution space. Different choices of lapse, shift, and matter gauge parameters correspond to different gauge choices and thus yield different points in the solution space that are all related by gauge transformations.

So, the mapping between the covariant phase space and the ADM phase space is a one-to-one correspondence between the gauge orbits on $\mathcal{C}_{\mathcal{H},D,G}$ and the points in the covariant phase space (i.e., the solution space modulo gauge redundancies). It is in this context that we can clearly explain what is meant by the statement that the Hamiltonian constraint shifts the interior of the Cauchy surfaces up or down in the bulk time.

By a direct computation, we can see how the metric changes under the gauge transformations generated by the Hamiltonian and diff constraints:
\begin{align}
    \delta_N g_{ab}(x) &= \{g_{ab}(x),\mathcal{H}[N]\} =2N  \frac{2\kappa}{\sqrt{g}}  \left(\Pi_{ab} -\frac{1}{d-1} g_{ab} \Pi \right) ,\label{metricchangeunderH}\\
    \delta_{N^a} g_{ab}(x) &= \{g_{ab}(x),D[N^c]\} = \nabla_a N_b + \nabla_b N_a = (\mathcal{L}_N g)_{ab}.
\end{align}
The second equation makes it clear that the diff constraint changes the metric by a diffeomorphism on $\Sigma$ (spatial diffeomorphism) generated by the vector field $N^a$ on $\Sigma$. We can give the first equation a clear interpretation by embedding $\Sigma$ in an on-shell spacetime configuration. Consider a point in the solution space, and take a Cauchy surface in $\Omega$ with induced metric $g_{ab}(x)$. The quantity $\Pi^{ab}$ on $\Sigma$ is related to the extrinsic curvature $K_{ab}$ of $\Sigma$ (as embedded in the on-shell spacetime configuration in $\Omega$) by
\begin{equation}
    \Pi^{ab} = \frac{\sqrt{g}}{2\kappa} \left(K^{ab} - g^{ab} K \right),
\end{equation}
and so the first equation \eqref{metricchangeunderH} becomes
\begin{align}
   \delta_N g_{ab}(x) &= 2N  K_{ab} = N (\mathcal{L}_n g)_{ab},
\end{align}
where $n$ is the unit normal to $\Sigma$ in $\Omega$. In plain terms, this means the metric changes exactly as it would if we had pushed the entire slice $\Sigma$ forward in time along $n$ at a “speed” $N$ in the chosen on-shell spacetime configuration. Since $N$ vanishes on $\partial\Sigma$, that boundary remains fixed while the interior of $\Sigma$ moves in time. This is not unexpected, because this evolution is nothing but time evolution using the Hamilton equations of motion with a particular choice of shift—namely, zero.

All other fields on $\Sigma$ change in a similar fashion. This is what is meant by the statement that the Hamiltonian constraint generates temporal diffeomorphisms. With this picture in mind, what the Hamiltonian constraint does is move $\Sigma$ “forward or backward in time,” and the diff constraint moves $\Sigma$ “sideways” (i.e., tangential to $\Sigma$) when embedded in an on-shell configuration in $\Omega$. These are the gauge orbits of these constraints.

\subsection{Gauge Fixing the Hamiltonian Constraint}


Fixing first-class (gauge) constraints involves two steps:
\begin{enumerate}[label=(\roman*)]
    \item Solving the constraint equations to restrict the system to the constraint surface;
    \item Quotienting the constraint surface by the gauge transformations, or equivalently, selecting a gauge slice within the constraint surface.
\end{enumerate}

A gauge slice is a submanifold in the phase space that intersects \emph{all} of the gauge orbits on the constraint surface \emph{exactly once}. To select a gauge slice, one introduces a gauge-fixing condition—an additional function on phase space, chosen by hand, whose vanishing defines the slice. The number of gauge-fixing conditions introduced must equal the number of first-class constraints in the theory, and they must also be linearly independent from each other; only then is all gauge freedom fully fixed. Otherwise, some residual gauge symmetry will remain, and the fixing will be only partial.

Once introduced, these gauge-fixing conditions are treated as additional constraints in phase space. As a result, the original first-class constraints, along with the newly added gauge-fixing conditions, all become second-class, since the Poisson brackets between the gauge constraints and the corresponding gauge-fixing conditions do not vanish on the constraint surface. This non-vanishing reflects the fact that the gauge-fixing conditions are constructed specifically to break the gauge symmetry: they are designed so that gauge transformations generated by the gauge constraints move the system off the gauge slice, ensuring a unique intersection.

Both of the above steps present challenges. In general, it is difficult to solve all the constraint equations: for instance, the full set of solutions to the Hamiltonian and diffeomorphism constraints is not known, so the complete structure of the constraint surface remains unknown. Furthermore, a global gauge-fixing condition may not even exist—and in fact, it does not, even in simpler and well-understood cases such as Yang–Mills theory. 

In such cases, one can only perform \emph{local} gauge fixing: that is, in a neighborhood of a particular point on the constraint surface, there exists a function whose vanishing defines a local gauge slice in phase space that intersects each gauge orbit uniquely within that neighborhood. However, the local gauge slice might intersect these gauge orbits again outside that neighborhood, or might fail to intersect some gauge orbits that do not pass through this neighborhood. These failures of the gauge slice are known as Gribov obstructions. This forces one to remain within the confines of perturbation theory around a fixed point, in which case it is acceptable to proceed with local gauge-fixing conditions.

Due to challenges like these—and the desire to maintain manifest gauge symmetry—one often avoids gauge fixing at the classical level and proceeds directly to quantization. However, this introduces other challenges, particularly those arising from the Hamiltonian constraint.

But it turns out that there exists a scenario in which one can partially gauge-fix the phase space by gauge-fixing only the Hamiltonian constraint, while retaining all other constraints as gauge constraints.

\begin{proposition} \label{Prop: MVS} Suppose that for every on-shell spacetime configuration in $\Omega$, there exists a unique maximal volume slice. Then the phase space functional  
\begin{equation}\label{gaugeconditionforHam}
\mathcal{K}[\omega] = \int_{\Sigma} d^d x \, \omega(x) \, \Pi(x)
\end{equation}
defines a valid gauge-fixing condition \emph{solely} for the Hamiltonian constraint, where $\omega(x)$ is an arbitrary compactly supported smearing function on $\Sigma$.
\end{proposition}

\noindent
Here, by “solely,” we mean that the rest of the gauge transformations (spatial diffeomorphisms and matter gauge transformations) are left unaffected by this gauge condition. Note that the above condition is just the smeared form of the local condition $\Pi(x) = 0$ everywhere on $\Sigma$. Let us now prove this proposition.
\begin{proof}
    Take any point $p$ on the constraint surface $\mathcal{C}_{\mathcal{H},D,G}$ and, with any choice of lapse, shift, and matter gauge parameter, evolve this initial data using the extended Hamiltonian to construct an on-shell spacetime configuration $s$ in $\Omega$. Let $\Sigma_p$ be the starting Cauchy surface in $\Omega$ on which the induced data from $s$ is $p$. As we explained in Subsection~\ref{Gauge orbits of H and D}, the gauge transformations generated by the Hamiltonian constraints time-translate the Cauchy surfaces in $\Omega$ along their normal direction. Therefore, the gauge orbit of $p$ is the set of all valid initial value data on all Cauchy surfaces (induced from $s$) in $\Omega$ that are time-translated along their normal direction starting from $\Sigma_p$.

    A maximal volume slice is a Cauchy surface on which $K = 0$ everywhere, or equivalently, $\Pi = 0$ everywhere—or, in smeared form, $\mathcal{K}[\omega] = 0$ for all $\omega$. Now, by our assumption, there exists a unique maximal volume slice $\Sigma_{K=0}$ in $\Omega$, and so the gauge-fixing condition is satisfied by exactly one point in the gauge orbit of $p$. This means the gauge slice defined by the above gauge-fixing condition intersects the gauge orbit of $p$ exactly once. Therefore, the condition \eqref{gaugeconditionforHam} is indeed a valid gauge condition to fix the temporal diffeomorphisms, i.e., the gauge transformations generated by the Hamiltonian constraint.

    If existence failed, then the gauge slice would not intersect the gauge orbit at all. If uniqueness failed, then the slice would intersect it more than once. Either case renders the condition invalid. This is why the existence and uniqueness of maximal volume slices are essential.

    After introducing this gauge-fixing condition into the phase space—now as an additional constraint— we must restrict to the new constraint surface $\mathcal{C}_{\mathcal{H},D,G,\mathcal{K}} \subset \mathcal{P}_{\text{ADM}}$, where $\mathcal{K}$ also vanishes in addition to $\mathcal{H}$, $D$, and $G$. Alongside \eqref{PBrelations}, the additional Poisson bracket relations involving $\mathcal{K}$ are:
    \begin{align}
        \{\mathcal{H}[N],\mathcal{K}[\omega]\} &\not\approx 0, \quad  
        \{D[N^a],\mathcal{K}[\omega]\} = \mathcal{K}[\mathcal{L}_N \omega], \quad   
        \{G[\alpha],\mathcal{K}[\omega]\} = 0, \quad 
        \{\mathcal{K}[\omega],\mathcal{K}[\tilde{\omega}]\} = 0,
    \end{align}
    where $\not\approx 0$ (respectively, $\approx 0$) indicates non-vanishing (respectively, vanishing) on the constraint surface $\mathcal{C}_{\mathcal{H},D,G,\mathcal{K}}$.

    Thus, $\mathcal{H}$ now becomes a second-class constraint along with $\mathcal{K}$, while $D$ and $G$ remain first-class constraints. Hence, the gauge condition solely gauge-fixes the Hamiltonian constraint, leaving the diffeomorphism and matter gauge constraints intact as gauge constraints. This completes the proof of the proposition.
\end{proof}

The next step is to solve the Hamiltonian constraint and the gauge condition in order to eliminate them from the phase space. Solving $\mathcal{K} = 0$ is trivial, as it simply restricts $\Pi^{ab}$ to be traceless, i.e., $\Pi = 0$. Solving the Hamiltonian constraint and understanding its solution space is a non-trivial task. That is, the surface $\mathcal{C}_{\mathcal{H}} \subset \mathcal{P}_{\text{ADM}}$, where only $\mathcal{H}$ vanishes, is not fully known.

However, it turns out that the solutions to the Hamiltonian constraint are well understood on maximal volume slices—precisely on the gauge slice. In other words, the Hamiltonian constraint can be solved in the special case $K = 0$, and the surface $\mathcal{C}_{\mathcal{H},\mathcal{K}}$ is well understood. To proceed, one first expresses the Hamiltonian constraint in a different set of variables obtained via the conformal decomposition.

For $d>2$, in the conformal decomposition, the metric is expressed as
\begin{equation}\label{conformal decomposition equation}
    g_{ab} = \phi^\alpha \gamma_{ab},
\end{equation}
where $\alpha = \frac{4}{d-2}$, $\phi$ represents the conformal factor (a positive scalar field), and $\gamma_{ab}$ is the conformal part of the metric on $\Sigma$, chosen in some specific way. The discussion for the $d = 2$ case is deferred to Appendix~\ref{appendix d=2} to keep the notation for the $d > 2$ case simple, but everything holds in a similar fashion for that case as well. The conformal factor $\phi$ is required to approach 1 at the boundary: $\phi|_{\partial \Sigma} = 1$. This is because $\Sigma$ is an AH manifold, so all metrics $g_{ab}$ have their Ricci scalar approaching $2\Lambda$ at the boundary $\partial \Sigma$, leaving no room for Weyl transformations at the boundary. Under this transformation, the Ricci scalar transforms as
\begin{align}\label{Weyl transformation of Ricci scalar}
    \phi^\alpha R_{[g]}  &= R_{[\gamma]} - \frac{4(d-1)}{\phi(d-2)} \nabla^2_{[\gamma]} \phi ,
\end{align}
where $R_{[g]}$ and $R_{[\gamma]}$ are the Ricci scalars of the metrics $g_{ab}$ and $\gamma_{ab}$ respectively, and $\nabla^2_{[\gamma]}$ is the Laplacian for the metric $\gamma_{ab}$.

Next, the conjugate momentum $\Pi^{ab}$ is split into a traceless part $\pi^{ab}$ and a trace part:
\begin{equation}\label{momentum split in conformal decomposition}
    \Pi^{ab} =\phi^{-\alpha} \pi^{ab} + \frac{1}{d}\Pi g^{ab}.
\end{equation}

Then, on the maximal volume slice ($\Pi = 0$), the Hamiltonian constraint $\mathcal{H} = 0$ becomes the Lichnerowicz equation:
\begin{align}\label{Lichnerowicz equation}
      \nabla^2_{[\gamma]} \phi 
      + \frac{(d-2)}{4(d-1)} \!|\pi|^2 \phi^{\frac{2-3d}{d-2}}
      - \frac{(d-2)}{4(d-1)}R_{[\gamma]}  \phi
      + \frac{\Lambda(d-2)}{2(d-1)}\phi^{\frac{d+2}{d-2}}
     + \frac{(d-2)}{4(d-1)}\frac{2\kappa}{\sqrt{g}}\mathcal{H}_{\text{matter}}\phi^{\frac{d+2}{d-2}}=0,
\end{align}
where $|\pi|^2 := \left(\frac{2\kappa}{\sqrt{\gamma}}\right)^2 \gamma_{ac} \gamma_{bd} \pi^{ab} \pi^{cd} \geq 0$.

Under certain assumptions on $\mathcal{H}_{\text{matter}}$, we can now show that there exists a unique solution to this equation with the boundary condition $\phi|_{\partial \Sigma} = 1$. First, note that $\mathcal{H}_{\text{matter}}$ depends on $\Phi^i$, $\Pi_{\Phi^i}$, $\gamma_{ab}$, and $\phi$. For generic matter fields, $\mathcal{H}_{\text{matter}}$ can also depend on derivatives of $\phi$, possibly arising from Weyl-transforming covariant derivatives of the matter fields. This would complicate the analysis of the Hamiltonian constraint. Therefore, we restrict to only those matter fields for which $\mathcal{H}_{\text{matter}}$ depends on $\phi$ algebraically.

We can then express $\frac{2\kappa}{\sqrt{g}}\mathcal{H}_{\text{matter}}$ as a power series in $\phi$:
\begin{align}
 \frac{2\kappa}{\sqrt{g}}\mathcal{H}_{\text{matter}} &= A_i \phi^{n_i} + B \phi^{-\alpha} + C_j \phi^{m_j},    
\end{align}
where $n_i < -\alpha < m_j$, and the coefficients $A_i$, $B$, and $C_j$ depend only on $\Phi^i$, $\Pi_{\Phi^i}$, and $\gamma_{ab}$.

\begin{proposition}\label{Prop: Lichnerowicz}
Assume $\mathcal{H}_{\text{matter}}$ depends on the conformal factor $\phi$ algebraically, and that $A_i \geq 0$, $B > 2\Lambda$, and $C_j \leq 0$, with the weaker bound $C_j < -2\Lambda$ permitted when $m_j = 0$. Then there exists a unique solution to the Lichnerowicz equation \eqref{Lichnerowicz equation} with boundary condition $\phi|_{\partial \Sigma} = 1$.
\end{proposition}

\noindent
We will use the method of sub- and supersolutions in the proof of this proposition. This method will be explained in Section~\ref{Section Lichnerowicz Equation}, and so we defer the proof to that section.


Now, as a consequence of Propositions~\ref{Prop: MVS} and~\ref{Prop: Lichnerowicz}, the gauge fixing of the Hamiltonian constraint is successfully achieved. The ADM phase space, which was initially coordinatized by 
\begin{equation}
(g_{ab},\Pi^{ab},\Phi^i,\Pi_{\Phi^i}),
\end{equation}
is, under the conformal decomposition, coordinatized by
\begin{equation}
(\phi,\Pi,\gamma_{ab},\pi^{ab},\Phi^i,\Pi_{\Phi^i}).
\end{equation}
Then, the gauge-fixing condition \eqref{gaugeconditionforHam} fixes $\Pi = 0$, and the Lichnerowicz equation \eqref{Lichnerowicz equation} (i.e., the Hamiltonian constraint expressed in terms of the conformally decomposed phase space variables) fixes $\phi$ to be the unique solution to that equation. This eliminates both $\phi$ and $\Pi$ as independent variables, yielding a reduced phase space coordinatized by
\begin{equation}
(\gamma_{ab},\pi^{ab},\Phi^i,\Pi_{\Phi^i}).
\end{equation}

The underlying manifold of this reduced phase space is the constraint surface $\mathcal{C}_{\mathcal{H},\mathcal{K}}$, and the above variables coordinatize this space. The symplectic form $\omega^{\mathrm{red}}_{\mathrm{ADM}}$ on $\mathcal{C}_{\mathcal{H},\mathcal{K}}$ is obtained by pulling back the ADM symplectic form $\omega_{\mathrm{ADM}}$ on $\mathcal{P}_{\mathrm{ADM}}$ to the constraint surface via the inclusion map $\iota : \mathcal{C}_{\mathcal{H},\mathcal{K}} \hookrightarrow \mathcal{P}_{\mathrm{ADM}}$:
\begin{equation}
\omega^{\mathrm{red}}_{\mathrm{ADM}} = \iota^* \omega_{\mathrm{ADM}}.
\end{equation}

The unfixed gauge constraints $D_a$ and $G^A$ now constrain this reduced phase space. Putting everything together, the reduced phase space, denoted by $\Gamma^{\text{red}}_{\text{ADM}}$, is
\begin{equation}
    \Gamma^{\text{red}}_{\text{ADM}} = \left(\,\mathcal{C}_{\mathcal{H},\mathcal{K}}, \,\omega^{\mathrm{red}}_{\mathrm{ADM}}\ ;\  D_a,\,G^A\,\right).
\end{equation}

In summary, we have gauge-fixed the Hamiltonian constraint to obtain the reduced phase space from the ADM phase space:
\begin{equation}
\Gamma_{\mathrm{ADM}} \xrightarrow{\text{Gauge fixing } \mathcal{H}} \Gamma^{\mathrm{red}}_{\mathrm{ADM}}.
\end{equation}

\subsection{Lichnerowicz Equation: Existence and Uniqueness of Solutions}\label{Section Lichnerowicz Equation}


The Lichnerowicz equation \eqref{Lichnerowicz equation} can be expressed as
\begin{equation}\label{general Lichnerowicz equation}
     \nabla^2_{[\gamma]} \phi - F(\phi,x) = 0,
\end{equation}
where
\begin{align} \label{specific F}
  \frac{4(d-1)}{(d-2)} F(\phi,x) &= -  |\pi|^2 \phi^{\frac{2-3d}{d-2}}
   -  A_i \phi^{\tilde{n}_i} 
    + (R_{[\gamma]}-B)  \phi
    -  C_j \phi^{\tilde{m}_j}
    - 2\Lambda \phi^{\frac{d+2}{d-2}},
\end{align}
with $\tilde{n}_i = n_i + \frac{d+2}{d-2}$ and $\tilde{m}_j = m_j + \frac{d+2}{d-2}$, so that $\tilde{n}_i < 1 < \tilde{m}_j$.

Let us now explain the method of sub- and supersolutions.\footnote{While the current discussion is for the $d > 2$ case, the $d = 2$ case is discussed in Appendix~\ref{appendix d=2}.} This is a standard method for proving the existence of solutions to Lichnerowicz-type equations. See its use in, for example, \cite{Isenberg_1995, Sakovich:2009nb, Andersson_Chrusciel, Witten:2022xxp}.

A positive function $\phi_{-}$ is called a \emph{subsolution} if
\begin{equation}
    \nabla^2_{[\gamma]} \phi_{-} - F(\phi_{-},x) \geq 0,
\end{equation}
and a positive function $\phi_{+}$ is called a \emph{supersolution} if
\begin{equation}
    \nabla^2_{[\gamma]} \phi_{+} - F(\phi_{+},x) \leq 0.
\end{equation}

\begin{proposition}\label{Prop: supersubsolution}
    If there exists a subsolution $\phi_{-}$ and a supersolution $\phi_{+}$ such that $\phi_{-} < \phi_{+}$, then there exists a solution $\phi$ of~\eqref{general Lichnerowicz equation} satisfying $\phi_{-} \leq \phi \leq \phi_{+}$.
\end{proposition}

\noindent
We do not prove this here; a rigorous version of this result, including a detailed proof, can be found in Theorem~5.2 of \cite{Sakovich:2009nb}. See also \cite{Isenberg_1995, Sakovich:2009nb, Andersson_Chrusciel, Witten:2022xxp} for related results and a review. This proposition is stated for generic equations of the form \eqref{general Lichnerowicz equation} with arbitrary $F(\phi,x)$, and will later be applied to the specific case given in equation~\eqref{specific F}.

Now let us also explain another important result. The Ricci scalar of a given metric can be sign-indefinite, and it changes under a Weyl transformation according to equation~\eqref{Weyl transformation of Ricci scalar}. Given some metric, one can ask whether it is always possible to perform a Weyl transformation to obtain a new metric with constant Ricci scalar, and whether this transformation is unique or admits multiple solutions. This is known as the \emph{Yamabe problem}.

\begin{proposition}\label{Prop: Yamabe}
    Let $(\Sigma, g_{ab})$ be any asymptotically hyperbolic manifold. Then there exists a unique solution $\phi$ to equation~\eqref{Weyl transformation of Ricci scalar}, with boundary condition $\phi|_{\partial \Sigma} = 1$, such that the Ricci scalar $R_{[\gamma]}$ of the Weyl-transformed metric $\gamma_{ab}$ satisfies $R_{[\gamma]} = 2\Lambda$.
\end{proposition}

\noindent
We do not prove this here; a rigorous version of this result, including a detailed proof, can be found in Theorem~1.2 of \cite{Andersson:1992yk}. This essentially means that any metric $g_{ab}$ on an AH manifold $\Sigma$ can be uniquely Weyl-transformed to a metric $\gamma_{ab}$ with constant Ricci scalar equal to $2\Lambda$. Another way to state the same result is that, for any metric $g_{ab}$ on an AH manifold $\Sigma$, the conformal class of that metric contains exactly one representative with Ricci scalar equal to $2\Lambda$.

Now we can proceed to prove Proposition~\ref{Prop: Lichnerowicz}. The proof below is a condensed review of the arguments in \cite{Sakovich:2009nb, Witten:2022xxp} (with slight generalization), included here for completeness.

\begin{proof}\textbf{Existence:} We begin by proving existence. Let us choose the conformal metric $\gamma_{ab}$ in the conformal decomposition \eqref{conformal decomposition equation} to be the unique metric in the conformal class of $g_{ab}$ with Ricci scalar $R_{[\gamma]} = 2\Lambda$, which is guaranteed by Proposition~\ref{Prop: Yamabe}. Then, from our assumption $B > 2\Lambda$, it follows that $R_{[\gamma]} - B = 2\Lambda - B < 0$. Thus, all terms in $F(\phi,x)$ from equation~\eqref{specific F} with powers of $\phi$ less than 1 are negative semidefinite (since each $A_i \geq 0$ by assumption), while the term with power 1 is negative definite. Additionally, all terms in $F(\phi,x)$ with powers of $\phi$ greater than 1 are positive semidefinite, and the term with power $\frac{d+2}{d-2}$ is positive definite (since $C_j \leq 0$, with the weaker bound $C_j < -2\Lambda$ permitted when $m_j = 0$). Therefore, for a sufficiently large positive constant $c_{+} \gg 1$, we have $F(c_{+},x) > 0$ everywhere on $\Sigma$, and for a sufficiently small positive constant $c_{-} \ll 1$, we have $F(c_{-},x) < 0$ everywhere on $\Sigma$. Hence, $c_{+}$ and $c_{-}$ serve as supersolution and subsolution, respectively. However, they do not satisfy the boundary condition (specifically, they do not tend to 1 at $\partial\Sigma$).

For some neighborhood near the boundary $\partial\Sigma$, we can express the AH metric in Fefferman–Graham coordinates as
\begin{equation}\label{FG spatial coordinates} 
ds^2 = \gamma_{ab}\, dx^a dx^b =\frac{L^2}{z^2} (dz^2 + \sigma_{ij}(y,z)dy^i dy^j),
\end{equation}
where the radial coordinate $z = 0$ corresponds to the boundary, $z > 0$ is the interior of $\Sigma$, and $y^i$ are coordinates on constant-$z$ slices with metric $\sigma_{ij}$, which has the expansion
\begin{equation}\label{sigmametric expansion}
   \sigma_{ij}(y,z) = \sigma_{ij}^{(0)}(y)+z^2 \sigma_{ij}^{(1)} (y)+z^4 \sigma_{ij}^{(2)} (y)+\cdots,
\end{equation}
and $L$ is the AdS length scale, given by $L^2 = -\frac{d(d-1)}{2 \Lambda}$. The Laplacian then takes the form
\begin{align}\label{Laplacian in FG}
\nabla^2_{[\gamma]} \phi
&= \frac{z^2}{L^2}\,\partial_z^2 \phi
\;+\frac{z^2}{L^2}\;\Bigl[\frac{1}{2}\,\sigma^{ij}\,\partial_z\sigma_{ij}
         \;-\;\frac{d-2}{z}\Bigr]\,\partial_z \phi
\;+\;\frac{z^2}{L^2}\,\nabla^2_{[\sigma]} \phi,
\end{align}
where $\nabla^2_{[\sigma]}$ denotes the Laplacian for the metric $\sigma_{ij}$.

Let us choose a small $\epsilon > 0$ and two smooth monotonic interpolating functions $u_{+}(z)$ and $u_{-}(z)$ such that $u_{+}(\epsilon) = c_{+}$, $u_{+}(0) = 1$, $u_{-}(\epsilon) = c_{-}$, and $u_{-}(0) = 1$. Then let us define $\phi_{+}$ and $\phi_{-}$ as follows:
\begin{equation}
\phi_{+}(z) =
\begin{cases}
c_{+} & \text{for } z \geq \epsilon, \\
u_{+}(z) & \text{for } 0 < z < \epsilon,
\end{cases}
\qquad
\phi_{-}(z) =
\begin{cases}
c_{-} & \text{for } z \leq \epsilon, \\
u_{-}(z) & \text{for } 0 < z < \epsilon.
\end{cases}
\end{equation}

Since $\phi_+$ and $\phi_-$ only depend on $z$, we have $\nabla^2_{[\sigma]} \phi_+ = \nabla^2_{[\sigma]} \phi_- = 0$.  
Remember that $z$ increases inward from the boundary, so $\partial_z \phi_{+}(z) \geq 0$ and $\partial_z \phi_{-}(z) \leq 0$ for $0 < z < \epsilon$.  
From \eqref{sigmametric expansion}, we see that $\sigma^{ij}\,\partial_z \sigma_{ij} = 2z\,\sigma^{ij}_{(0)}\,\sigma_{ij}^{(1)} + \cdots$, where $\sigma^{ij}_{(0)}$ is the inverse of $\sigma_{ij}^{(0)}$.  
Therefore, when $0 < z < \epsilon$, for small enough $\epsilon$, the dominant term in \eqref{Laplacian in FG} is  
\[
\frac{z^2}{L^2}\,\partial_z^2 \phi - \frac{(d-2)}{L^2} z\,\partial_z \phi.
\]
Now we can choose the interpolating functions to be close enough to a linear function of $z$ and make $\partial^2_z \phi$ arbitrarily small, so that $-\frac{(d-2)}{L^2} z\,\partial_z \phi$ dominates.  
This implies $\nabla^2_{[\gamma]} \phi_{+} \leq 0$ and $\nabla^2_{[\gamma]} \phi_{-} \geq 0$ when $0 < z < \epsilon$.  

Next, for any $z$ in $0 < z < \epsilon$, by choosing a large enough (but finite) $c_+$, we can make $u_{+}(z)$ sufficiently large to have $F(u_+,x) > 0$.  
Similarly, in this interval, by choosing a small enough $c_-$, we can make $u_{-}(z)$ sufficiently close to $0$ to have $F(u_-,x) < 0$.  
So we have 
\[
\nabla^2_{[\gamma]} \phi_{+} - F(\phi_{+},x) \leq 0 \quad \text{and} \quad \nabla^2_{[\gamma]} \phi_{-} - F(\phi_{-},x) \geq 0
\]
everywhere on $\Sigma$.  

Also, $\phi_{+}|_{\partial\Sigma} = \phi_{-}|_{\partial\Sigma} = 1$.  
Thus, $\phi_{+}$ and $\phi_{-}$ are a supersolution and a subsolution, respectively, and both approach $1$ at the boundary with $\phi_{-} < \phi_{+}$.  
Then, from Proposition~\ref{Prop: supersubsolution}, it follows that there exists a solution $\phi$ to the Lichnerowicz equation \eqref{general Lichnerowicz equation} satisfying the boundary condition $\phi|_{\partial\Sigma} = 1$.  
This completes the existence proof.

\textbf{Uniqueness:} Now that we have proved existence, let us next prove uniqueness.  
Let $\phi_1$ and $\phi_2$ be two solutions to \eqref{general Lichnerowicz equation}, both approaching $1$ at the boundary.  
Then the metrics $\tilde{h}_{ab} = \phi_1^\alpha \gamma_{ab}$ and $h_{ab} = \phi_2^\alpha \gamma_{ab}$ will both satisfy the Hamiltonian constraint.  

Now consider the equation
\begin{equation}
     \nabla^2_{[h]} \tilde{\phi} - F_{h}(\tilde{\phi},x) = 0,
\end{equation}
where $\nabla^2_{[h]}$ is the Laplacian of the metric $h_{ab}$ and $F_h$ takes the same form as in equation \eqref{specific F}, but with the metric everywhere in it replaced by $h_{ab}$.  
That is, $F_h$ contains $R_{[h]}$ and terms like $A_i$ are evaluated using the metric $h$.  
This equation is just the Hamiltonian constraint evaluated for the metric $\tilde{\phi}^\alpha h_{ab}$.  
Since both $h_{ab}$ and $\tilde{h}_{ab}$ satisfy the Hamiltonian constraint, this equation has two solutions: $\tilde{\phi} = 1$ and $\tilde{\phi} = \phi_0 := \phi_1/\phi_2$.  
So $F_h(1,x) = 0$ and $\nabla^2_{[h]} \phi_0 - F_{h}(\phi_0,x) = 0$.  
This gives
\begin{align}
 0&= \frac{4(d-1)}{(d-2)} \left(\nabla^2_{[h]} \phi_0 - F_{h}(\phi_0,x) +  \phi_0 F_h(1,x)\right)\label{adding together to prove uniqueness}\\
 &= \frac{4(d-1)}{(d-2)} \nabla^2_{[h]} \phi_0 
 + \!|\pi|^2 \left(\phi_0^{\frac{2-3d}{d-2}}-\phi_0\right)
   +  A_i \left(\phi_0^{\tilde{n}_i} -\phi_0\right)
    +  C_j\left( \phi_0^{\tilde{m}_j}-\phi_0\right)
    + 2\Lambda \left(\phi_0^{\frac{d+2}{d-2}} -\phi_0\right).
\end{align}

If the maximum value of $\phi_0$ is greater than $1$, then at this maximum the right-hand side of the above equation is negative definite, violating the equation.  
So the maximum value of $\phi_0$ cannot be greater than $1$.  
If the minimum value of $\phi_0$ is less than $1$, then at this minimum the right-hand side is positive definite, again violating the equation.  
Hence, the minimum value of $\phi_0$ cannot be less than $1$.  
Therefore, $\phi_0$ must be equal to $1$ everywhere, implying $\phi_1 = \phi_2$.  
Thus, the solution to the Lichnerowicz equation \eqref{general Lichnerowicz equation} is unique.  
This completes the proof of Proposition~\ref{Prop: Lichnerowicz}.
\end{proof}

\section{Phase Space Enlargement}\label{Phase Space Enlargement}


In the previous section, we gauge-fixed the Hamiltonian constraint in the ADM phase space $\Gamma_{\text{ADM}}$ to obtain the reduced phase space $\Gamma^{\text{red}}_{\text{ADM}}$. The remaining gauge symmetries in the reduced phase space are only the spatial diffeomorphisms and matter gauge transformations. Nevertheless, constructing quantum gravity states—i.e., wavefunctionals that satisfy the quantum versions of the spatial diffeomorphism and matter-gauge constraints obtained by quantizing this classical reduced phase space—is highly non-trivial, especially if one aims to go beyond perturbative approaches, where the quantum constraint equations are solved only perturbatively. We will therefore now enlarge the phase space again by adding pure gauge degrees of freedom. At first glance, this might seem counterintuitive: why enlarge the phase space just after reducing it? The reason is that this enlargement will allow us to replace the Hamiltonian constraint with a new constraint—one that is more suitable for constructing candidate quantum gravity states. Specifically, this new constraint, together with the quantum versions of the spatial diffeomorphism and matter-gauge constraints, will allow us to write down  wavefunctionals satisfying the operator gauge constraints. In this section, we will construct this alternative phase space purely at the classical level. Then, in the next section~\ref{Quantum Gravity Wave Function}, we will carry out the quantization and identify the wavefunctionals that satisfy the full set of constraints in this alternative phase space, thereby yielding candidate quantum gravity states.

Let us briefly outline what is to come in this section. In Section~\ref{section: Weyl-Anomaly Constraint}, we introduce the Weyl-anomaly constraint $\mathcal{W} + \mathcal{A}$, which will be imposed in the alternative phase space, and explain its relation to the Hamiltonian constraint. In Section~\ref{section: Poisson Algebra in alternative phase space}, we show that the Poisson algebra among $\mathcal{W} + \mathcal{A}$, $D_a$, and $G^A$ closes among themselves, thereby validating $\mathcal{W} + \mathcal{A}$ as a first-class constraint generating gauge transformations. In Section~\ref{section: Alternative Phase Space}, we formally introduce the alternative phase space and describe all its components. Section~\ref{section: Gauge orbits of Weyl-Anomaly constraint} discusses the gauge transformations generated by $\mathcal{W} + \mathcal{A}$ and the structure of their gauge orbits. In Section~\ref{section: Gauge Fixing the Weyl-Anomaly Constraint}, we introduce a valid gauge-fixing condition for $\mathcal{W} + \mathcal{A}$. Section~\ref{section: Covariant Conformal Decomposition} introduces the covariant conformal decomposition. We then proceed, in Section~\ref{Gauge Fixing the Weyl-Anomaly Constraint}, to gauge fix the Weyl-anomaly constraint and obtain a reduced phase space from the alternative phase space. We prove that this reduced phase space is symplectomorphic to the reduced phase space $\Gamma^{\text{red}}_{\text{ADM}}$ originally obtained from the ADM phase space, thereby establishing Theorem~\ref{theorem: phase space equivalence}. In Section~\ref{section: Symplectomorphism between the two Phase Spaces}, we present a more abstract and elegant proof that the alternative phase space and the ADM phase space are symplectomorphic to each other, meaning they are physically equivalent at the kinematic level. Section~\ref{section: Translating between the two Phase Spaces} explains how to explicitly translate between the ADM and alternative phase space descriptions. Finally, in Section~\ref{section: Inheriting Dynamics from the Boundary ADM Hamiltonian}, we discuss the principle that determines the extended Hamiltonian governing dynamics in the alternative phase space.

\subsection{Real Weyl-Anomaly Constraint \texorpdfstring{$\mathcal{W}+\mathcal{A}$}{}}\label{section: Weyl-Anomaly Constraint}

Let us take a detour to AdS/CFT and then return to classical physics. In the AdS/CFT correspondence, the conformal anomaly (also known as the Weyl anomaly) of the dual $d$-dimensional CFT (living on $\partial\mathcal{M}$) is related to the Hamiltonian constraint in the $(d+1)$-dimensional bulk gravity theory. Let us elaborate on how this connection arises.

\begin{figure}[htbp]
\centering
\begin{tikzpicture}[scale=1.5]
\draw (0,0) ellipse (1.25 and 0.3);
\draw (-1.25,0) -- (-1.25,-3.5);
\draw (1.25,-3.5) -- (1.25,0) node [midway] (N) {}; 
\draw (0,-3.5) ellipse (1.25 and 0.3);
\draw (0,0) ellipse (1.25*1.5 and 0.3*1.5);
\draw (-1.25*1.5,0) -- (-1.25*1.5,-3.5);
\draw (1.25*1.5,-3.5) -- (1.25*1.5,0) node [midway] (M) {}; 
\draw (0,-3.5) ellipse (1.25*1.5 and 0.3*1.5);
\fill [gray,opacity=0.2] (-1.25,0) -- (-1.25,-3.5) arc (180:360:1.25 and 0.3) -- (1.25,0) arc (0:180:1.25 and 0.3);
\draw (M) to  (N)[->,>=stealth'];
\node at (M) [label={[xshift=-4.5mm, yshift=-1mm]$\epsilon$}] {};
\node at (M) [label={[xshift=4mm, yshift=7mm]$\partial\mathcal{M}$}] {};
\node at (N) [label={[xshift=4.5
mm, yshift=7mm]$\partial\mathcal{M}_\epsilon$}] {};
\node at (N) [label={[xshift=-18
mm, yshift=7mm]$\mathcal{M}_\epsilon$}] {};
\node at (N) [label={[xshift=-18
mm, yshift=-15mm]$\mathbf{g},\mathbf{\Phi}$}] {};
\node at (N) [label={[xshift=4
mm, yshift=-15mm]$\bar{g},\bar{\Phi}$}] {};
\end{tikzpicture}
\caption{\small The shaded region denotes the finite spacetime region $\mathcal{M}_\epsilon$ with a timelike boundary $\partial\mathcal{M}_\epsilon$ that approaches the conformal boundary $\partial\mathcal{M}$ as $\epsilon \to 0$. The spacetime metric $\mathbf{g}$ and matter fields $\mathbf{\Phi}$ in $\mathcal{M}_\epsilon$ satisfy Dirichlet boundary conditions on $\partial\mathcal{M}_\epsilon$, with the induced metric and matter fields fixed to $\bar{g}$ and $\bar{\Phi}$, respectively.}
\label{radial flow}
\end{figure}
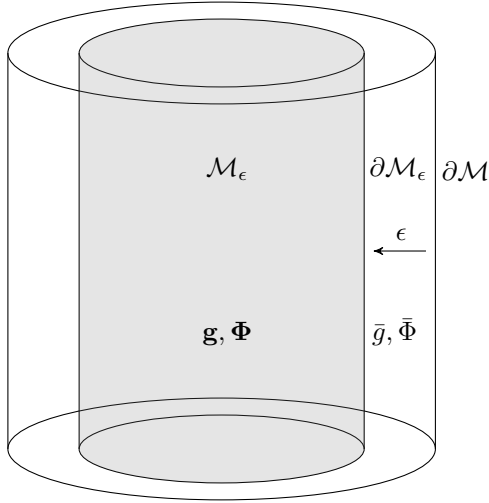

Consider a finite region of spacetime $\mathcal{M}_\epsilon \subset \mathcal{M}$, with boundary $\partial\mathcal{M}_\epsilon$. By “finite,” we mean finite in space, not in time. As $\epsilon \to 0$, the finite boundary $\partial\mathcal{M}_\epsilon$ approaches the asymptotic boundary $\partial\mathcal{M}$. See Figure~\ref{radial flow}. Consider the gravitational path integral:
\begin{align}\label{gravitational path integral definition}
    Z_{\text{grav}}[\bar{g},\bar{\Phi}] = \int \frac{D\mathbf{g} \, D\mathbf{\Phi}}{\text{G}(\mathcal{M}_\epsilon)} e^{iS[\mathbf{g},\mathbf{\Phi}]} \quad \text{with} \quad \mathbf{g}|_{\partial\mathcal{M}_\epsilon} = \bar{g}, \ \mathbf{\Phi}|_{\partial\mathcal{M}_\epsilon} = \bar{\Phi},
\end{align}
where $\mathbf{g}$ and $\mathbf{\Phi}$ now represent the spacetime metric and matter fields (let this collectively denote all the matter fields in consideration) in $\mathcal{M}_\epsilon$, satisfying the Dirichlet boundary conditions: $\mathbf{g}|_{\partial\mathcal{M}_\epsilon} = \bar{g}, \ \mathbf{\Phi}|_{\partial\mathcal{M}_\epsilon} = \bar{\Phi}$ with boundary metric $\bar{g}$ and boundary matter fields $\bar{\Phi}$ on $\partial\mathcal{M}_\epsilon$. The volume of the gauge group (spacetime diffeomorphisms and any matter gauge transformations in $\mathcal{M}_\epsilon$) is formally denoted by $\text{G}(\mathcal{M}_\epsilon)$, and $S[\mathbf{g},\mathbf{\Phi}]$ is the total action. This gravitational path integral satisfies the radial WDW equation as a consequence of invariance under radial diffeomorphisms:
\begin{align}\label{radialWDWequation}
    \widehat{\mathcal{H}}^{\text{radial}} Z_{\text{grav}}[\bar{g},\bar{\Phi}] = 0.
\end{align}

When the finite boundary $\partial\mathcal{M}_\epsilon$ approaches the asymptotic boundary $\partial\mathcal{M}$ in the limit $\epsilon \to 0$, the action $S[\mathbf{g},\mathbf{\Phi}]$ diverges and the gravitational path integral $Z_{\text{grav}}$ becomes divergent. To absorb these divergences, holographic counterterms $\widehat{CT}$ are introduced, yielding the renormalized path integral:
\begin{align}\label{renormalised gravitational path integral definition}
    Z^{\text{ren}}_{\text{grav}}[\bar{g},\bar{\Phi}] = e^{-\widehat{CT}[\bar{g},\bar{\Phi},\widehat{\Pi}_{\bar{g}},\widehat{\Pi}_{\bar{\Phi}}]} Z_{\text{grav}}[\bar{g},\bar{\Phi}].
\end{align}
The holographic counterterms depend on $\bar{g}, \bar{\Phi}$ and, in full generality, may also depend on $\widehat{\Pi}_{\bar{g}}^{\bar{a}\bar{b}} = -i \frac{\delta}{\delta \bar{g}_{\bar{a}\bar{b}}}$ and $\widehat{\Pi}_{\bar{\Phi}^i} = -i \frac{\delta}{\delta \bar{\Phi}^i}$. Here, $\bar{a}$ and $\bar{b}$ label the boundary coordinates. Equivalently, one can think of the holographic counterterms as being added directly to the action as $S_{\text{CT}}[\bar{g},\bar{\Phi},\Pi_{\bar{g}},\Pi_{\bar{\Phi}}]$, where $\Pi_{\bar{g}}$ and $\Pi_{\bar{\Phi}}$ depend on the boundary values of derivatives of the metric and matter fields with respect to the normal to the boundary.

One of the main statements of the AdS/CFT conjecture is that the renormalised gravitational path integral, in the limit where the finite boundary is taken to infinity, equals the partition function of the dual CFT:
\begin{align}\label{main statement of AdS/CFT duality}
   \lim_{\epsilon \to 0} \widetilde{Z}_{\text{grav}}[\epsilon ^{\Delta_g} g_0,\epsilon ^{\Delta_\Phi} \Phi_0] = Z_{\text{CFT}}[g_0,\Phi_0].
\end{align}

When the finite boundary $\partial\mathcal{M}_\epsilon$ is taken to infinity, the metric and matter fields on it reach their asymptotic scaling form:
$\bar{g} = \epsilon^{\Delta_g} g_0$ and $\bar{\Phi} = \epsilon^{\Delta_\Phi} \Phi_0$.
Here, $\Delta_g$ and $\Delta_\Phi$ are the conformal dimensions of the metric and matter fields, respectively. By convention, we take $\Delta_g = -2$.
$g_0$ and $\Phi_0$ are the metric and matter field defined on the conformal boundary (i.e., the boundary after conformal compactification used to construct Penrose diagrams). 
$g_0$ and $\Phi_0$ now act as background sources for the dual $d$-dimensional CFT.\footnote{While the background metric is commonly recognized as a source for the CFT, other sources are often misunderstood. 
It is a common misconception that turning on sources other than the metric breaks conformal symmetry and renders the theory non-conformal. This is incorrect. 
When coupling the CFT to a matter source $\Phi_0$, one must rescale $\Phi_0$ appropriately under Weyl transformations—just as one does with the metric $g_0$. 
The metric is not unique in this respect; all sources must transform consistently to preserve conformal symmetry.} 

The CFT may not have a Lagrangian description; nonetheless, for the sake of explanation, let us assume it has a classical action $S_{\text{CFT}}[\chi;g_0,\Phi_0]$, 
where $\chi$ denotes the dynamical field content of the CFT, coupled to the background metric $g_0$ and other sources $\Phi_0$. 
By conformal invariance at the classical level, one means that when the dynamical fields $\chi$ are locally rescaled, and all sources $g_0$ and $\Phi_0$ are simultaneously rescaled, the classical action remains invariant:
\begin{equation}
    S_{\text{CFT}}[\chi;g_0,\Phi_0] =  S_{\text{CFT}}[e^{\Delta_\chi \omega}\chi;e^{\Delta_{g} \omega}g_0,e^{\Delta_{\Phi} \omega}\Phi_0],
\end{equation}
where $\Delta_\chi$ is the conformal dimension of $\chi$, and $\omega$ is an arbitrary smooth function on the conformal boundary. 
This conformal symmetry is broken at the quantum level by the path integral measure, leading to the Weyl-anomaly equation:
\begin{align}
  \left(\widehat{\mathcal{W}}_0(x)-i\mathcal{A}_0(x)\right) Z_{\text{CFT}}[g_0,\Phi_0] = 0.  
\end{align}
Here, the Weyl generator is 
$\widehat{\mathcal{W}}_0 = -\Delta_g \,\widehat{\Pi}_{g_0} -\Delta_{\Phi^i}\, \Phi^i_0 \,\widehat{\Pi}_{\Phi^i_0}
= 2 \,\widehat{\Pi}_{g_0} -\Delta_{\Phi^i}\, \Phi^i_0 \,\widehat{\Pi}_{\Phi^i_0}$, 
since $\Delta_g=-2$ by convention, and $\mathcal{A}_0$ is the conformal anomaly. 
The subscript $0$ simply denotes that these are boundary quantities. Note that the Weyl generator also locally rescales the matter sources in the CFT. 
In the CFT partition function $Z_{\text{CFT}}$, the dynamical field contents $\chi$ are integrated over, 
and it is the measure of this path integral that breaks conformal symmetry, giving rise to the conformal anomaly. 
Even for CFTs that do not have a Lagrangian description, they still satisfy this equation.

The conformal anomaly $\mathcal{A}_0$ is some local function of the metric $g_0$ and matter sources $\Phi_0$. 
Also, observe the presence of the factor ``$i$" in this equation. 
Later, we will encounter a similar-looking equation without this ``$i$", 
so let us refer to the above equation as the ``imaginary'' Weyl-anomaly equation.

While the gravitational path integral $Z_{\text{grav}}$ satisfies the radial WDW equation \eqref{radialWDWequation}, the renormalised gravitational path integral $Z^{\text{ren}}_{\text{grav}}$ satisfies the modified radial WDW equation:
\begin{align}\label{modifiedradialWDWequation}
\widehat{\mathcal{H}}^{\text{radial}}_{\text{modified}} Z^{\text{ren}}_{\text{grav}}[\bar{g},\bar{\Phi}] = 0,
\end{align}
where
\begin{equation}            \widehat{\mathcal{H}}^{\text{radial}}_{\text{modified}} = e^{-\widehat{CT}}\widehat{\mathcal{H}}^{\text{radial}} e^{\widehat{CT}}.
\end{equation}
When the finite boundary $\partial\mathcal{M}_\epsilon$ is taken to infinity, while the metric and matter fields scale as 
$\bar{g}=\epsilon ^{\Delta_g} g_0$ and $\bar{\Phi}=\epsilon ^{\Delta_\Phi} \Phi_0$, their conjugate momentum operators scale oppositely as $\widehat{\Pi}^{\bar{a}\bar{b}}_{\bar{g}}=\epsilon ^{-\Delta_g} \widehat{\Pi}^{\bar{a}\bar{b}}_{g_0}$ and $\widehat{\Pi}_{\bar{\Phi}}=\epsilon ^{-\Delta_\Phi} \widehat{\Pi}_{\Phi_0}$. Under this rescaling, in the limit $\epsilon \to 0$, the terms in $\widehat{\mathcal{H}}^{\text{radial}}$ that blow up are called relevant terms, those that vanish are called irrelevant terms, and those that remain unchanged are called marginal terms. The holographic counterterms are precisely designed to kill all the relevant terms in $\widehat{\mathcal{H}}^{\text{radial}}$, leaving only the irrelevant and marginal terms surviving in $\widehat{\mathcal{H}}^{\text{radial}}_{\text{modified}}$, such that the marginal term becomes proportional to the Weyl-anomaly term $\left(\widehat{\mathcal{W}}_0(x)-i\mathcal{A}_0(x)\right)$.

So, in the limit $\epsilon \to 0$, only this marginal term in $\widehat{\mathcal{H}}^{\text{radial}}_{\text{modified}}$ survives while the irrelevant terms die out, thereby yielding the Weyl-anomaly equation—but now satisfied by the renormalised gravitational path integral:
\begin{align}
  \left(\widehat{\mathcal{W}}_0(x)-i\mathcal{A}_0(x)\right)  \lim_{\epsilon \to 0} \widetilde{Z}_{\text{grav}}[\epsilon ^{\Delta_g} g_0,\epsilon ^{\Delta_\Phi} \Phi_0] = 0.  
\end{align}
This equation simply reflects the fact that the two quantities in equation \eqref{main statement of AdS/CFT duality}, which are related to each other by the duality, satisfy the same Weyl-anomaly equation. This is exactly how the Weyl-anomaly equation is related to the modified radial WDW equation, which in turn is related to the radial WDW equation.

So, to summarise in a nutshell, the holographic counterterms modify the radial WDW equation into the modified radial WDW equation, which then becomes the Weyl-anomaly equation when the boundary is taken to infinity:
\begin{equation}
\widehat{\mathcal{H}}^{\text{radial}} \xrightarrow{\widehat{CT}}\widehat{\mathcal{H}}^{\text{radial}}_{\text{modified}} \xrightarrow{\partial\mathcal{M}_\epsilon \to \partial\mathcal{M}} \propto \left(\widehat{\mathcal{W}}_0(x)-i\mathcal{A}_0(x)\right).
\end{equation}
The ``proportional'' ($\propto$) symbol above is to emphasize that the marginal term in the modified radial WDW equation is only proportional to the Weyl-anomaly equation. All of the statements made above are not special to the radial WDW equation. Everything follows for the temporal WDW equation \eqref{ham} too in an identical fashion, albeit with some sign differences. While the rescaling in the radial case had the interpretation of sending the finite boundary to infinity, in the temporal case, just think of it as an abstract rescaling of variables by some abstract parameter $\epsilon$ which will be sent to $0$. Then, by similarly designing holographic counterterms to kill the relevant terms in $\widehat{\mathcal{H}}$, one obtains the modified WDW equation with $\widehat{\mathcal{H}}_{\text{modified}}$, which then has $\left(\widehat{\mathcal{W}}(x)-i\mathcal{A}(x)\right)$ in its marginal term and some irrelevant terms which vanish in the limit $\epsilon \to 0$:
\begin{equation}
\widehat{\mathcal{H}} \xrightarrow{\widehat{CT}}\widehat{\mathcal{H}}_{\text{modified}} \xrightarrow{\epsilon\to0}\propto \left(\widehat{\mathcal{W}}(x)-i\mathcal{A}(x)\right).
\end{equation}
All the subscripts $0$ are now dropped to denote that these are quantities on the Cauchy slice and not on the boundary. Then, when the Lorentzian CFT $Z_{\text{CFT}}[g_0,\Phi_0]$ is Wick-rotated and placed on the Cauchy slice $\Sigma$ with background metric $g$ and matter sources $\Phi^i$, its partition function satisfies a similar equation:
\begin{align}
  \left(\widehat{\mathcal{W}}(x)-i\mathcal{A}(x)\right) Z_{\text{CFT}}[g,\Phi^i;\psi_{\text{CFT}}] = 0, 
\end{align}
where now, since $\Sigma$ is an open manifold with a boundary $\partial\Sigma$, one needs to impose a boundary condition for the CFT on $\Sigma$ to define its partition function, which is done by inputting any CFT state $\psi_{\text{CFT}}$ on the boundary $\partial\Sigma$.

Let us now conclude this detour and return to classical physics, focusing again on the phase space and functions defined on it. All of the above quantum statements have classical analogues. We will now explain how to choose appropriate counterterms to modify the Hamiltonian constraint $\mathcal{H}$ into a new constraint function $\mathcal{H}_{\text{modified}}$, and show that the imaginary Weyl-anomaly function appears in its marginal term. At this stage, we are working with functions on the classical phase space, not quantum operators. However, incorporating these counterterms can lead to expressions involving complexified quantities. Thus, we need not interpret these transformations as canonical transformations in the usual sense, but rather treat them as algebraic manipulations aimed at constructing a well-defined real-valued function on phase space—one that we will ultimately impose as a constraint.

Now start with the classical Hamiltonian constraint function $\mathcal{H}$
\begin{equation}
 \mathcal{H} =    \frac{2\kappa}{\sqrt{g}}\!\Big(\Pi_{ab}\Pi^{ab}-\frac{1}{d-1}\Pi^2\Big)\!-\frac{\sqrt{g}}{2\kappa}(R-2\Lambda) + \mathcal{H}_{\text{matter}}.
\end{equation}
Next rescale the variables as below:
\begin{equation}\label{rescaling phase space variables}
  \text{Rescaling:} \quad  g_{ab} \rightarrow \epsilon^{\Delta_g} g_{ab},  \quad \Pi^{ab} \rightarrow \epsilon^{-\Delta_g} \Pi^{ab}, \quad \Phi^i \rightarrow \epsilon^{\Delta_{\Phi^i}} \Phi^i, \quad \Pi_{\Phi^i} \rightarrow \epsilon^{-\Delta_{\Phi^i}} \Pi_{\Phi^i}.
\end{equation}
To keep the notation clean, we represent the rescaled fields using the same symbols, with the understanding that the context will make it clear whether the variables are rescaled or not. This rescaling leaves the symplectic form—and hence the Poisson brackets—unchanged, since each variable is rescaled inversely relative to its canonically conjugate partner. Let us also recall that $\Delta_g = -2$ by convention; we have kept it explicit above but will substitute this value in what follows. Then, under this global rescaling, we obtain the rescaled Hamiltonian constraint $\mathcal{H}^{\text{rescaled}}$:
\begin{equation}
    \mathcal{H}^{\text{rescaled}} =  \epsilon^{d}\frac{2\kappa}{\sqrt{g}}\!\Big(\Pi_{ab}\Pi^{ab}-\frac{1}{d-1}\Pi^2\Big)\!- \epsilon^{2-d}\frac{\sqrt{g}}{2\kappa} R+\epsilon^{- d} \frac{\sqrt{g}}{2\kappa} 2\Lambda + \mathcal{H}_{\text{matter}}^{\text{rescaled}},
\end{equation}
where $\mathcal{H}_{\text{matter}}^{\text{rescaled}}$ is obtained from $\mathcal{H}_{\text{matter}}$ by the same rescaling but we have kept it general in this form. Now similar to the quantum counterparts, the terms are classified as relevant, irrelevant, or marginal based on them blowing up, vanishing, or non-changing in the limit $\epsilon\to0$. So one can see that in the gravitational part, the first term is irrelevant, the $2\Lambda \sqrt{g}$ term is relevant, and the $\sqrt{g} R$ term is relevant when $d>2$ and is marginal when $d=2$. 

Now we will explain how to choose holographic counterterms to kill the relevant terms. For the purpose of explanation, we don't have to have all the counterterms in one shot, rather we can choose them one by one to eliminate the relevant terms one by one. First focus on the most relevant term $2\Lambda \sqrt{g}$. A strategy is to choose the counterterm to be proportional to the relevant term that we want to eliminate. To eliminate the $2\Lambda\sqrt{g}$ term, choose the first counterterm to be
\begin{equation}
    CT_{\Lambda} = a_\Lambda \int_{\Sigma} d^dx \sqrt{g},
\end{equation}
where $a_\Lambda$ will later be set to be just the value it needs to be to kill the $2\Lambda\sqrt{g}$ term. The finite transformation generated by some function $C$ acting on some function $F$ is given by:
\begin{equation}
\exp\bigl(\,\{\cdot\,,C\}\bigr)\,F
\;=\;
\sum_{n=0}^{\infty}\frac{1}{n!}\,
\underbrace{\bigl\{\!\cdots\{\,F,\,C\},\dots,C\bigr\}}_{n\text{ times}} = F + \{F,C\} + \frac{1}{2} \{\{F,C\},C\}\cdots.
\end{equation}

The finite transformation generated by the counterterm $CT_{\Lambda}$ only changes $\Pi^{ab}$
\begin{equation}
  e^{(\{\cdot,CT_{\Lambda}\})} \Pi^{ab} = \Pi^{ab} + \{\Pi^{ab},CT_{\Lambda}\} = \Pi^{ab} -  \frac{a_\Lambda}{2}\sqrt{g} g^{ab},
\end{equation}
and leaves the rest of the variables $g_{ab}$, $\Phi^i$, and $\Pi_{\Phi^i}$ unchanged. So then the $\mathcal{H}^{\text{rescaled}}$ changes by:
\begin{align}
e^{(\{\cdot,CT_{\Lambda}\})} \mathcal{H}^{\text{rescaled}} &= +\epsilon^{d}\frac{2\kappa a_\Lambda}{2(d-1)}2\Pi +\epsilon^{d}\frac{2\kappa}{\sqrt{g}}\!\Big(\Pi_{ab}\Pi^{ab}-\frac{1}{d-1}\Pi^2\Big)\!- \epsilon^{2-d}\frac{\sqrt{g}}{2\kappa} R + \mathcal{H}_{\text{matter}}^{\text{rescaled}}\nonumber\\
&-\left(a_\Lambda^2 \epsilon^{d}\frac{2\kappa  d}{4 (d-1)}-\epsilon^{- d} \frac{2\Lambda}{2\kappa}\right)\sqrt{g}. 
\end{align}
By setting\footnote{Notice that there is a sign choice here. For example, we could choose $a_\Lambda$ with the opposite sign, and it would also eliminate the relevant term. However, flipping this sign would eventually lead to a Weyl-anomaly-like term with the opposite sign for the anomaly. We need to make the sign choice that yields the conformal anomaly with the correct sign. We do not wish to delve into this technicality here, so we refer the reader to \cite{Freidel:2008sh,CSH} for more details on why there are two sign choices in the holographic counterterms and which one to choose.}
\begin{equation}
   a_\Lambda =i\epsilon^{- d} \sqrt{\frac{|2\Lambda|}{\kappa^2} \frac{(d-1)}{d}},
\end{equation}
the relevant term $2\Lambda\sqrt{g}$ can be eliminated, giving
\begin{align}
e^{(\{\cdot,CT_{\Lambda}\})} \mathcal{H}^{\text{rescaled}} &= i \sqrt{\frac{|2\Lambda|}{d(d-1)}} \ 2\Pi - \epsilon^{2-d}\frac{\sqrt{g}}{2\kappa} R+\epsilon^{d}\frac{2\kappa}{\sqrt{g}}\!\Big(\Pi_{ab}\Pi^{ab}-\frac{1}{d-1}\Pi^2\Big)\! + \mathcal{H}_{\text{matter}}^{\text{rescaled}}. 
\end{align}
Notice that in the process, we also generate the Weyl generator term for the pure metric part, $2\Pi$. Next, this process must simply be repeated. For example, if $d > 2$, then the $\sqrt{g}R$ term is relevant, and another counterterm $CT_{R} = a_R \int d^dx \sqrt{g} R$ is introduced, with the coefficient $a_R$ chosen in a similar fashion. Depending on the details of the matter Hamiltonian, additional counterterms are also introduced (again following the same strategy: choosing them to be proportional to the relevant terms we wish to eliminate and fixing the proportionality constant accordingly). This also generates the Weyl generator for the matter part, giving the total Weyl generator
\begin{equation}
    \mathcal{W} = 2\Pi -\Delta_{\Phi^i}\Phi^i\Pi_{\Phi^i},
\end{equation}
and if this term does not appear for some fields, it simply means that the conformal dimension for those fields is zero. The other marginal terms generated are then grouped into the anomaly $\mathcal{A}$. If $d = 2$, the $\sqrt{g}R$ term is marginal and becomes part of the anomaly. If we are considering only pure gravity without any matter fields, then this concludes the process, and we obtain
\begin{align}
    e^{(\{\cdot,CT_{\Lambda}\})} \mathcal{H}^{\text{rescaled}}_{\text{pure gravity in }d=2} &=  i\sqrt{|\Lambda|} \ \left(2\Pi - i (-1) \frac{1}{2\kappa\sqrt{|\Lambda|}} \sqrt{g} R\right)+\epsilon^{d}\frac{2\kappa}{\sqrt{g}}\!\Big(\Pi_{ab}\Pi^{ab}-\frac{1}{d-1}\Pi^2\Big)\!. 
\end{align}
Notice that the marginal term on the right-hand side of the above equation is nothing but the classical counterpart of the Weyl-anomaly equation for a $d=2$ CFT:
\begin{equation}
   \left( 2\widehat{\Pi} - i \mathcal{A}_g\right) Z_{\text{CFT}}=0,
\end{equation}
with the conformal anomaly given by $\mathcal{A}_g = -\frac{c}{24 \pi}\sqrt{g}R$, and the central charge $c = \frac{12 \pi}{\kappa\sqrt{|\Lambda|}}$.

Similarly, in full generality, after repeating this process to eliminate all relevant terms, one obtains the modified Hamiltonian function:
\begin{align}\label{modifiedHamiltonianequation}
 \mathcal{H}_{\text{modified}} =  e^{(\{\cdot,CT\})} \mathcal{H}^{\text{rescaled}} &= i \sqrt{\frac{|2\Lambda|}{d(d-1)}} \ \left(\mathcal{W} - i \mathcal{A}\right)+X_{\text{irrelevant}}. 
\end{align}
Thus, we can express the imaginary Weyl-anomaly function as
\begin{equation}\label{imaginary weyl anomaly function as a limit}
         \left(\mathcal{W} - i \mathcal{A}\right) = - i \sqrt{\frac{d(d-1)}{|2\Lambda|}} \ \lim_{\epsilon \to 0}  \ e^{(\{\cdot,CT\})} \mathcal{H}^{\text{rescaled}}.
\end{equation}

Both the Weyl generator $\mathcal{W}$ and the conformal anomaly $\mathcal{A}$ are real functions on the phase space. So, the imaginary Weyl-anomaly function $\left(\mathcal{W} - i \mathcal{A}\right)$ is complex-valued. But we are after a real-valued function on the phase space. To obtain this, we define the following phase space function:
\begin{equation}
    \text{\emph{Real Weyl-anomaly constraint:}} \quad  W\!A[\omega] = W[\omega] + A[\omega] = \int_{\Sigma} d^dx \, \omega(x) \left(\mathcal{W}(x) + \mathcal{A}(x)\right),
\end{equation}
where we have simply removed by hand the factor of $``-i"$ in the imaginary Weyl-anomaly function to obtain the real-valued function $\left(\mathcal{W}(x) + \mathcal{A}(x)\right)$, which we refer to as the real Weyl-anomaly constraint. Here, $\omega$ is a compactly supported smooth smearing function, $W[\omega] = \int_{\Sigma} d^dx \, \omega(x) \mathcal{W}(x)$, and $A[\omega] = \int_{\Sigma} d^dx \, \omega(x) \mathcal{A}(x)$. It is this real-valued function that we will later impose as a constraint on the phase space. This concludes the explanation of how the real Weyl-anomaly constraint is extracted and how exactly it is related to the Hamiltonian constraint $\mathcal{H}$.

Let us give another example. For Einstein gravity with a minimally coupled massive scalar field in $(4+1)$-spacetime dimensions, the conformal anomaly is
\begin{align}
\mathcal{A} &= -  \frac{c}{8\pi^2 } \,\sqrt{g}\, 
  \Big(    G^{ab}    G_{ab} - \frac{1}{3} G^2\Big),
\end{align}
when the conformal dimension of the scalar field lies in the range $\Delta_\Phi \in (1,2)$. The central charge $c$ here is
\begin{equation}
 c= \frac{\pi L_{AdS}^3}{8 G_N }   = \frac{\pi^2 }{\kappa} \left(\frac{6}{|\Lambda|} \right)^{3/2}.
\end{equation}
When $\Delta_\Phi=1$, the conformal anomaly is
\begin{equation}\label{anomaly 4d and DeltaPhi1}
    \mathcal{A} = - \frac{c}{8\pi^2} \sqrt{g}\Bigg(  
  \Big(    G^{ab}    G_{ab} - \frac{1}{3} G^2\Big) -  \frac{2\kappa|\Lambda|}{3} \left( g^{ab}\nabla_a\Phi\nabla_b\Phi  
 +  \frac{1}{6}        \Phi^2  R \right)
 -     \frac{\kappa^2 |\Lambda|^2 }{27}     \Phi^4  \Bigg).
\end{equation}
See Appendix~\ref{appendix d=4 massive scalar} for its derivation. The factors of $\kappa$ appear in the above anomaly because the scalar field was canonically normalized in the action (i.e., the coefficient in front of the kinetic term is $\frac{1}{2}$). If we instead normalize it with a factor of $\frac{1}{\kappa}$ in front of the matter action (as is usually done in string theory), then all factors of $\sqrt{\kappa}$ in the above anomaly would be absorbed into the scalar field. With this particular normalization convention, and after also absorbing the factors of $\sqrt{|\Lambda|}$ into the bulk fields, the rescaled bulk fields become the sources of the dual CFT, and the CFT satisfies the imaginary Weyl-anomaly constraint with these sources.

\subsection{Poisson Algebra of \texorpdfstring{$\mathcal{W}+\mathcal{A},\, D_a, \, \text{and} \ G^A$}{}}\label{section: Poisson Algebra in alternative phase space}

Now that we have introduced the real Weyl-anomaly constraint in the previous section, let us analyze its Poisson bracket with the spatial diffeomorphism constraint $D_a$ and the matter gauge constraints $G^A$. In order for us to be able to impose $\mathcal{W} + \mathcal{A}$ as a first-class constraint alongside $D_a$ and $G^A$, it must be true that all these constraints close among themselves—and, in fact, they do. This is the content of the following proposition, which we prove below.

\begin{proposition}\label{Prop: Poisson closure of Weyl-anomaly constraint}
The real Weyl-anomaly constraint $W\!A[\omega]$, the spatial diffeomorphism constraint $D[N^a]$, and the matter-gauge constraint $G[\alpha]$ form a closed Poisson algebra among themselves. The Poisson brackets involving $W\! A[\omega]$ are given by:
\begin{subequations}\label{weylanomalyPBrelations}
\begin{align} 
\{ W\!A[\omega],\,W\!A[\widetilde{\omega}]\}
&= 0,\label{PBWW} \\[6pt]
\{D[N^{a}],\,W\!A[\omega]\} 
&= W\!A\!\bigl[\mathcal L_{N}\omega\bigr],\label{PBDW} \\[6pt]
\{W\!A[\omega],\,G[\alpha]\} 
&=0.\label{PBWG}
\end{align}
\end{subequations}
\end{proposition}
\noindent
This proposition is actually trivial once you understand the relation between the real Weyl-anomaly constraint and the Hamiltonian constraint. It just follows as a consequence of equations \eqref{PBHH}, \eqref{PBDH}, and \eqref{PBHG}. Nonetheless, we are stating this as a proposition and will prove it because it is important. These relations will later allow us to impose $\mathcal{W}+\mathcal{A}$ as a first-class constraint along with the other constraints and interpret it as a gauge generator. The first equation \eqref{PBWW} is also referred to as the holographic Wess–Zumino consistency condition and has been discussed in \cite{Shyam:2017qlr,Shyam:2016zuk}.
\begin{proof}
Expand the Poisson bracket of the real Weyl-anomaly constraint with itself:
\begin{equation}
   \{ W\!A[\omega],\,W\!A[\widetilde{\omega}]\} =  \{W[\omega],W[\widetilde{\omega}]\} +\{W[\omega],A[\widetilde{\omega}]\} +\{A[\omega],W[\widetilde{\omega}]\} +\{A[\omega],A[\widetilde{\omega}]\}.
\end{equation}
The first term $\{W[\omega],W[\widetilde{\omega}]\}$ is trivially zero because all the terms in the Weyl generator $\mathcal{W}=2\Pi-\Delta_{\Phi^i}\Phi^i \Pi_{\Phi^i}$ have coordinate variables and momenta variables occurring together, and they get Weyl-rescaled oppositely and hence are invariant. The last term $\{A[\omega],A[\widetilde{\omega}]\}$ is also trivially zero because the conformal anomaly, being a local function of $g_{ab}$ and $\Phi^i$ only, has no momenta terms in it. So we have
\begin{equation}\label{realimaginarweylanomalyPB}
   \{ W\!A[\omega],\,W\!A[\widetilde{\omega}]\} = \{W[\omega],A[\widetilde{\omega}]\} +\{A[\omega],W[\widetilde{\omega}]\} = i\,\{W[\omega]-i A[\omega] , W[\widetilde{\omega}]-i A[\widetilde{\omega}] \}.
\end{equation}

Smear the modified Hamiltonian in equation \eqref{modifiedHamiltonianequation} as
\begin{align}
\mathcal{H}_{\text{modified}}[\omega] &= \int_{\Sigma} d^dx \, \omega(x) \,\mathcal{H}_{\text{modified}}(x),\nonumber\\
&=\int_{\Sigma} d^dx \, \omega(x) \,e^{(\{\cdot,CT\})} \mathcal{H}^{\text{rescaled}}(x),\nonumber\\
&=e^{(\{\cdot,CT\})}\int_{\Sigma} d^dx \, \omega(x) \, \mathcal{H}^{\text{rescaled}}(x),\nonumber\\
&=e^{(\{\cdot,CT\})}\, \mathcal{H}^{\text{rescaled}}[\omega],
\end{align}
and take its Poisson bracket with itself to get
\begin{align}
    \{\mathcal{H}_{\text{modified}}[\omega] ,\mathcal{H}_{\text{modified}}[\widetilde{\omega}] \} &=\{e^{(\{\cdot,CT\})}\, \mathcal{H}^{\text{rescaled}}[\omega] , e^{(\{\cdot,CT\})}\, \mathcal{H}^{\text{rescaled}}[\widetilde{\omega}]\},\nonumber\\
    &=e^{(\{\cdot,CT\})}\, \{ \mathcal{H}^{\text{rescaled}}[\omega] ,  \mathcal{H}^{\text{rescaled}}[\widetilde{\omega}]\}.
\end{align}

Under the rescaling \eqref{rescaling phase space variables}, the Poisson bracket of the Hamiltonian constraint with itself in equation \eqref{PBHH} becomes
\begin{equation}
    \{ \mathcal{H}^{\text{rescaled}}[\omega] ,  \mathcal{H}^{\text{rescaled}}[\widetilde{\omega}]\} = D\!\bigl[\epsilon^{2} g^{ab}(\omega\,\partial_{b}\widetilde{\omega} - \widetilde{\omega}\,\partial_{b}\omega)\bigr] =\epsilon^{2} D\!\bigl[ g^{ab}(\omega\,\partial_{b}\widetilde{\omega} - \widetilde{\omega}\,\partial_{b}\omega)\bigr].
\end{equation}
The local diffeomorphism constraints $D_a(x)$ do not rescale, because they contain only terms where coordinate variables and momenta variables appear together, and thus rescale oppositely, leaving the combination invariant. Although covariant derivatives appear in these terms, they are themselves invariant under the global rescaling. However, the smearing function on the right-hand side of the above equation depends on the inverse metric $g^{ab}$, and it is only through this dependence that the smeared diffeomorphism constraint acquires an $\epsilon$ scaling as shown above. Then we get
\begin{equation}
    \{\mathcal{H}_{\text{modified}}[\omega] ,\mathcal{H}_{\text{modified}}[\widetilde{\omega}] \} = \epsilon^{2} e^{(\{\cdot,CT\})}\,  D\!\bigl[ g^{ab}(\omega\,\partial_{b}\widetilde{\omega} - \widetilde{\omega}\,\partial_{b}\omega)\bigr] =  \epsilon^{2} \,  D\!\bigl[ g^{ab}(\omega\,\partial_{b}\widetilde{\omega} - \widetilde{\omega}\,\partial_{b}\omega)\bigr],\label{vanishing hh term}
\end{equation}
where the last equality is a consequence of the counterterms being spatial diffeomorphism invariant.\footnote{There are no smearing functions in $CT$; in other words, the smearing function in $CT$ is just a constant, and so its Poisson bracket with the diffeomorphism constraint vanishes.} Now, again from \eqref{modifiedHamiltonianequation}, we have
\begin{align}
     \{\mathcal{H}_{\text{modified}}[\omega] ,\mathcal{H}_{\text{modified}}[\widetilde{\omega}] \} &= \frac{2\Lambda}{d(d-1)} \ \left\{ \left(W[\omega] - i A[\omega]\right) ,   \left(W[\widetilde{\omega}] - i A[\widetilde{\omega}]\right)\right\}\label{surviving first term}\\
     + i \sqrt{\frac{|2\Lambda|}{d(d-1)}} \ \big(\left\{  \left(W[\omega] - i A[\omega]\right) , X_{\text{irrelevant}}[\widetilde{\omega}]\right\} &+   \left\{X_{\text{irrelevant}}[\omega],   \left(W[\widetilde{\omega}] - i A[\widetilde{\omega}]\right) \right\}\big)\label{vanishig second term}\\
     &+ \left\{ X_{\text{irrelevant}}[\omega] , X_{\text{irrelevant}}[\widetilde{\omega}]\right\}.\label{vanishing third term}
\end{align}
Now, in the limit $\epsilon \to 0$, the second term \eqref{vanishig second term} and the third term \eqref{vanishing third term} vanish, leaving only the first term in \eqref{surviving first term} as the surviving term. But then, in equation \eqref{vanishing hh term}, the RHS also vanishes in the limit $\epsilon \to 0$. So this gives
\begin{equation}
    \left\{ \left(W[\omega] - i A[\omega]\right) ,   \left(W[\widetilde{\omega}] - i A[\widetilde{\omega}]\right)\right\}=0,
\end{equation}
which, via \eqref{realimaginarweylanomalyPB}, leads to
\begin{equation}
     \{ W\!A[\omega],\,W\!A[\widetilde{\omega}]\} = 0,
\end{equation}
thus proving equation \eqref{PBWW}.

Next, using equation \eqref{PBDH}, we have
\begin{equation}
    \{D[N^{a}],\,\mathcal H[\omega]\} 
= \mathcal H\!\bigl[\mathcal L_{N}\omega\bigr].
\end{equation}
The smearing vector $N^a$ and smearing function $\omega$ are independent of the phase space variables and hence invariant under the rescaling \eqref{rescaling phase space variables}. So the smearing function on the RHS, $\mathcal{L}_N \omega = N^a \partial_a \omega$, is also invariant under this rescaling. As we previously mentioned, $D_a$ is also invariant. Only $\mathcal{H}$ changes under the rescaling, and so the above equation becomes
\begin{equation}
    \{D[N^{a}],\,\mathcal H^{\text{rescaled}}[\omega]\} 
= \mathcal H^{\text{rescaled}}\!\bigl[\mathcal L_{N}\omega\bigr].
\end{equation}
Then this gives
\begin{align}
  e^{(\{\cdot,CT\})}\,  \{D[N^{a}],\,\mathcal H^{\text{rescaled}}[\omega]\} 
&= e^{(\{\cdot,CT\})}\, \mathcal H^{\text{rescaled}}\!\bigl[\mathcal L_{N}\omega\bigr],\\ \implies
  \{e^{(\{\cdot,CT\})}\,D[N^{a}],\,e^{(\{\cdot,CT\})}\,\mathcal H^{\text{rescaled}}[\omega]\} 
&=  e^{(\{\cdot,CT\})}\,\mathcal H^{\text{rescaled}}\!\bigl[\mathcal L_{N}\omega\bigr],\\ \implies
  \{D[N^{a}],\,\mathcal{H}_{\text{modified}}[\omega]\} 
&= \mathcal{H}_{\text{modified}}\!\bigl[\mathcal L_{N}\omega\bigr],
\end{align}
which, in the limit $\epsilon \to 0$, gives
\begin{equation}
  \{D[N^{a}],W[\omega] -i A[\omega]\} 
= W\!\bigl[\mathcal L_{N}\omega\bigr] - i A\!\bigl[\mathcal L_{N}\omega\bigr] , 
\end{equation}
which in turn gives
\begin{equation}
\{D[N^{a}],\,W\!A[\omega]\} 
= W\!A\!\bigl[\mathcal L_{N}\omega\bigr].
\end{equation}

There is a better way to arrive at the above equation. Since the terms in the Hamiltonian constraint are built from covariant terms (covariant under spatial diffeomorphisms), the holographic counterterms are consequently also built from covariant terms. Therefore, both the Weyl generator $\mathcal{W}$ and the conformal anomaly $\mathcal{A}$ are built from covariant terms, and so they satisfy $\{D[N^{a}],W[\omega]\} = W\!\bigl[\mathcal L_{N}\omega\bigr]$ and $\{D[N^{a}],A[\omega]\} = A\!\bigl[\mathcal L_{N}\omega\bigr]$, thus giving \eqref{PBDW}. 

Similarly, since all the terms in the Hamiltonian constraint $\mathcal{H}$ are invariant under matter-gauge transformations, the holographic counterterms are consequently also invariant:
\begin{equation}
    \{CT,G[\alpha]\}=0,
\end{equation}
and so all terms in the modified Hamiltonian function are invariant. Therefore, under matter-gauge transformations, the Weyl generator and conformal anomaly are also invariant: 
\begin{align}
    \{W[\omega],\,G[\alpha]\}&=0\label{PBweylG},\\
    \{A[\omega],\,G[\alpha]\}&=0,
\end{align}
thus leading to the equation \eqref{PBWG}:
\begin{equation}
    \{W\!A[\omega],\,G[\alpha]\}=0.
\end{equation}
This completes the proof of this proposition.
\end{proof}

Let us now consider some examples. For the case of pure Einstein gravity in $d=2$, the Poisson bracket of the real Weyl-anomaly constraint with itself is
\begin{align}
\{W\!A[\omega],\,W\!A[\widetilde{\omega}]\} &= \int d^2x d^2y \omega(x) \widetilde{\omega}(y) \{2\Pi(x) -\frac{c}{24 \pi}\sqrt{g}R(x) , 2\Pi(y) -\frac{c}{24 \pi}\sqrt{g}R(y)\},\nonumber\\
&= \frac{2c}{24 \pi}\int d^2x d^2y \left(\omega(x) \widetilde{\omega}(y)
-\omega(y) \widetilde{\omega}(x)
\right)g_{ab}(x) \frac{\delta(\sqrt{g}R(y))}{\delta g_{ab}(x)} ,\nonumber\\
&= \frac{c}{12 \pi}\int d^2x   \sqrt{g}\left(\widetilde{\omega}\nabla^2 \omega-\omega\nabla^2 \widetilde{\omega} \right) = 0,
\end{align}
where integration by parts is used in the last equality. The Poisson bracket with the diffeomorphism constraint is
\begin{align}
    \{D[N^a],W\!A[\omega]\} &= \int d^2x d^2y N^a(x)\omega(y)  \{-2\nabla_b \Pi^{b}_a(x) , 2\Pi(y) -\frac{c}{24 \pi}\sqrt{g}R(y)\},\nonumber\\
    &= \int d^2x  (\mathcal{L}_N\omega )\left( 2\Pi     -\frac{c}{24 \pi}    \sqrt{g} R  \right) = W\!A[\mathcal{L}_N\omega],
\end{align}
as expected.

Next, for pure Einstein gravity in $d=4$, the Poisson bracket of the real Weyl-anomaly constraint with itself is
\begin{align}
\{W\!A[\omega],\,W\!A[\widetilde{\omega}]\} &=    \frac{c}{2\pi^2 } \int d^4x   \sqrt{g}\,G^{ab}\left(\widetilde{\omega}\nabla_a\nabla_b \omega-\omega\nabla_a\nabla_b \widetilde{\omega} \right) = 0.
\end{align}
The explicit computation for the above is too long to be displayed here, but the reader can perform it themselves using standard Mathematica packages (we used \texttt{xTras}).

Similarly, for Einstein gravity in $d=4$ with a minimally coupled massive scalar field of conformal dimension $\Delta_\Phi=1$, the Poisson bracket of the real Weyl-anomaly constraint with itself also vanishes, provided the correct anomaly in equation \eqref{anomaly 4d and DeltaPhi1} is used.

But the point we want to emphasize is that the closure holds only for the correct anomaly, and by the correct anomaly we mean the holographic conformal anomaly. If one uses any other arbitrary function in place of the anomaly in the Weyl-anomaly function, then the closure can fail. This is because such functions might not originate from the Hamiltonian constraint in the way the holographic conformal anomaly does, as explained in the previous subsection.

To give an example where the closure fails with an incorrect anomaly, consider just the $\sqrt{g} R \Phi^2$ term in \eqref{anomaly 4d and DeltaPhi1}:
\begin{align}
    \int d^4x \int d^4y & \,\omega(x) \tilde{\omega}(y)\, \{2\Pi(x)- \Phi\Pi_\Phi(x) -\sqrt{g} R \Phi^2(x) , 2\Pi(y)- \Phi\Pi_\Phi(y)- \sqrt{g} R \Phi^2(y)\} \nonumber\\
    &= 6\int d^4x \,\sqrt{g}\,\Phi \,
\Bigl(
\tilde\omega\,\Phi\,\nabla_{a}\nabla^{a}\omega
-\omega\,\Phi\,\nabla_{a}\nabla^{a}\tilde\omega
+4\,\nabla_{a}\Phi\,
\bigl(\tilde\omega\,\nabla^{a}\omega-\omega\,\nabla^{a}\tilde\omega\bigr)
\Bigr) \neq 0,
\end{align}
and only after taking into account all the terms in \eqref{anomaly 4d and DeltaPhi1} does the closure hold. So it is important that the anomaly used is the correct holographic anomaly, and its derivation from the Hamiltonian constraint is what guarantees the closure.

\subsection{Alternative Phase Space}\label{section: Alternative Phase Space}

Now that we have all the ingredients in place, we proceed to construct the alternative phase space, denoted by $\Gamma_{\text{ALT}}$. Let $\mathcal{P}_{\text{ALT}}$ denote the underlying manifold of this phase space. As in the ADM formulation, we introduce the canonical variables: the spatial metric $\mathfrak{g}_{ab}$, the matter fields $\Phi^i$, and their respective conjugate momenta $\mathsf{\Pi}^{ab}$ and $\Pi_{\Phi^i}$, all defined on the Cauchy slice $\Sigma$. These variables coordinatize $\mathcal{P}_{\text{ALT}}$. The symplectic structure on this phase space is given by the symplectic potential
\begin{equation}
\theta_{\text{ALT}} = \int_{\Sigma} d^dx (\mathsf{\Pi}^{ab} \delta \mathfrak{g}_{ab} + \Pi_{\Phi^i} \delta \Phi^i),
\end{equation}
and equivalently by the symplectic two-form
\begin{equation}
\omega_{\text{ALT}} = \delta\theta_{\text{ALT}} = \int_{\Sigma} d^d x \left( \delta \mathsf{\Pi}^{ab} \wedge \delta \mathfrak{g}_{ab} + \delta \Pi_{\Phi^i} \wedge \delta \Phi^i \right).
\end{equation}

So far, this construction mirrors the ADM phase space, differing only in notation. The key distinction arises in the set of constraints we impose. In the alternative phase space, we impose the \emph{real Weyl-anomaly constraint}, the diffeomorphism constraints, and the matter-gauge constraints:
\begin{subequations}
\begin{align}
    \text{Real Weyl-anomaly constraint:} \quad W\!A[\omega] &= 0, \\
    \text{Diffeomorphism constraint:} \quad D[N^a] &= 0, \\
    \text{Matter-gauge constraint:} \quad G[\alpha] &= 0.
\end{align}
\end{subequations}
Thanks to Proposition~\ref{Prop: Poisson closure of Weyl-anomaly constraint}, all of these constraints are first-class and thus generate gauge transformations in the alternative phase space. Whenever we refer to constraints or any quantity in this alternative phase space, they are understood to be expressed in terms of the phase space variables of the alternative phase space. For example, the Weyl generator $\mathcal{W}$ in the alternative phase space is given by
\begin{equation}
    \mathcal{W} = 2\mathsf{\Pi} - \Delta_{\Phi^i} \Phi^i \Pi_{\Phi^i},
\end{equation}
where $\mathsf{\Pi} = \mathfrak{g}_{ab}\mathsf{\Pi}^{ab}$.

Let us now summarize the structure of the alternative phase space $\Gamma_{\text{ALT}}$ by expressing all of its components together as
\begin{equation}
   \Gamma_{\text{ALT}} = \left(\,\mathcal{P}_{\text{ALT}}, \,\omega_{\text{ALT}}\ ;\ \mathcal{W}+\mathcal{A}, \, D_a,\,G^A\,\right).
\end{equation}

Our aim is to prove that the alternative phase space $\Gamma_{\text{ALT}}$ is physically equivalent to the ADM phase space $\Gamma_{\text{ADM}}$ at the kinematical level. This is precisely the content of Theorem~\ref{theorem: phase space equivalence}. Of course, the underlying phase space manifolds and symplectic structures—$\left(\mathcal{P}_{\text{ALT}}, \omega_{\text{ALT}}\right)$ and $\left(\mathcal{P}_{\text{ADM}}, \omega_{\text{ADM}}\right)$—are the same. However, this does not imply that $\Gamma_{\text{ALT}}$ and $\Gamma_{\text{ADM}}$ are identical, as they differ in the constraints imposed. To establish the equivalence of $\Gamma_{\text{ALT}}$ and $\Gamma_{\text{ADM}}$, we must show that their reduced phase spaces coincide. This is exactly what we will prove.

\subsection{Gauge Orbits of \texorpdfstring{$\mathcal{W}+\mathcal{A}$}{}}\label{section: Gauge orbits of Weyl-Anomaly constraint}

Let us now analyze the gauge transformations generated by the real Weyl-anomaly constraint. Taking the Poisson bracket of $W\!A[\omega]$ with the phase space variables, we get
\begin{subequations}
\begin{align}
    \delta_{\omega} \mathfrak{g}_{ab}(x) &= \{\mathfrak{g}_{ab}(x), W\!A[\omega]\} = 2 \,  \omega(x) \, \mathfrak{g}_{ab}(x),\\
    \delta_{\omega} \Phi^i(x) &= \{\Phi^i(x), W\!A[\omega]\} = -\Delta_{\Phi^i} \, \omega(x) \, \Phi^i(x),\\
    \delta_{\omega} \mathsf{\Pi}^{ab}(x) &= \{\mathsf{\Pi}^{ab}(x), W\!A[\omega]\}  =-2 \,\omega(x) \,  \mathsf{\Pi}^{ab}(x) - \int d^dy \, \omega(y) \,  \frac{\delta \mathcal{A}(y)}{\delta \mathfrak{g}_{ab}(x)},\\
    \delta_{\omega} \Pi_{\Phi^i}(x) &= \{\Pi_{\Phi^i}(x), W\!A[\omega]\}  =\Delta_{\Phi^i} \,\omega(x) \,  \Pi_{\Phi^i}(x) - \int d^dy \, \omega(y) \,  \frac{\delta \mathcal{A}(y)}{\delta \Phi^i(x)}.
\end{align}
\end{subequations}
So the gauge transformations generated by $W\!A[\omega]$ perform local rescalings (Weyl transformations) of the metric $\mathfrak{g}_{ab}$ and the matter fields $\Phi^i$. In addition to rescaling $\mathsf{\Pi}^{ab}$ and $\Pi_{\Phi^i}$, these gauge transformations also translate them by an amount proportional to the derivative of the conformal anomaly.

Therefore, the gauge orbit of a point in phase space containing the metric $\mathfrak{g}_{ab}$ corresponds to the conformal class of that metric. A similar statement holds for $\Phi^i$, while the behavior of $\mathsf{\Pi}^{ab}$ and $\Pi_{\Phi^i}$ is more involved, as it depends on the specific structure of the conformal anomaly.

\subsection{Gauge-Fixing Condition for the Weyl-Anomaly Constraint}\label{section: Gauge Fixing the Weyl-Anomaly Constraint}

Since the real Weyl-anomaly constraint gauge-transforms the metric $\mathfrak{g}_{ab}$ to any other metric in its conformal class, choosing a gauge is equivalent to selecting a representative element from that conformal class. From Proposition~\ref{Prop: Yamabe}, we know that in the conformal class of any metric, there always exists a unique metric whose Ricci scalar equals $2\Lambda$. Therefore, the following condition serves as a valid gauge-fixing condition for the real Weyl-anomaly constraint:
\begin{equation}\label{Ricci constraint}
    \mathcal{R}[\varrho] = \int d^dx\, \sqrt{\mathfrak{g}} \, \varrho(x) \, \left(R_{[\mathfrak{g}]}-2\Lambda \right).
\end{equation}

After introducing this gauge-fixing condition into the alternative phase space as an additional constraint, the real Weyl-anomaly constraint and $\mathcal{R}$ together form a set of second-class constraints, while $D$ and $G$ remain first-class constraints. This can be seen from the Poisson brackets involving $\mathcal{R}$:
\begin{align}
\{W\!A[\omega],\mathcal{R}[\varrho]\} &\not\approx 0, \quad  \{D[N^a],\mathcal{R}[\varrho]\} = \mathcal{R}[\mathcal{L}_N \varrho], \quad   \{G[\alpha],\mathcal{R}[\varrho]\} = 0, \quad \{\mathcal{R}[\varrho],\mathcal{R}[\tilde{\varrho}]\} = 0,
\end{align}
where $\not\approx 0$ (respectively, $\approx 0$) denotes non-vanishing (respectively, vanishing) on the constraint surface $\mathcal{C}_{\mathcal{W}+\mathcal{A},D,G,\mathcal{R}}$. This surface, $\mathcal{C}_{\mathcal{W}+\mathcal{A},D,G,\mathcal{R}} \subset \mathcal{P}_{\text{ALT}}$, is defined as the surface where all constraints $\mathcal{W}+\mathcal{A}$, $D_a$, $G^A$, and $\mathcal{R}$ vanish. 

The first equation follows from the fact that the Ricci scalar changes under Weyl transformations. The second holds because the gauge condition only involves spatially covariant quantities. The third is trivial since $\mathcal{R}$ contains no matter fields, and the fourth is also trivial. 

Thus, this gauge-fixing condition solely fixes the real Weyl-anomaly constraint, leaving the diffeomorphism and matter-gauge constraints intact as gauge constraints. The next step is to solve the real Weyl-anomaly constraint and this gauge-fixing condition to obtain the reduced phase space. This is best done after performing a covariant conformal decomposition, which we now introduce.

\subsection{Covariant Conformal Decomposition}\label{section: Covariant Conformal Decomposition}

The metric $\mathfrak{g}_{ab}$ is decomposed in the same way as before in equation \eqref{conformal decomposition equation} for $d>2$:\footnote{While the current discussion is for the $d>2$ case, the $d=2$ case is discussed in Appendix~\ref{appendix d=2}.}
\begin{equation}
    \mathfrak{g}_{ab} = \varphi^\alpha \, \gamma_{ab},
\end{equation}
where the conformal factor is now denoted by $\varphi$ to distinguish it from its counterpart $\phi$ in the ADM phase space, $\alpha = \frac{4}{d-2}$ as before, and the conformal metric is defined by requiring the Ricci scalar to equal $2\Lambda$:
\begin{equation}
    R_{[\gamma]} = 2\Lambda.
\end{equation}
In these variables, the gauge fixing condition in equation \eqref{Ricci constraint} becomes equivalent to choosing the gauge fixing condition below (for which we continue to use the same notation):\footnote{Technically, using \eqref{Weyl transformation of Ricci scalar} in \eqref{Ricci constraint} gives
\begin{equation}
    \mathcal{R}[\varrho] = -\int d^dx\, \varphi \sqrt{\gamma} \, \varrho(x) \,   \,\left( \frac{4(d-1)}{(d-2)} \nabla^2_{[\gamma]} \varphi  -2\Lambda \varphi 
    +2\Lambda \varphi^{\alpha+1}\right), 
\end{equation} 
which imposes the term in the bracket to be zero. This is also of the form of the Lichnerowicz equation, and our previous proof of existence and uniqueness of solutions holds for this equation as well (since the coefficients satisfy the required assumptions). So the only solution is $\varphi=1$, and therefore we can equivalently replace this gauge fixing condition by \eqref{Ricci constraint in covariant conformal decomposition form}.}
\begin{equation}\label{Ricci constraint in covariant conformal decomposition form}
    \mathcal{R}[\varrho] = \int d^dx\, \sqrt{\gamma} \, \varrho(x) \, \left(\varphi - 1 \right).
\end{equation}

Next, the conjugate momentum $\mathsf{\Pi}^{ab}$ is split into a traceless part $\Sigma^{ab}$ and a trace part as follows:
\begin{equation}\label{momentum split in covariant conformal decomposition}
    \mathsf{\Pi^{ab}} = \varphi^{-\alpha}\, \Sigma^{ab} + \sqrt{\gamma}\, \varphi^{-\alpha}\, Y^{ab}_{\gamma} f,
\end{equation}
where 
\begin{align}
 &Y^{ab}_{\gamma} :=  -R_{[\gamma]}^{ab} +\nabla_{[\gamma]}^a \nabla_{[\gamma]}^b - \gamma^{ab} \nabla_{[\gamma]}^2 , \\
 &Y_{\gamma} := \gamma_{ab}Y^{ab}_{\gamma} = -2\Lambda  - (d-1) \nabla_{[\gamma]}^2 ,
\end{align}
and $f$ is defined as
\begin{equation}
    f:= \lim_{\epsilon\to0} f_\epsilon,
\end{equation}
where $f_\epsilon$ is the unique solution to the equation
\begin{equation}
    Y_{\gamma} f_\epsilon = \frac{1}{\sqrt{\gamma}} \mathsf{\Pi} \quad \forall x \in \Sigma_\epsilon  \quad \text{satisfying the boundary condition} \quad f|_{\partial\Sigma_\epsilon} = -\frac{1}{2\Lambda}\frac{\mathsf{\Pi}}{\sqrt{\gamma}} \Bigg|_{\partial\Sigma_\epsilon},
\end{equation}
where $\partial\Sigma_\epsilon$ is the constant-$\epsilon$ slice embedded in $\Sigma$ in the FG coordinates \eqref{FG spatial coordinates}, and $\Sigma_\epsilon \subset \Sigma$ is the interior part of $\Sigma$ with its only boundary being $\partial\Sigma_\epsilon$. So the function $f$ satisfies
\begin{equation}
    Y_{\gamma} f = \frac{1}{\sqrt{\gamma}} \mathsf{\Pi} \quad \forall x \in \Sigma,
\end{equation}
noting that $\mathsf{\Pi}$ is still defined as $\mathsf{\Pi} = \mathfrak{g}_{ab} \mathsf{\Pi}^{ab}$. Thus, for a given $\frac{1}{\sqrt{\gamma}} \mathsf{\Pi}$, there exists a unique $f$. The reason for this split will become clear soon. We can see that $\Sigma^{ab}$ is traceless:
\begin{equation}
   \gamma_{ab} \Sigma^{ab} = \varphi^\alpha \gamma_{ab} \mathsf{\Pi}^{ab} - \sqrt{\gamma}\,  \gamma_{ab} Y^{ab}_{\gamma} f =  \mathsf{\Pi} - \sqrt{\gamma}\,  Y_{\gamma} f = 0.
\end{equation}

After this decomposition, we have
\begin{align}
     \mathsf{\Pi}^{ab}  \delta \mathfrak{g}_{ab} &= (\varphi^{-\alpha}\, \Sigma^{ab} + \sqrt{\gamma}\, \varphi^{-\alpha}\, Y^{ab}_{\gamma} f)  (\varphi^{\alpha} \delta \gamma_{ab} + \alpha \varphi^{\alpha-1} \gamma_{ab} \delta \varphi),\nonumber\\
     &=  \Sigma^{ab}  \delta \gamma_{ab}  + \frac{\alpha}{\varphi} \sqrt{\gamma}   (Y_{\gamma} f)    \delta \varphi + \sqrt{\gamma}\,  (Y^{ab}_{\gamma} f)   \delta \gamma_{ab} ,\nonumber\\
     &=  \Sigma^{ab}  \delta \gamma_{ab}  + \frac{\alpha}{\varphi} \mathsf{\Pi}    \delta \varphi + \sqrt{\gamma}\,  (Y^{ab}_{\gamma} f)   \delta \gamma_{ab} .
\end{align}
Defining $\Pi_\varphi:= \frac{\alpha}{\varphi} \mathsf{\Pi}$, we have
\begin{align}\label{midstepsymplecticpotential}
   \int_\Sigma d^dx \,  \mathsf{\Pi}^{ab}  \delta \mathfrak{g}_{ab} &= \int_\Sigma d^dx \, (\Sigma^{ab}  \delta \gamma_{ab}  + \Pi_\varphi    \delta \varphi) + \int_\Sigma d^dx \, \sqrt{\gamma}\,  (Y^{ab}_{\gamma} f)   \delta \gamma_{ab} .
\end{align}

$\delta\gamma_{ab}(x)$ are one-forms in $\mathcal{P}_{\text{ALT}}$, and they are not all independent from each other because $\gamma_{ab}$ is required to satisfy the condition $R_{[\gamma]} = 2\Lambda$. So the variation of $R_{[\gamma]}$ with respect to $\gamma_{ab}$ must vanish:
\begin{equation}
    0 = \delta R_{[\gamma]}(x) = \int d^dy \, \frac{\delta R_{[\gamma]}(x)}{\delta \gamma_{ab}(y)} \delta\gamma_{ab}(y) = \left(-R_{[\gamma]}^{ab} +\nabla_{[\gamma]}^a \nabla_{[\gamma]}^b - \gamma^{ab} \nabla_{[\gamma]}^2 \right) \delta\gamma_{ab}(x) .
\end{equation}
So this gives the specific linear combinations of the one-forms $\delta\gamma_{ab}$ that vanish:
\begin{equation}
  Y^{ab}_{\gamma}\delta\gamma_{ab}(x) = 0.  
\end{equation}

Using integration by parts, we can write \eqref{midstepsymplecticpotential} as
\begin{align}
   \int_\Sigma d^dx \,  \mathsf{\Pi}^{ab}  \delta \mathfrak{g}_{ab} &= \int_\Sigma d^dx \, (\Sigma^{ab}  \delta \gamma_{ab}  + \Pi_\varphi    \delta \varphi) + \int_\Sigma d^dx \, \sqrt{\gamma}\,   f \,  Y^{ab}_{\gamma}  \delta \gamma_{ab} ,
\end{align}
and the boundary terms arising from this integration by parts vanish as a consequence of the Dirichlet boundary conditions for both $\gamma_{ab}$ and $f$. So the last term vanishes, giving us the symplectic potential in this alternative phase space to be
\begin{equation}
    \theta_{\text{ALT}} = \int_\Sigma d^dx \, (\Sigma^{ab}  \delta \gamma_{ab}  + \Pi_\varphi    \delta \varphi + \Pi_{\Phi^i} \delta \Phi^i),
\end{equation}
and so the symplectic two-form expressed in the covariant conformal decomposed coordinates is
\begin{equation}\label{symplectic two form in covariant conformal decomposed coordinates}
    \omega_{\text{ALT}} = \delta  \theta_{\text{ALT}} =  \int_\Sigma d^dx \, (\delta\Sigma^{ab} \wedge \delta \gamma_{ab}  + \delta\Pi_\varphi  \wedge  \delta \varphi + \delta\Pi_{\Phi^i} \wedge \delta \Phi^i).
\end{equation}

Now it becomes evident that $\Sigma^{ab}$ is canonically conjugate to $\gamma_{ab}$ and $\Pi_{\varphi}$ is canonically conjugate to $\varphi$. Achieving this was the reason we specifically designed this decomposition. This change of variables from $(\mathfrak{g}_{ab},\mathsf{\Pi}^{ab},\Phi^i,\Pi_{\Phi^i})$ to $(\varphi,\Pi_\varphi,\gamma_{ab},\Sigma^{ab},\Phi^i,\Pi_{\Phi^i})$ is what we refer to as the \emph{covariant conformal decomposition}.

Had we not done this and instead performed the usual conformal decomposition, in which $\mathsf{\Pi}^{ab}$ is split as
\begin{equation}
    \mathsf{\Pi}^{ab} = \varphi^{-\alpha} \pi^{ab} + \frac{1}{d} \,\mathsf{\Pi} \, \mathfrak{g}^{ab},
\end{equation}
instead of using \eqref{momentum split in covariant conformal decomposition}, then the symplectic form in these variables would be
\begin{align}
   \omega_{\text{ALT}} &=  \int_\Sigma d^dx \, (  \delta \pi^{ab} \wedge \delta \gamma_{ab}  + \delta \Pi_{\varphi}  \wedge  \delta \varphi + \frac{1}{d} \,\delta \mathsf{\Pi} \wedge \gamma^{ab}   \delta \gamma_{ab} +\delta\Pi_{\Phi^i} \wedge \delta \Phi^i).
\end{align}
Now you can see that due to the presence of the third term, $\frac{1}{d} \,\delta \mathsf{\Pi} \wedge \gamma^{ab}   \delta \gamma_{ab}$, the symplectic form in these variables is not in canonical form because $\gamma^{ab} \delta \gamma_{ab}$ is not zero unless $\gamma_{ab}$ is required to satisfy a different condition—such as fixing its determinant to 1—which would then be non-covariant and therefore undesirable. Moreover, when we restrict the symplectic form to the constraint surface $\mathcal{C}_{\mathcal{W}+\mathcal{A},\mathcal{R}}$ (defined by the vanishing of $\mathcal{W}+\mathcal{A}$ and $\mathcal{R}$), the conformally decomposed variables lead to complicated Poisson brackets that are not in canonical form. This is why the covariant conformal decomposition is better.

\subsection{Gauge Fixing the Weyl-Anomaly Constraint}\label{Gauge Fixing the Weyl-Anomaly Constraint}


In the covariant conformal decomposed coordinates, the alternative phase space is coordinatized by
\begin{equation}    (\varphi,\Pi_\varphi,\gamma_{ab},\Sigma^{ab},\Phi^i,\Pi_{\Phi^i}).
\end{equation}
Then the real Weyl-anomaly constraint fixes $\mathsf{\Pi}$ to be
\begin{equation}
    \mathsf{\Pi} = \frac{\Delta_{\Phi^i}}{2} \Phi^i \Pi_{\Phi^i} - \frac{1}{2}\mathcal{A},
\end{equation}
which in turn fixes $\Pi_{\varphi}$, and the gauge fixing condition sets the conformal factor to $\varphi = 1$, thereby eliminating these variables and yielding a reduced phase space coordinatized by
\begin{equation}
    (\gamma_{ab}, \Sigma^{ab}, \Phi^i, \Pi_{\Phi^i}).
\end{equation}
The underlying manifold of this reduced phase space is the constraint surface $\mathcal{C}_{\mathcal{W}+\mathcal{A},\mathcal{R}}$, and the above variables coordinatize this space. This constraint surface $\mathcal{C}_{\mathcal{W}+\mathcal{A},\mathcal{R}} \subset \mathcal{P}_{\text{ALT}}$ is defined as the submanifold on which the constraints $\mathcal{W}+\mathcal{A}$ and $\mathcal{R}$ vanish. The symplectic form $\omega^{\mathrm{red}}_{\mathrm{ALT}}$ on $\mathcal{C}_{\mathcal{W}+\mathcal{A},\mathcal{R}}$ is obtained by pulling back the symplectic form $\omega_{\mathrm{ALT}}$ on $\mathcal{P}_{\mathrm{ALT}}$ to the constraint surface via the inclusion map $\iota : \mathcal{C}_{\mathcal{W}+\mathcal{A},\mathcal{R}} \hookrightarrow \mathcal{P}_{\mathrm{ALT}}$:
\begin{equation}
\omega^{\mathrm{red}}_{\mathrm{ALT}} = \iota^* \omega_{\mathrm{ALT}}.
\end{equation}
The remaining first-class constraints $D_a$ and $G^A$ now act as gauge constraints on this reduced phase space. Writing everything together, the reduced phase space, denoted by $\Gamma^{\text{red}}_{\text{ALT}}$, is
\begin{equation}
    \Gamma^{\text{red}}_{\text{ALT}} = \left(\,\mathcal{C}_{\mathcal{W}+\mathcal{A},\mathcal{R}}, \,\omega^{\mathrm{red}}_{\mathrm{ALT}}\ ;\  D_a,\,G^A\,\right).
\end{equation}
This reduced phase space $\Gamma^{\mathrm{red}}_{\mathrm{ALT}}$ has the same set of gauge transformations as $\Gamma^{\mathrm{red}}_{\mathrm{ADM}}$, namely the spatial diffeomorphisms generated by $D_a$ and the matter-gauge transformations generated by $G^A$.

In summary, we have gauge-fixed the real Weyl-anomaly constraint to obtain the reduced phase space from the alternative phase space:
\begin{equation}
\Gamma_{\mathrm{ALT}} \xrightarrow{\text{Gauge fixing } \mathcal{W}+\mathcal{A}} \Gamma^{\mathrm{red}}_{\mathrm{ALT}}.
\end{equation}

We can easily see the explicit form of $\omega^{\mathrm{red}}_{\mathrm{ALT}}$. 
On the constraint surface $\mathcal{C}_{\mathcal{W}+\mathcal{A},\mathcal{R}}$, we have $\varphi = 1$ and hence $\delta \varphi = 0$. Therefore, when $\omega_{\text{ALT}}$ is pulled back to $\mathcal{C}_{\mathcal{W}+\mathcal{A},\mathcal{R}}$, we obtain
\begin{equation}
    \omega^{\mathrm{red}}_{\mathrm{ALT}} = \int_\Sigma d^dx \, (\delta\Sigma^{ab} \wedge \delta \gamma_{ab}  + \delta\Pi_{\Phi^i} \wedge \delta \Phi^i),
\end{equation}
which shows that $\Sigma^{ab}$ remains canonically conjugate to $\gamma_{ab}$ even in the reduced phase space.

Let us also obtain the explicit form of $\omega^{\mathrm{red}}_{\mathrm{ADM}}$. On the constraint surface $\mathcal{C}_{\mathcal{H},\mathcal{K}}$, we have $\Pi = 0$ and hence $\delta\Pi = 0$. Therefore, when $\omega_{\text{ADM}}$ is pulled back to $\mathcal{C}_{\mathcal{H},\mathcal{K}}$, we obtain
\begin{equation}
    \omega^{\mathrm{red}}_{\mathrm{ADM}} = \int_\Sigma d^dx \, (\delta\pi^{ab} \wedge \delta \gamma_{ab}  + \delta\Pi_{\Phi^i} \wedge \delta \Phi^i),
\end{equation}
where $\pi^{ab}$ is defined as in equation~\eqref{momentum split in conformal decomposition}.

We can now see that the underlying manifolds $\mathcal{C}_{\mathcal{H},\mathcal{K}}$ and $\mathcal{C}_{\mathcal{W}+\mathcal{A},\mathcal{R}}$ of the reduced phase spaces $\Gamma^{\mathrm{red}}_{\mathrm{ADM}}$ and $\Gamma^{\mathrm{red}}_{\mathrm{ALT}}$ are the same. They also share identical symplectic two-forms $\omega^{\mathrm{red}}_{\mathrm{ADM}}$ and $\omega^{\mathrm{red}}_{\mathrm{ALT}}$, and therefore possess the same symplectic structure. Finally, the gauge symmetries on both reduced phase spaces---namely, spatial diffeomorphisms and matter-gauge transformations---are also the same. Thus, the two reduced phase spaces are symplectomorphic:
\begin{equation}
    \Gamma^{\mathrm{red}}_{\mathrm{ADM}} \cong_{\text{symp}} \Gamma^{\mathrm{red}}_{\mathrm{ALT}},
\end{equation}
and this symplectomorphism maps $\Sigma^{ab}$ to $\pi^{ab}$ bijectively, while acting trivially on the remaining variables $(\gamma_{ab}, \Phi^i, \Pi_{\Phi^i})$. Therefore, the ADM phase space $\Gamma_{\mathrm{ADM}}$ and the alternative phase space $\Gamma_{\mathrm{ALT}}$ are physically equivalent.

This completes the proof of Theorem~\ref{theorem: phase space equivalence}.

\subsection{Symplectomorphism between the two Phase Spaces}\label{section: Symplectomorphism between the two Phase Spaces}

Although we proved Theorem~\ref{theorem: phase space equivalence} in the previous subsection, we now present a different proof based on an alternative line of argument. This proof is more elegant and, in fact, reflects our original insight. It served as the guiding principle in identifying the required symplectomorphism and ultimately motivated the development of the covariant conformal decomposition. In what follows, we present this alternative argument, which relies on the following proposition.

\begin{proposition}\label{Prop: two gauge slices are same}
    In any gauge theory, any two distinct gauge slices on the same constraint surface are symplectomorphic to each other.
\end{proposition}

This proposition is essentially trivial, but holds under the assumption that well-defined gauge slices exist—i.e., there are no Gribov obstructions and all constraints involved are genuinely gauge constraints. The reasoning is straightforward: the symplectic form $\omega$ on the full phase space is non-degenerate, but when pulled back to the constraint surface, it becomes degenerate and is referred to as the \emph{pre-symplectic form} $\omega_{\text{pre}}$. The degenerate directions of $\omega_{\text{pre}}$ correspond precisely to the gauge directions, and so the form remains unchanged along them. Moreover, both the symplectic and pre-symplectic forms are closed: $\mathrm{d}\omega = 0$ and $\mathrm{d}\omega_{\text{pre}} = 0$. Using Cartan's magic formula for the Lie derivative $\mathcal{L}_X$ of a differential form along a vector field $X$,
\begin{equation}
    \mathcal{L}_X \omega_{\text{pre}} = \iota_X \mathrm{d}\omega_{\text{pre}} + \mathrm{d}(\iota_X \omega_{\text{pre}}),
\end{equation}
we see that for a degenerate direction $X$ (i.e., $\iota_X \omega_{\text{pre}} = 0$) and a closed form $\mathrm{d}\omega_{\text{pre}} = 0$, it follows that
\begin{equation}
    \mathcal{L}_X \omega_{\text{pre}} = 0.
\end{equation}
Hence, the pre-symplectic form is preserved under gauge transformations generated by the constraints, and its pullback to any gauge slice yields the same symplectic structure. This shows that different gauge slices are symplectomorphic.

Let us now proceed with the second proof of Theorem~\ref{theorem: phase space equivalence}.

\begin{proof}
Consider the two phase spaces
\begin{equation}
    \Gamma_1 = \left(\,\mathcal{P}, \,\omega\ ;\ \mathsf{\Pi}, \, D_a,\,G^A\,;\,\right)   \quad \text{and} \quad
    \Gamma_2 = \left(\,\mathcal{P}, \,\omega\ ;\ \sqrt{\gamma}(\varphi - 1), \, D_a,\,G^A\,;\,\right),
\end{equation}
where $\mathcal{P}$ and $\omega$ are the same as the alternative phase space manifold $\mathcal{P}_{\text{ALT}}$ and symplectic form $\omega_{\text{ALT}}$, respectively, which are also the same as the ADM phase space manifold $\mathcal{P}_{\text{ADM}}$ and symplectic form $\omega_{\text{ADM}}$. We will express everything in covariant conformal decomposed coordinates, and so $\omega$ is expressed as in equation \eqref{symplectic two form in covariant conformal decomposed coordinates}. We drop the subscript $\text{ALT}$ here, as we will be considering different sets of gauge constraints imposed on the same underlying phase space manifold and symplectic structure in various contexts.

Also, the notation here is such that the first set of constraints listed after the symplectic two-form are first-class constraints, followed by a semicolon and then any second-class constraints, if present. If there are no second-class constraints, that space is left empty. All constraints in either of these two phase spaces are clearly first-class.

In $\Gamma_1$, $\mathsf{\Pi}$ generates gauge transformations that shift $\varphi$. One choice of gauge fixing condition is $\varphi=1$. Another choice is where $\varphi$ is any function of the other phase space variables $(\gamma_{ab},\Sigma^{ab},\Phi^i,\Pi_{\Phi^i})$. One such choice is the solution to the Lichnerowicz equation itself, denoted by $\varphi_L$. So the other gauge fixing condition is $\varphi=\varphi_L$. This gauge condition is equivalent to forcing $\varphi$ to be the unique solution to the Lichnerowicz equation and hence equivalent to just imposing the LHS of the Lichnerowicz equation to vanish as a constraint—i.e., the Hamiltonian constraint $\mathcal{H}=0$ expressed in these variables.

So the following two phase spaces (obtained by imposing the gauge fixing conditions as additional constraints)
\begin{equation}
    \Gamma_3 := \left(\,\mathcal{P}, \,\omega\ ;\, D_a,\,G^A\,; \, \mathsf{\Pi}, \, \sqrt{\gamma}(\varphi-1)\,\right)   \quad  \text{and} \quad 
    \Gamma_4 := \left(\,\mathcal{P}, \,\omega\ ;\, D_a,\,G^A\,; \, \mathsf{\Pi}, \, \mathcal{H}\,\right),
\end{equation}
are symplectomorphic to each other:
\begin{equation}
    \Gamma_3   \cong_{\text{symp}} \Gamma_4.
\end{equation}

Similarly, in $\Gamma_2$, $\sqrt{\gamma}(\varphi - 1)$ generates gauge transformations that shift $\mathsf{\Pi}$. One choice of gauge fixing condition is $\mathsf{\Pi} = 0$. Another choice is where $\mathsf{\Pi}$ is any function of the other phase space variables $(\gamma_{ab},\Sigma^{ab},\Phi^i,\Pi_{\Phi^i})$. One such choice is $\mathsf{\Pi} = \frac{\Delta_{\Phi^i}}{2} \Phi^i \Pi_{\Phi^i} - \frac{1}{2}\mathcal{A}$, which is nothing but the real Weyl-anomaly constraint $\mathcal{W}+\mathcal{A}=0$.

So the following two phase spaces (obtained by imposing the gauge fixing conditions as additional constraints)
\begin{equation}
    \Gamma_3 = \left(\,\mathcal{P}, \,\omega\ ;\, D_a,\,G^A\,; \, \mathsf{\Pi}, \, \sqrt{\gamma}(\varphi-1)\,\right)   \quad  \text{and} \quad 
    \Gamma_5 := \left(\,\mathcal{P}, \,\omega\ ;\, D_a,\,G^A\,; \, \mathcal{W}+\mathcal{A}, \, \sqrt{\gamma}(\varphi-1)\,\right),
\end{equation}
are symplectomorphic to each other:
\begin{equation}
    \Gamma_3   \cong_{\text{symp}} \Gamma_5.
\end{equation}

It then follows that
\begin{equation}
    \Gamma_4   \cong_{\text{symp}} \Gamma_5.
\end{equation}

$\Gamma_4$ is nothing but the ADM phase space, and $\Gamma_5$ is nothing but the alternative phase space. This completes the second proof of Theorem~\ref{theorem: phase space equivalence}.
\end{proof}

Note that we can explicitly see how Proposition~\ref{Prop: two gauge slices are same} holds in the two instances used in the proof. In the first instance, used in $\Gamma_1$, on the constraint surface $\mathcal{C}_{\mathsf{\Pi}}$, we have $\delta\mathsf{\Pi} = 0$, and so the $\varphi$ direction becomes degenerate when $\omega$ is induced on $\mathcal{C}_{\mathsf{\Pi}}$. In the second instance, used in $\Gamma_2$, on the constraint surface $\mathcal{C}_{\varphi - 1}$, we have $\delta\varphi = 0$, and so the $\mathsf{\Pi}$ direction becomes degenerate when $\omega$ is induced on $\mathcal{C}_{\varphi - 1}$.

The symplectomorphism $\mathscr{S}_{\text{ADM} \leftrightarrow \text{ALT}}$ between the ADM phase space $\Gamma_4$ and the alternative phase space $\Gamma_5$ is given by the composition of the symplectomorphisms $\mathscr{S}_{4 \leftrightarrow 3}$ between $\Gamma_4$ and $\Gamma_3$, and $\mathscr{S}_{3 \leftrightarrow 5}$ between $\Gamma_3$ and $\Gamma_5$:
\begin{equation}
    \mathscr{S}_{\text{ADM} \leftrightarrow \text{ALT}} = \mathscr{S}_{3 \leftrightarrow 5} \circ \mathscr{S}_{4 \leftrightarrow 3}.
\end{equation}
The map $\mathscr{S}_{4 \leftrightarrow 3}$ is simply the finite gauge transformation generated by $\mathsf{\Pi}$ that relates the two gauge slices $\sqrt{\gamma}(\varphi - 1) = 0$ and $\mathcal{H} = 0$ in $\Gamma_1$. Similarly, $\mathscr{S}_{3 \leftrightarrow 5}$ is the finite gauge transformation generated by $\sqrt{\gamma}(\varphi - 1)$ that relates the two gauge slices $\mathsf{\Pi} = 0$ and $\mathcal{W} + \mathcal{A} = 0$ in $\Gamma_2$.

The smearing functions are the gauge parameters that dictate which gauge transformations are performed and by how much. To relate the two gauge slices (in either of the two cases above) via a gauge transformation, a specific choice of smearing function is used. It is important to note that the smearing functions employed in these gauge transformations may themselves depend on the phase space variables. This is entirely permissible, as smearing functions can, in principle, be arbitrary. To illustrate this, consider a gauge theory with a constraint $C(x)$ (in local form), and a smearing function $c(x)$. The integrated form of the constraint is given by 
\[ C[c] = \int d^dx\, c(x)\, C(x),\]
where $C(x)$ is assumed to be a scalar density (as it is generally best to represent all local constraints in this way). Then, for any function $F$ on phase space, the gauge transformation of $F$ generated by $C[c]$ is
\[\delta_c F = \{F, C[c]\}.\]
Now, if the smearing function $c(x)$ depends on the phase space variables, then the bracket becomes
\[\delta_c F = \{F, C[c]\} = \int d^dx\, c(x) \{F, C(x)\} + \int d^dx\, C(x) \{F, c(x)\}.\]
However, on the constraint surface, the second term vanishes because $C(x) \approx 0$. Thus, it does not matter whether the smearing functions (gauge parameters) depend on the phase space variables or not; either is valid. They may be chosen arbitrarily and still generate valid gauge transformations.

$\mathscr{S}_{4 \leftrightarrow 3}$ leaves the variables $(\gamma_{ab}, \Sigma^{ab}, \Phi^i, \Pi_{\Phi^i})$ unchanged, and $\mathscr{S}_{3 \leftrightarrow 5}$ leaves the variables $(\gamma_{ab}, \Phi^i, \Pi_{\Phi^i})$ unchanged while only changing $\Sigma^{ab}$. Therefore, under the symplectomorphism between the ADM phase space $\Gamma_4$ and the alternative phase space $\Gamma_5$, the variables $(\gamma_{ab}, \Phi^i, \Pi_{\Phi^i})$ remain unchanged, and only $\Sigma^{ab}$ is modified—just as we observed in the previous subsection.

Strictly speaking, we glossed over a subtle point. The transformation that maps the gauge slice $\mathsf{\Pi} = 0$ to the gauge slice $\mathcal{W} + \mathcal{A} = 0$ in $\Gamma_2$ is actually a \emph{large} gauge transformation, since the required smearing function does not vanish at infinity. Nonetheless, as explained just after the proof, we explicitly saw that the pre-symplectic form is preserved under this transformation, and hence the argument remains valid.

Thus, we have now provided two independent proofs of Theorem~\ref{theorem: phase space equivalence}.

\subsection{Translating between the two Phase Spaces}\label{section: Translating between the two Phase Spaces}

Let us now explain how to translate between the ADM phase space and the alternative phase space in a physically relevant context. Since we have already established a symplectomorphism between them, the translation is, in principle, known. However, it is helpful to emphasize a few key aspects of this correspondence and streamline the discussion—especially in case the details of the symplectomorphisms in the previous two subsections were overwhelming.

From this point onward, we refer to the metric $g_{ab}$ in the ADM phase space as the \emph{physical metric}, and similarly, all other variables in this phase space as \emph{physical quantities}. In the alternative phase space, we are unable to come up with a better name for the quantities, so we shall simply refer to the metric $\mathfrak{g}_{ab}$ as the \emph{alternative metric}. The remaining variables in the alternative phase space will likewise be called \emph{alternative quantities}.

The physical and alternative metrics are not directly related: in fact, the conformal factor of the alternative metric is a pure gauge degree of freedom. This implies that ``distances'' measured using the alternative metric are gauge-dependent and therefore unphysical. However, the conformal class of the alternative metric—encoding angular structure—is physical. The conformal class of the physical metric and that of the alternative metric are equivalent, since the symplectomorphism between the two phase spaces maps the conformal parts of the two metrics trivially. The same applies to the matter fields and their conjugate momenta, which are also mapped identically between the two formulations.

On the other hand, the momentum conjugate to the conformal part of the metric transforms nontrivially under the symplectomorphism. Specifically, to go from the alternative to the ADM phase space, one must compute $\Sigma^{ab}$ from $\mathsf{\Pi}^{ab}$ using equation~\eqref{momentum split in covariant conformal decomposition}; this $\Sigma^{ab}$ then becomes $\pi^{ab}$ in the ADM phase space.

\medskip

\noindent
In summary, given the alternative quantities $(\mathfrak{g}_{ab},\mathsf{\Pi}^{ab},\Phi^i,\Pi_{\Phi^i})$, the translation proceeds as follows:
\begin{itemize}
    \item Extract the conformal class of $\mathfrak{g}_{ab}$ and select the unique representative $\gamma_{ab}$ in that class whose Ricci scalar is equal to $2\Lambda$.
    \item Compute $\Sigma^{ab}$ from $\mathsf{\Pi}^{ab}$ via equation~\eqref{momentum split in covariant conformal decomposition}.
    \item Assemble $(\gamma_{ab},\Sigma^{ab},\Phi^i,\Pi_{\Phi^i})$, which are then identified as the physical quantities in the ADM phase space. The variable $\Sigma^{ab}$ is identified with $\pi^{ab}$ in the ADM phase space, and the other variables are identified trivially.
    \item Solve the Lichnerowicz equation~\eqref{Lichnerowicz equation} formed from this assembled data to obtain the conformal factor $\phi$ of the physical metric.
    \item Weyl transform $\gamma_{ab}$ to obtain the physical metric: $g_{ab} = \phi^\alpha \gamma_{ab}$.
    \item Define the physical momentum $\Pi^{ab}$ conjugate to $g_{ab}$ as $\Pi^{ab} = \phi^{-\alpha} \pi^{ab}$.
    \item Finally, assemble $(g_{ab},\Pi^{ab},\Phi^i,\Pi_{\Phi^i})$. This is now a point on the constraint surface $\mathcal{C}_{\mathcal{H},\mathcal{K}} \subset \mathcal{P}_{\text{ADM}}$.
\end{itemize}

\noindent
This completes the reconstruction of all physical quantities in the ADM phase space.

\medskip

\noindent
Observables in both phase spaces are functions on their respective constraint surfaces that are invariant under their respective gauge transformations. They are related to each other via the symplectomorphism.

\subsection{Inheriting Dynamics from the Boundary ADM Hamiltonian}\label{section: Inheriting Dynamics from the Boundary ADM Hamiltonian}

So far, the discussion of physical equivalence between the ADM and alternative phase spaces has been purely kinematical. A physical theory, however, consists not just of a phase space but also of a Hamiltonian that generates dynamics. This Hamiltonian is a function on the phase space. In the ADM phase space, the extended Hamiltonian $H_{\text{ADM}}$ is given by equation~\eqref{extendedHamiltonian} and governs the boundary dynamics. We represent this physical system $\mathcal{S}_{\text{ADM}}$ by a tuple consisting of the ADM phase space $\Gamma_{\text{ADM}}$ and the extended Hamiltonian $H_{\text{ADM}}$:
\begin{equation}
    \mathcal{S}_{\text{ADM}} = \big(\Gamma_{\text{ADM}}, H_{\text{ADM}}\big).
\end{equation}

For the equivalence to extend to the dynamical level, both phase spaces must be equipped with an extended Hamiltonian, and the symplectomorphism between them must preserve the Hamiltonian. However, no such Hamiltonian has yet been defined on the alternative phase space. A natural approach is to inherit the extended Hamiltonian from the ADM phase space by pulling it back through the symplectomorphism. This ensures that the dynamics on the alternative phase space precisely mirror those of the ADM formulation.

Let us elaborate on what we mean here. Both $\Gamma_{\text{ADM}}$ and $\Gamma_{\text{ALT}}$ are phase spaces equipped with different sets of gauge constraints. By the symplectomorphism between them, we mean the symplectomorphism $\mathscr{S}_{\text{ADM} \leftrightarrow \text{ALT}}$ between the reduced phase spaces $\Gamma^{\text{red}}_{\text{ADM}}$ and $\Gamma^{\text{red}}_{\text{ALT}}$.

Although $H_{\text{ADM}}$ is defined on the whole of $\mathcal{P}_{\text{ADM}}$, it is its restriction to the constraint surface $\mathcal{C}_{\mathcal{H},\mathcal{K}} \subset \mathcal{P}_{\text{ADM}}$ that is physically meaningful. Denote this restriction by $H_{\text{ADM}}|_{\mathcal{C}_{\mathcal{H},\mathcal{K}}}$. This function is then pushed forward via the symplectomorphism $\mathscr{S}_{\text{ADM} \leftrightarrow \text{ALT}}$ to define a corresponding function $h_{\mathcal{C}_{\mathcal{W}+\mathcal{A},\mathcal{R}}}$ on the constraint surface $\mathcal{C}_{\mathcal{W}+\mathcal{A},\mathcal{R}}$:
\begin{equation}
h_{\mathcal{C}_{\mathcal{W}+\mathcal{A},\mathcal{R}}} =H_{\text{ADM}}|_{\mathcal{C}_{\mathcal{H},\mathcal{K}}} \circ \mathscr{S}_{\text{ADM} \leftrightarrow
\text{ALT}}.
\end{equation}

The diffeomorphism constraints $D_a$ and matter-gauge constraints $G^A$, originally expressed in the ADM variables, are transformed by the symplectomorphism into the same equations, now expressed in the alternative phase space variables. The Hamiltonian constraint $\mathcal{H}$ appearing in $H_{\text{ADM}}$ vanishes on the constraint surface $\mathcal{C}_{\mathcal{H},\mathcal{K}}$. Thus, the function $h_{\mathcal{C}_{\mathcal{W}+\mathcal{A},\mathcal{R}}}$ becomes
\begin{equation}
h_{\mathcal{C}_{\mathcal{W}+\mathcal{A},\mathcal{R}}} =H_{\text{bdy}}|_{\mathcal{C}_{\mathcal{H},\mathcal{K}}} \circ \mathscr{S}_{\text{ADM} \leftrightarrow
\text{ALT}} + D[N^a] + G[\alpha] . 
\end{equation}

Then one can extend this function to a function $h_{\mathcal{C}_{\mathcal{W}+\mathcal{A}}}$ on $\mathcal{C}_{\mathcal{W}+\mathcal{A}}$ in such a way that it is gauge-invariant under the gauge transformations generated by $W\!A[\omega]$ for any $\omega$:
\begin{equation}
h_{\mathcal{C}_{\mathcal{W}+\mathcal{A}}} =H_{\text{bdy}}|_{\mathcal{C}_{\mathcal{H},\mathcal{K}}} \circ \mathscr{S}_{\text{ADM} \leftrightarrow
\text{ALT}} \circ \mathscr{W} + D[N^a] + G[\alpha] , 
\end{equation}
where $\mathscr{W}$ maps any point in $\mathcal{C}_{\mathcal{W}+\mathcal{A}}$ to its gauge-equivalent point in $\mathcal{C}_{\mathcal{W}+\mathcal{A},\mathcal{R}}$ under a gauge transformation generated by $W\!A[\omega]$. Then one can extend this function arbitrarily away from the surface $\mathcal{C}_{\mathcal{W}+\mathcal{A}}$ to obtain a function $H_{\text{ALT}}$ on $\mathcal{P}_{\text{ALT}}$:
\begin{equation}\label{inherited hamiltonian}
H_{\text{ALT}}=H_{\text{bdy}}|_{\mathcal{C}_{\mathcal{H},\mathcal{K}}} \circ \mathscr{S}_{\text{ADM} \leftrightarrow
\text{ALT}} \circ \mathscr{W} + W\!A[\omega] + D[N^a] + G[\alpha]. 
\end{equation}

This is how the ADM Hamiltonian must be inherited to the alternative phase space in order to define dynamics there, thereby providing an alternative description of the bulk gravity plus matter fields as a physical system:
\begin{equation}
    \mathcal{S}_{\text{ALT}} = \big(\Gamma_{\text{ALT}},H_{\text{ALT}}\big).
\end{equation}

In this sense, asserting that the two phase spaces—with their respective extended Hamiltonians—are physically equivalent even at the dynamical level becomes tautological, since the dynamics on the alternative phase space are defined by construction to match those of the ADM phase space, and we have already established their physical equivalence at the kinematical level. Although we have explained how to do this formally, we will not carry it out explicitly in this paper and instead aim to do so in a second paper.

\section{Quantum Gravity Wave Function: \texorpdfstring{$\Psi_{\text{QG}} = Z_{\text{CFT}}^{(ic)}$}{Quantum Gravity Wave Function}}\label{Quantum Gravity Wave Function}

Now that we have established the physical equivalence between the ADM phase space and the alternative phase space—at least at the kinematical level—we can exploit this correspondence to our advantage in the quantum theory. We now turn to quantization in the alternative phase space. In Section~\ref{Quantization in the Alternative Phase Space}, we explain Dirac's constraint quantization procedure in the alternative phase space. In Section~\ref{CFT Partition Functions as Quantum Gravity States in the Alternative Phase Space}, we show that certain CFT partition functions defined on bulk Cauchy slices satisfy the quantum constraints, thereby yielding candidate quantum gravity states in the alternative formulation—this is the content of Theorem~\ref{theorem: QG states}, which we will thus prove.

\subsection{Quantization in the Alternative Phase Space}\label{Quantization in the Alternative Phase Space}

Let us start with the classical alternative phase space, which we rewrite here while reminding the reader of some of its features:
\begin{equation}
   \Gamma_{\text{ALT}} = \left(\,\mathcal{P}_{\text{ALT}}, \,\omega_{\text{ALT}}\ ;\ \mathcal{W}+\mathcal{A}, \, D_a,\,G^A\,\right).
\end{equation}
The phase space variables here consist of the set of geometries $\mathfrak{g}_{ab}$, matter fields $\Phi^i$, and their respective canonically conjugate momenta $\mathsf{\Pi}^{ab}$ and $\Pi_{\Phi^i}$. These coordinatise the underlying manifold $\mathcal{P}_{\text{ALT}}$. The symplectic two-form $\omega_{\text{ALT}}$ is
\begin{equation}
\omega_{\text{ALT}} = \int_{\Sigma} d^d x \left( \delta \mathsf{\Pi}^{ab} \wedge \delta \mathfrak{g}_{ab} + \delta \Pi_{\Phi^i} \wedge \delta \Phi^i \right).
\end{equation}
This gives the canonical Poisson brackets among the fields:
\begin{subequations}
\begin{align}
\{\mathfrak{g}_{ab}(x),\,\mathsf{\Pi}^{cd}(y)\} 
&= \delta_{ab}^{cd}\,\delta^{(d)}(x,y), \\[6pt]
\{\Phi^i(x),\,\Pi_{\Phi^j}(y)\} 
&= \delta^i_{\;j}\,\delta^{(d)}(x,y),
\end{align}
\end{subequations}
where $\delta_{ab}^{cd} = \frac{1}{2}(\delta_a^c \delta_b^d + \delta_a^d \delta_b^c)$, and all other Poisson brackets among the phase space variables vanish. The real Weyl-anomaly constraint $\mathcal{W}+\mathcal{A}$, the spatial diffeomorphism constraint $D_a$, and the matter-gauge constraints $G^A$ form a set of first-class (gauge) constraints with a closed Poisson bracket algebra. Let us write these gauge constraints again:
\begin{subequations}
\begin{align}
     \mathcal{W}+\mathcal{A} &= 2\mathsf{\Pi} - \Delta_{\Phi^i} \Phi^i \Pi_{\Phi^i} + \mathcal{A}=0,\\
     D_a(x) &= -2 \mathfrak{g}_{ab} \nabla_c \mathsf{\Pi}^{bc} + \mathfrak{g}_{ab} D^b_{\text{matter}}(x)=0,\\
     G^A &= 0,
\end{align}
\end{subequations}
where $\nabla$ is the connection compatible with $\mathfrak{g}_{ab}$. Let us also recall that the conformal anomaly $\mathcal{A}$ is of the form
\begin{equation}\label{form of anolamy}
    \mathcal{A} = (\text{central charge}) \ \times \ (\text{some local function of } \mathfrak{g}_{ab} \text{ and } \Phi^i).
\end{equation}

In the quantum theory, the phase space variables are upgraded to quantum operators\footnote{We are working in the metric and matter field representation. This is analogous to the position representation in standard quantum mechanics, where states are wavefunctions $\psi(x)$ of position $x$, the position operator is $\widehat{x}=x$, and the momentum operator is $\widehat{p}=-i\frac{\partial}{\partial x}$. We set $\hbar = 1$.}
\begin{subequations}
\begin{align}
   \mathfrak{g}_{ab} \to \widehat{\mathfrak{g}}_{ab} = \mathfrak{g}_{ab}  \ &, \quad \mathsf{\Pi}^{ab} \to \widehat{\mathsf{\Pi}}^{ab} = -i \frac{\delta}{\delta \mathfrak{g}_{ab}}
 \ ,\\ \Phi^i \to \widehat{\Phi}^i = \Phi^i \ &, \quad \Pi_{\Phi^i} \to \widehat{\Pi}_{\Phi^i} = -i \frac{\delta}{\delta \Phi^i}
    \ ,
\end{align}
\end{subequations}
satisfying the canonical commutation relations:
\begin{subequations}
\begin{align}
[\mathfrak{g}_{ab}(x),\,\widehat{\mathsf{\Pi}}^{cd}(y)] 
&= i\,\delta_{ab}^{cd}\,\delta^{(d)}(x,y), \\[6pt]
[\Phi^i(x),\,\widehat{\Pi}_{\Phi^j}(y)] 
&=i \, \delta^i_{\;j}\,\delta^{(d)}(x,y),
\end{align}
\end{subequations}
obtained from replacing the Poisson brackets with commutators along with a factor of $i$. The gauge constraints are then imposed as operator constraints on the quantum states. Quantum gravity states are wavefunctionals $\Psi_{\text{QG}}[\mathfrak{g},\Phi^i]$ of the metric $\mathfrak{g}_{ab}$ and matter fields $\Phi^i$, satisfying the operator gauge constraints:
\begin{align}\label{operator constraints in alternative phase space formulation}
    \left( \widehat{\mathcal{W}}+\mathcal{A}\right) \ \Psi_{\text{QG}}[\mathfrak{g},\Phi^i] &=0 \ , \quad \quad
     \widehat{D}_a \ \Psi_{\text{QG}}[\mathfrak{g},\Phi^i] =0 \ , \quad \quad
     \widehat{G}^A \ \Psi_{\text{QG}}[\mathfrak{g},\Phi^i] = 0 \ .
\end{align}
Any wavefunctional satisfying the above operator constraints is a valid quantum gravity state.

\subsection{CFT Partition Functions as Candidate Quantum Gravity States in the Alternative Phase Space}\label{CFT Partition Functions as Quantum Gravity States in the Alternative Phase Space}

Now we will construct candidate quantum gravity states in the alternative phase space formulation. Let us begin by recalling that, initially, the holographic CFT lives on the conformal boundary of AAdS with background metric $g_0$ and background source $\Phi_0$. Its partition function $Z_{\text{CFT}}[g_0,\Phi_0]$ is a functional of the background sources $g_0$ and $\Phi_0$. The central charge $c$ of this holographic CFT is related to a dimensionless free parameter of the bulk theory by the holographic duality. To make explicit that the CFT partition function depends on the central charge, we denote it as $Z^{(c)}_{\text{CFT}}[g_0,\Phi_0]$, thereby treating it as a function of $c$.\footnote{Of course, for any given CFT, the central charge $c$ is a fixed number. However, what we mean here is to consider a family of CFTs parametrized by $c$, and then view the partition function as a function over this family.}

This theory is then Wick-rotated and placed on an asymptotically hyperbolic Riemannian manifold $\Sigma$ with metric $\mathfrak{g}_{ab}$ and matter fields $\Phi^i$ (which now act as sources in the CFT). Since $\Sigma$ has a boundary $\partial\Sigma$, to define the partition function of this CFT, one needs to specify boundary conditions. This is naturally done by inputting a CFT state $\psi_{\text{CFT}}$, living on $\partial\Sigma$, as the boundary condition to the CFT partition function.

Inputting a CFT state as a boundary condition is done in the following way. We choose a complete commuting set of operators $\{\chi\}$ acting on the CFT Hilbert space associated with the boundary segment $\partial\Sigma|_{\partial\mathcal{M}}$, that is, $\partial\Sigma$ as embedded on the boundary $\partial\mathcal{M}$. Crucially, this does not necessarily coincide with evaluating the same operators on $\partial\Sigma|_{\Sigma}$, that is, $\partial\Sigma$ as embedded on $\Sigma$, because the sharp geometric corner at the interface between the bulk and boundary can cause certain operators—such as the stress tensor—to exhibit discontinuities across the junction (a point we previously explained in \cite{CSH}). This operator set $\{\chi\}$ defines a basis of states labeled by their eigenvalues $\{\chi\}$,\footnote{We are using the same symbol $\{\chi\}$ to denote both the complete set of operators and the set of eigenvalues used as boundary conditions for the partition function. The context should make this distinction clear.} and any general state on $\partial\Sigma|_{\partial\mathcal{M}}$ can be expressed as a superposition: $\psi_{\text{CFT}}[\{\chi\}]$.\footnote{These operators need not be local or fundamental fields, which often fail to form well-defined operators in strongly coupled theories. Instead, ${\chi}$ may include eigenvalues of nonlocal observables, such as total energy or angular momentum.} Then by the junction conditions\footnote{In \cite{CSH}, the junction conditions were derived from the Hamiltonian and diffeomorphism constraints but here they will be consequences of the imaginary Weyl-anomaly constraint, diff constraints and matter gauge constraints.} the operator set $\{\chi\}$ maps to the operator set $\{\widetilde{\chi}\}$ now defined on $\partial\Sigma|_{\Sigma}$, giving a state $\widetilde{\psi}_{\text{CFT}}[\{\widetilde{\chi}\}]$ on $\partial\Sigma|_{\Sigma}$, obtained from evolving $\psi_{\text{CFT}}[\{\chi\}]$ across a smooth regulated strip interpolating from $\partial\Sigma|_{\partial\mathcal{M}}$ to $\partial\Sigma|_{\Sigma}$, and then taking the limit where the regulator is sent to zero. See section 7.2 in \cite{CSH} for details of this interpolating strip. The partition function $Z^{(c)}_{\text{CFT}}[\mathfrak{g}_{ab},\Phi^i,\{\widetilde{\chi}\}]$ is then a functional of both the source fields ($\mathfrak{g}_{ab},\Phi^i$) and the boundary data $\{\widetilde{\chi}\}$. Then the state can be inputted as a boundary condition as
\begin{equation}
  Z^{(c)}_{\text{CFT}}[\mathfrak{g},\Phi^i;\psi_{\text{CFT}}] = \int d\{\widetilde{\chi}\}   \, Z^{(c)}_{\text{CFT}}[\mathfrak{g}_{ab},\Phi^i,\{\widetilde{\chi}\}] \, \widetilde{\psi}_{\text{CFT}}[\{\widetilde{\chi}\}].
\end{equation}

This now defines the partition function $Z^{(c)}_{\text{CFT}}[\mathfrak{g},\Phi^i;\psi_{\text{CFT}}]$ of this holographic CFT living on $\Sigma$. This CFT partition function is a wavefunctional of $\mathfrak{g}_{ab}$ and $\Phi^i$, and a function of the central charge $c$. Moreover, this CFT partition function $Z^{(c)}_{\text{CFT}}[\mathfrak{g},\Phi^i;\psi_{\text{CFT}}]$ satisfies the operator constraints 
\begin{align}\label{operator constraints satsified by CFT}
    \left( \widehat{\mathcal{W}} - i \mathcal{A}\right) \ Z^{(c)}_{\text{CFT}}[\mathfrak{g},\Phi^i;\psi_{\text{CFT}}] &=0 \ , \quad 
     \widehat{D}_a \ Z^{(c)}_{\text{CFT}}[\mathfrak{g},\Phi^i;\psi_{\text{CFT}}] =0 \ , \quad 
     \widehat{G}^A \ Z^{(c)}_{\text{CFT}}[\mathfrak{g},\Phi^i;\psi_{\text{CFT}}] = 0 \ .
\end{align}
Let us elaborate on these equations one by one.

The first equation of \eqref{operator constraints satsified by CFT} is the standard Weyl-anomaly equation (which we refer to as the \emph{imaginary Weyl-anomaly equation} to distinguish it from the real Weyl-anomaly equation), and it is satisfied by all conformal field theories with their respective anomaly. If this equation is unfamiliar, let us explain it in a more familiar form for $d=2$ CFTs coupled only to a metric source and no other source fields.

Consider a $d=2$ classical CFT with an Euclidean action $S_{\text{CFT}}[\chi,\mathfrak{g}]$ on a curved background $\mathfrak{g}_{ab}$, where $\chi$ denotes the dynamical fields. The stress-energy tensor is defined as
\begin{equation}
T^{ab} = \frac{2}{\sqrt{\mathfrak{g}}} \frac{\delta S_{\text{CFT}}}{\delta \mathfrak{g}_{ab}},
\end{equation} 
and is classically traceless due to conformal symmetry:
\begin{equation}
T =\mathfrak{g}_{ab}T^{ab}=0.
\end{equation}
However, after quantization, the expectation value $\langle \widehat{T} \rangle$ of the trace of the stress-energy tensor operator $\widehat{T}$ in any CFT state can become non-zero:
\begin{equation}\label{conformal anomaly for T}
    \langle \widehat{T} \rangle = -\frac{c}{24\pi} R_{[\mathfrak{g}]}.
\end{equation}
$ \langle \widehat{T} \rangle$ can also be expressed as
\begin{align}
     \langle \widehat{T} \rangle 
     &= \frac{1}{Z_{\text{CFT}}} \int D\chi e^{-S_{\text{CFT}}} T 
     = \frac{1}{Z_{\text{CFT}}} \int D\chi e^{-S_{\text{CFT}}} \mathfrak{g}_{ab} \frac{2}{\sqrt{\mathfrak{g}}} \frac{\delta S_{\text{CFT}}}{\delta \mathfrak{g}_{ab}} ,\nonumber\\
     &= -\frac{1}{Z_{\text{CFT}}} \mathfrak{g}_{ab} \frac{2}{\sqrt{\mathfrak{g}}} \frac{\delta }{\delta \mathfrak{g}_{ab}} Z_{\text{CFT}} = -\frac{i}{Z_{\text{CFT}}}  \frac{2}{\sqrt{\mathfrak{g}}} \widehat{\mathsf{\Pi}} Z_{\text{CFT}}.
\end{align}
Thus, equation \eqref{conformal anomaly for T} can be written as
\begin{equation}
       \left(2\widehat{\mathsf{\Pi}}   + i \frac{c}{24\pi} \sqrt{\mathfrak{g}} R_{[\mathfrak{g}]} \right) Z_{\text{CFT}} = 0.
\end{equation}
This is precisely the imaginary Weyl-anomaly equation for the $d=2$ case with only a background metric. The Weyl generator and the conformal anomaly in this case are given by:
\begin{equation}
\widehat{\mathcal{W}}=2\widehat{\mathsf{\Pi}} \ , \quad \mathcal{A} = - \frac{c}{24\pi} \sqrt{\mathfrak{g}} \, R_{[\mathfrak{g}]}.
\end{equation}
The central charge $c$ depends on the theory. For a holographic $d=2$ CFT, it is related to bulk parameters as:
\begin{equation}\label{d=2 c interms of bulk free parameters}
    c= \frac{12 \pi}{\kappa\sqrt{|\Lambda|}}. 
\end{equation}

Let us now give another example: a $d=4$ CFT coupled to a background metric $\mathfrak{g}_{ab}$ and no other sources. The expectation value of the trace of the stress-energy tensor in any state satisfies
\begin{equation}
     \langle \widehat{T} \rangle =- \frac{c}{16\pi^2} C^{abcd}C_{abcd}+\frac{a}{16\pi^2}E,
\end{equation}
where $C$ is the Weyl tensor and $E$ is the Euler density. The coefficients $c$ and $a$ are model-dependent and may differ, but for a holographic theory, they are equal: $a = c$, and are related to bulk parameters via:
\begin{equation}
 c= \frac{\pi L_{AdS}^3}{8 G_N }   = \frac{\pi^2 }{\kappa} \left(\frac{6}{|\Lambda|} \right)^{3/2}.
\end{equation}
This holds for $\mathcal{N}=4$ $SU(N)$ Super Yang-Mills theory. The trace anomaly becomes:
\begin{equation}
     \langle \widehat{T} \rangle = -\frac{c}{8\pi^2} \left(G^{ab}G_{ab} - \frac{1}{3}G^2\right),
\end{equation}
which can be expressed as the imaginary Weyl-anomaly equation:
\begin{equation}
   \left( 2\widehat{\mathsf{\Pi}}  -i\mathcal{A} \right)Z_{\text{CFT}} = 0,
\end{equation}
where the anomaly is:
\begin{equation}
    \mathcal{A} = -\frac{c}{8\pi^2} \sqrt{\mathfrak{g}}\left(G^{ab}G_{ab} - \frac{1}{3}G^2\right).
\end{equation}

In general, for any $d$-dimensional holographic CFT coupled to a background metric $\mathfrak{g}_{ab}$ and matter sources $\Phi^i$, the CFT partition function satisfies the imaginary Weyl-anomaly equation, and the conformal anomaly is of the form given in \eqref{form of anolamy}.

Next, the second equation of \eqref{operator constraints satsified by CFT} is trivial because it is simply a consequence of the CFT partition function being a covariant functional of $\mathfrak{g}_{ab}$ and $\Phi^i$, meaning it does not depend on the coordinate system used on $\Sigma$. The diffeomorphism constraint, when imposed as an operator constraint, essentially enforces this covariance. This property is not unique to CFTs; in fact, any QFT partition function $Z_{\text{QFT}}$ is covariant and thus satisfies this equation: $\widehat{D}_a Z_{\text{QFT}} = 0$. If there are no matter sources, then this equation reduces to the covariant conservation of the stress-energy tensor in the quantum theory:
\begin{equation}
\langle \nabla_a \widehat{T}^{ab} \rangle = 0.
\end{equation}
When the QFT is coupled to other sources $\Phi^i$, the stress-energy tensor alone is not covariantly conserved; instead, the full expression $\langle \widehat{D}_a \rangle = 0$ holds.

Finally, the third equation of \eqref{operator constraints satsified by CFT} follows from the holographic duality as precisely stated in equation \eqref{main statement of AdS/CFT duality}. Let us explain how. The gravitational path integral $Z_{\text{grav}}$ defined in equation \ref{gravitational path integral definition} is manifestly invariant under matter-gauge transformations, and therefore satisfies
\begin{equation}
\widehat{G}^A Z_{\text{grav}} = 0.
\end{equation}
The holographic counterterms are also invariant under matter-gauge transformations, as they arise directly from terms in the Hamiltonian constraint. We previously explained why every term in the Hamiltonian constraint is invariant under such transformations. It follows that the renormalised gravitational path integral $Z^{\text{ren}}_{\text{grav}}$, defined in equation \ref{renormalised gravitational path integral definition}, also satisfies
\begin{equation}
    \widehat{G}^A Z^{\text{ren}}_{\text{grav}} = 0.
\end{equation}
Moreover, from equation \eqref{PBweylG}, it follows that $\widehat{G}^A$ is invariant under the rescalings in equation \eqref{rescaling phase space variables} (interpreted now at the quantum level via the commutator analog of the classical Poisson bracket). Therefore, even after rescaling the arguments of $Z^{\text{ren}}_{\text{grav}}$ as done in equation \eqref{main statement of AdS/CFT duality} and sending the finite boundary to infinity (i.e., $\epsilon \to 0$), the resulting object still satisfies
\begin{equation}
\widehat{G}^A \lim_{\epsilon\to 0} Z^{\text{ren}}_{\text{grav}} = 0.
\end{equation}
By the equality in \eqref{main statement of AdS/CFT duality}, it follows that the boundary Lorentzian CFT partition function also satisfies
\begin{equation}
    \widehat{G}^A Z_{\text{CFT}} = 0,
\end{equation}
i.e., the Lorentzian CFT partition function is invariant under matter-gauge transformations of its sources $\Phi^i$. Wick rotation does not affect this property, so the Euclidean CFT placed on $\Sigma$ also satisfies
\begin{equation}
    \widehat{G}^A Z^{(c)}_{\text{CFT}}[\mathfrak{g},\Phi^i;\psi_{\text{CFT}}] = 0.
\end{equation}
Thus, $Z^{(c)}_{\text{CFT}}[\mathfrak{g},\Phi^i;\psi_{\text{CFT}}]$ satisfies all of the operator constraints in equation \eqref{operator constraints satsified by CFT}.

The operator constraints \eqref{operator constraints satsified by CFT} satisfied by $Z^{(c)}_{\text{CFT}}[\mathfrak{g},\Phi^i;\psi_{\text{CFT}}]$ are almost the quantum operator constraints \eqref{operator constraints in alternative phase space formulation} that a quantum gravity state must satisfy—but not quite. The only difference lies in the factor of $``i"$ in the imaginary Weyl-anomaly equation. While $Z^{(c)}_{\text{CFT}}$ satisfies the imaginary Weyl-anomaly constraint in \eqref{operator constraints satsified by CFT}, we require wavefunctionals that satisfy the real Weyl-anomaly constraint \eqref{operator constraints in alternative phase space formulation} in order for them to be candidate quantum gravity states.

This is precisely where we invoke the assumption of theorem \ref{theorem: QG states}. Namely, we assume that the CFT partition function $Z^{(c)}_{\text{CFT}}$ is an analytic function of its central charge $c$. We then analytically continue this function to imaginary central charge $c \to ic$ to obtain $Z^{(ic)}_{\text{CFT}}[\mathfrak{g},\Phi^i;\psi_{\text{CFT}}]$.\footnote{Under this analytic continuation, any correlation function in the CFT that is expressed as a function of $c$ also gets analytically continued to $ic$. This naively just means replacing all appearances of $c$ with $ic$. This yields the correlation functions in the analytically continued CFT.}

Since the conformal anomaly $\mathcal{A}$ is proportional to $c$ as per equation \eqref{form of anolamy}, it transforms as $\mathcal{A} \to i\mathcal{A}$ under this continuation. Therefore, the operator $\widehat{\mathcal{W}} - i \mathcal{A}$ becomes $\widehat{\mathcal{W}} + \mathcal{A}$, and the equation
\begin{equation}
\left( \widehat{\mathcal{W}} - i \mathcal{A}\right) \ Z^{(c)}_{\text{CFT}}[\mathfrak{g},\Phi^i;\psi_{\text{CFT}}] =0
\end{equation}
becomes
\begin{equation}
\left( \widehat{\mathcal{W}} + \mathcal{A}\right) \ Z^{(ic)}_{\text{CFT}}[\mathfrak{g},\Phi^i;\psi_{\text{CFT}}] =0.
\end{equation}

The other operator constraint equations remain unaffected since the central charge $c$ does not appear in $D_a$ or $G^A$. Analytically continuing $c$ does not spoil spatial covariance or invariance under matter-gauge transformations. Thus, the analytically continued CFT partition function $Z^{(ic)}_{\text{CFT}}$ satisfies all of the quantum operator constraints \eqref{operator constraints in alternative phase space formulation}:
\begin{align}
    \left( \widehat{\mathcal{W}} + \mathcal{A}\right) \ Z^{(ic)}_{\text{CFT}}[\mathfrak{g},\Phi^i;\psi_{\text{CFT}}] &=0 \ , \quad 
     \widehat{D}_a \ Z^{(ic)}_{\text{CFT}}[\mathfrak{g},\Phi^i;\psi_{\text{CFT}}] =0 \ , \quad 
     \widehat{G}^A \ Z^{(ic)}_{\text{CFT}}[\mathfrak{g},\Phi^i;\psi_{\text{CFT}}] = 0 \ .
\end{align}
Therefore, it provides us with a candidate quantum gravity state:
\begin{equation}\label{CFT partition function as QG states}
     \Psi_{\text{QG}}[\mathfrak{g}_{ab},\Phi^i] = Z^{(ic)}_{\text{CFT}}[\mathfrak{g}_{ab},\Phi^i;\psi_{\text{CFT}}].
\end{equation}
This completes the proof of theorem \ref{theorem: QG states}.

Let us now briefly comment on the significance of this result. The above equation defines different bulk candidate quantum gravity states $\Psi_{\text{QG}}$ for different boundary CFT states $\psi_{\text{CFT}}$. This construction provides an explicit, non-perturbative, and background-independent realization of candidate quantum gravity states. These states arise from CFT partition functions that are themselves well defined and UV complete. At no point in our construction did we invoke a perturbative expansion around a fixed background geometry, nor did we select any particular classical spacetime to quantize around. Instead, the candidate quantum gravity states emerge naturally and directly from the CFT partition function, without appealing to any specific background structure. Moreover, the entire construction avoids the need for gauge fixing, and therefore evades the Gribov problem—an obstruction that arises even in simple gauge theories like Yang–Mills. In our earlier work \cite{CSH}, Wheeler–DeWitt states were constructed using partition functions of the $T^2$ theory, but this approach requires a UV completion of the $T^2$ theory to move beyond an effective description of quantum gravity. Here, by contrast, the candidate quantum gravity states are expressed directly as partition functions of conformal field theories, which are already UV complete. We will return in the discussion section to elaborate further on the interpretation and broader implications of this result.

\section{Discussion}\label{Discussion}

In this section, we explore several broader implications of the alternative phase space formulation. We begin by outlining current limitations, open questions, and directions for future work. We then contrast WDW states with quantum gravity states in the alternative phase space formulation, emphasizing how the latter provides a sharper notion of bulk time beyond the semiclassical regime. Next, we discuss the advantage of having QG states realized as CFT partition functions: it provides a pathway toward a UV-complete description of bulk quantum gravity states. We then conclude with a brief summary of the main results.

\subsection{Outstanding Issues and Future Work}

Throughout the paper, we emphasized that the analytically continued holographic CFT partition function $Z^{(ic)}_{\text{CFT}}$ serves as a “candidate” quantum gravity state. This phrasing reflects the fact that additional criteria must be met to establish such a functional as a valid quantum gravity state. Beyond satisfying the operator gauge constraints, one expects valid quantum gravity states to be normalizable with respect to a suitably defined inner product—something we have not yet constructed, but intend to address in future work. Another important direction is to determine the inherited Hamiltonian in the alternative phase space and study its spectrum by identifying its eigenstates.

Note that in equation \eqref{CFT partition function as QG states}, different choices of boundary CFT states $\psi_{\text{CFT}}$—used to impose boundary conditions for defining the CFT partition function on the open manifold $\Sigma$—lead to different quantum gravity states in this alternative phase space picture. This defines a map $\mathcal{Z}$ from the boundary CFT Hilbert space $\mathscr{H}_{\text{CFT}}$ to a space of bulk quantum gravity states:
\begin{equation}\label{tentative dictionary}
    \begin{aligned}
        \mathcal{Z}: \quad & \mathscr{H}_{\text{CFT}} \longrightarrow \{\text{Bulk Quantum Gravity States}\} \\
                 & \psi_{\text{CFT}} \longmapsto \Psi_{\text{QG}}[\mathfrak{g}_{ab},\Phi^i] = Z^{(ic)}_{\text{CFT}}[\mathfrak{g}_{ab},\Phi^i;\psi_{\text{CFT}}].
    \end{aligned}
\end{equation}
This potentially opens the door to a new AdS/CFT dictionary, though further work is required to determine whether it provides the correct correspondence. By an AdS/CFT dictionary, we mean a one-to-one map $\mathcal{Z}$ between the boundary CFT Hilbert space $\mathscr{H}_{\text{CFT}}$ and the bulk quantum gravity Hilbert space $\mathscr{H}_{\text{QG}}$.

An obvious requirement for the map is that it should be linear and preserve the inner product. The map proposed in equation~\eqref{tentative dictionary} is indeed linear. The main requirement we must ensure is that the map $\mathcal{Z}$ preserves the dynamics—meaning the boundary CFT dynamics must be consistent with those of the dual bulk theory. As a first step, this involves identifying the Hamiltonian in the alternative phase space, $H_{\text{ALT}}$, which is inherited from the ADM Hamiltonian as described in Section~\ref{section: Inheriting Dynamics from the Boundary ADM Hamiltonian}. This Hamiltonian must then be promoted to a quantum operator $\widehat{H}_{\text{ALT}}$. Now, $\widehat{H}_{\text{ALT}}$ depends on the boundary lapse and shift, because the first term in equation~\eqref{inherited hamiltonian} contains the boundary term of the ADM Hamiltonian, $H_{\text{bdy}}$, which itself depends on these boundary quantities. While the bulk terms in $\widehat{H}_{\text{ALT}}$ generate gauge transformations in the bulk, its boundary term generates the boundary dynamics, which must be consistent with the boundary dynamics via the dictionary. Therefore, for any chosen value of the boundary lapse and shift, if the conformal boundary metric is ADM decomposed using those values, then the CFT Hamiltonian operator $\widehat{H}_{\text{CFT}}$ in that boundary coordinate system must satisfy
\begin{equation}
    \widehat{H}_{\text{ALT}}\, \mathcal{Z}(\psi_{\text{CFT}}) = \mathcal{Z}(\widehat{H}_{\text{CFT}}\, \psi_{\text{CFT}}),
\end{equation}
in order for the dynamics to be consistent. The holographic duality~\eqref{main statement of AdS/CFT duality} gives us reason to expect this to hold, but we still need to determine exactly how this happens and whether it holds for the map given in~\eqref{tentative dictionary} or for some modified map $\mathcal{Z}$. We aim to explore this further in future work.

Once the correct dictionary is established, one can always evolve boundary CFT states and map them to bulk quantum gravity states to study how these bulk states evolve, thereby gaining insight into dynamical quantum gravitational processes in the bulk. However, to fully extract physical information, one must also map the alternative quantities to physical quantities. All of the bulk quantum gravity data would then be encoded in, and extractable from, the correlation functions of this analytically continued CFT evaluated on bulk maximal volume slices.

\subsection{WDW states vs QG states in the Alternative Phase Space Formulation}

In the WDW formulation, the WDW equation (the Hamiltonian constraint imposed as a quantum operator equation $\widehat{H} \Psi_{\text{WDW}} = 0$) encodes temporal diffeomorphism invariance of the quantum gravity states $\Psi_{\text{WDW}}$. The WDW state $\Psi_{\text{WDW}}$ encodes physics throughout the entire bulk domain of dependence $\Omega$ (a.k.a.~WDW patch) of some boundary Cauchy surface $\partial\Sigma$. One should think of WDW states as “living” on $\partial\Sigma$, analogous to quantum mechanical states $\psi(x,t)$ of a particle living on a time slice. The full set of bulk Cauchy slices within $\Omega$ are related by gauge transformations, as bulk time translations are gauge redundancies.

Since temporal gauge transformations were not fixed prior to quantization, extracting bulk physics from the WDW state at some chosen “time” requires conditioning the arguments of the WDW state appropriately. For this, clocks and observers must be defined using the arguments of the WDW state itself. One approach is to set the metric to a chosen configuration and then, if the WDW state admits a semiclassical description in a neighborhood of a bulk Cauchy surface with that metric, we can interpret its correlation functions (i.e., derivatives of the WDW state with respect to its arguments) as encoding physics on that slice, as explained in~\cite{Khan:Semiclassical}.

\begin{figure}[ht]
    \centering
\begin{tikzpicture}[x=0.75pt,y=0.75pt,yscale=-1,xscale=1]


\draw (0,22.54) -- (1.38,249.96);
\draw (159.23,22.54) -- (160.62,249.96);

\draw (0,22.54) .. controls (0,17) and (35.65,12.5) .. (79.62,12.5) .. controls (123.59,12.5) and (159.23,17) .. (159.23,22.54) .. controls (159.23,28.09) and (123.59,32.59) .. (79.62,32.59) .. controls (35.65,32.59) and (0,28.09) .. (0,22.54) -- cycle;
\draw (1.38,249.96) .. controls (1.38,244.41) and (37.03,239.91) .. (81,239.91) .. controls (124.97,239.91) and (160.62,244.41) .. (160.62,249.96) .. controls (160.62,255.5) and (124.97,260) .. (81,260) .. controls (37.03,260) and (1.38,255.5) .. (1.38,249.96) -- cycle;

\filldraw[gray,opacity=0.2] (78.12,62.72) -- (159.23,129.69) -- (78.12,196.65) -- (0,129.69) -- cycle;

\draw[line width=2pt] (0,129.69) .. controls (40,99.69) and (122.5,156) .. (159.23,129.69); 
\draw (0,129.69) .. controls (40,95.69) and (87,101) .. (159.23,129.69);
\draw (0,129.69) .. controls (37.5,130) and (96.5,177) .. (159.23,129.69);

\draw (166,121.86) node [anchor=north west] {$\partial\Sigma$};
\draw (166,40) node [anchor=north west] {$\partial\mathcal{M}$};
\draw (65,40) node [anchor=north west] {$\mathcal{M}$};
\draw (92,115) node [anchor=north west] {$\Sigma_{K=0}$};
\draw (73,88) node [anchor=north west] {$\Sigma_f$};
\draw (73,153) node [anchor=north west] {$\Sigma_p$};


\draw (240,22.54) -- (241.38,249.96);
\draw (399.23,22.54) -- (400.62,249.96);

\draw (240,22.54) .. controls (240,17) and (275.65,12.5) .. (319.62,12.5) .. controls (363.59,12.5) and (399.23,17) .. (399.23,22.54) .. controls (399.23,28.09) and (363.59,32.59) .. (319.62,32.59) .. controls (275.65,32.59) and (240,28.09) .. (240,22.54) -- cycle;
\draw (241.38,249.96) .. controls (241.38,244.41) and (277.03,239.91) .. (321,239.91) .. controls (364.97,239.91) and (400.62,244.41) .. (400.62,249.96) .. controls (400.62,255.5) and (364.97,260) .. (321,260) .. controls (277.03,260) and (241.38,255.5) .. (241.38,249.96) -- cycle;

\draw[line width=2pt] (240,129.69) -- (399.23,129.69);

\draw (406,121.86) node [anchor=north west] {$\partial\Sigma$};
\draw (406,40) node [anchor=north west] {$\partial\mathcal{M}$};
\draw (305,40) node [anchor=north west] {$\mathcal{M}$};
\draw (295,108) node [anchor=north west] {$\Sigma_{K=0}$};

\end{tikzpicture}
\caption{\small Two side-by-side geometries. On the left, the shaded region denotes the bulk domain of dependence, where the physics is emergent from the WDW state. All information in this region is encoded in correlation functions of the WDW state. If a semiclassical bulk exists for a WDW state, one can condition the arguments of the WDW wavefunctional—e.g., by specifying the metric—yielding the quantum gravity information on a bulk Cauchy surface with that metric. However, this procedure becomes ambiguous when distinct slices $\Sigma_f$ and $\Sigma_p$ in the semiclassical bulk share the same induced metric. On the right, the maximal slice $\Sigma_{K=0}$ is depicted as a straight line. In the alternative phase space formulation, the quantum gravity state is defined directly on this slice—even in the absence of a semiclassical bulk—and its correlation functions are encoded in the analytically continued CFT. This approach provides a more objective notion of bulk “time,” one that remains well-defined even at the full quantum level.}
\label{fig:AdS_geometries}
\end{figure}
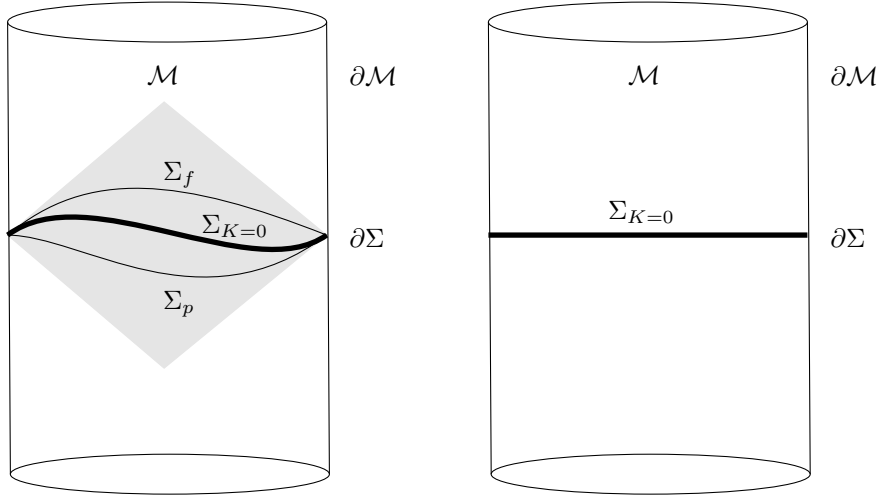

However, if there exist multiple Cauchy surfaces in the bulk spacetime with the same intrinsic metric—say, two distinct slices $\Sigma_f$ and $\Sigma_p$ (see Figure~\ref{fig:AdS_geometries})—this interpretation may require separating the WDW state into distinct WKB branches, each associated with one of those slices. This issue, referred to as the branching problem, was discussed in~\cite{Khan:Semiclassical}. Moreover, such an interpretation is only valid when the WDW state admits a semiclassical bulk geometry. If the state is highly quantum, e.g., a superposition of vastly different geometries, then no semiclassical bulk description exists. In such cases, the WDW state must be understood as encoding physics on the entire WDW patch $\Omega$, and the very notions of bulk clocks and local observables can break down.

In summary, it is generally difficult to condition WDW states to extract a description of physics at a specific bulk “time.” This remains a deep and intriguing problem, one that we began to explore in~\cite{CSH}.

Now, all of this is circumvented in the alternative phase space formulation. Since the temporal diffeomorphisms are gauge-fixed prior to quantization, the quantum gravity states $\Psi_{\text{QG}}$ in equation \eqref{CFT partition function as QG states} in the alternative phase space formulation encode quantum gravity information precisely on the bulk maximal volume slice $\Sigma_{K=0}$. See the second figure in Figure~\ref{fig:AdS_geometries}. Thus, we can interpret the quantum gravity state as “living” on a specific time slice in the bulk. Importantly, even if the quantum gravity state here is highly quantum mechanical, this interpretation remains valid—unlike in the WDW case.

\subsection{UV Completeness of Bulk Quantum Gravity}

In \cite{CSH}, we expressed WDW states as the partition function of the $T^2$ theory:
\begin{equation}
    \Psi_{\text{WDW}} = Z_{T^2}.
\end{equation}
The $T^2$ theory was obtained by Wick rotating the holographic CFT and defining it on Cauchy surfaces to get $Z_\text{CFT}[g,\Phi^i;\psi_{\text{CFT}}]$, then deforming this theory via the $T^2$ deformation, and finally transforming it using the holographic counterterms:
\begin{align} 
Z_{T^2}[g,\Phi^i;\psi_{\text{CFT}}] = e^{\widehat{CT}(\mu)} \left( \text{P} \exp \int_{\epsilon}^{\mu}\frac{d\lambda}{\lambda} \widehat{O}(\lambda) \right) Z_\text{CFT}[g,\Phi^i;\psi_{\text{CFT}}].
\end{align} 
This $T^2$ deformation operator $\widehat{O}(\lambda)$ is an irrelevant deformation, and the flow it induces satisfies the renormalization group equations, i.e., the Callan--Symanzik equations. Hence, the $T^2$ theory is an effective quantum field theory. Thus, in \cite{CSH}, we interpreted the WDW states obtained this way as states in an effective theory of quantum gravity, akin to effective QFTs in the Wilsonian sense. To extract any UV-sensitive information about the bulk from the WDW state, we argued in \cite{CSH} that the $T^2$ theory must be UV-completed---a very daunting task.

This need for UV completion can be circumvented in the alternative phase space formulation. In this framework, the quantum gravity states
\begin{equation}
    \Psi_{\text{QG}}=Z^{(ic)}_{\text{CFT}}
\end{equation}
are given directly by the CFT partition functions and are therefore tautologically UV complete. Consequently, we can expect that the correlation functions of the CFT partition function, defined on the bulk maximal volume slice, encode UV-sensitive information about quantum gravity.

The bulk theory appearing in a concrete $\mathrm{AdS/CFT}$ pair is usually far richer than pure Einstein gravity: string states, higher–derivative interactions, and additional matter multiplets all conspire to make the ultraviolet theory complicated but, by construction, UV–complete.\footnote{Moreover, the bulk has extra compact dimensions, but they can be reinterpreted as an infinite tower of fields in the noncompact $d+1$ dimensions.  For instance, a scalar on $\mathcal M_{d+1}\times\mathcal F$, with $\mathcal F$ compact, can be expanded in harmonics on $\mathcal F$, and each harmonic mode appears as a distinct scalar field on $\mathcal M_{d+1}$.} The phase–space reduction reviewed in section~\ref{Phase Space Reduction} is, however, proven only for the \emph{low–energy} Einstein sector. What happens when one includes the genuinely high–energy (higher–curvature) terms? Two logically distinct scenarios can occur:

\medskip
\noindent\textbf{(i) Full reduction.}  
If the complete bulk action—higher curvature terms and all—admits the same gauge fixing $K=0$ and a unique solution of the Lichnerowicz equation, then the full UV theory reduces to a cotangent bundle over conformal classes of geometries (now with a much larger collection of matter gauge constraints). Enlarging that reduced phase space by adding back the conformal factor and its momentum in a similar fashion, along with the real Weyl-anomaly constraint, produces an \emph{alternative} phase space that is still symplectomorphic to the original. In this favorable case, the exact CFT partition function on each maximal-volume slice, $\Psi_{\text{QG}}=Z^{(ic)}_{\text{CFT}}$, furnishes a \emph{fully UV–complete, background–independent} description of quantum gravity in bulk variables.

\medskip
\noindent\textbf{(ii) Partial reduction.}  
More conservatively, the phase-space reduction may fail once the full tower of higher–curvature operators is included; the Hamiltonian constraint might no longer admit a unique solution. Suppose, however, that the \emph{low–energy} sector (Einstein gravity plus light matter) still reduces cleanly. Then the alternative phase space provides an \emph{effective} bulk description valid below a cutoff, and the wavefunctionals
$\Psi_{\text{QG}}=Z^{(ic)}_{\text{CFT}}$
are best interpreted as \emph{effective} quantum-gravitational states. They remain manifestly background-independent, but any question probing curvatures above the cutoff will require the full, unreduced UV theory.

\medskip
\noindent\textbf{Exact vs.~approximate CFTs.}  
Just as many inequivalent UV-complete bulk theories can flow to the same low-energy effective field theory—namely, Einstein gravity coupled to light fields—their holographic duals, which are distinct exact CFTs, can exhibit identical correlators within a restricted infrared (IR) window. In this low-energy regime, any one of these exact CFTs—or even an arbitrary linear combination of them—functions equally well as an approximate CFT, dual to the low-energy sector of the bulk. The corresponding wavefunctional, \(\Psi_{\text{QG}}=Z^{(ic)}_{\text{CFT}}\), then defines a consistent effective bulk quantum gravity state. As long as we are concerned only with bulk dynamics governed by Einstein gravity and light fields—or more generally, any low-energy sector for which the phase space reduction is well understood—such an approximate CFT suffices to describe quantum gravity states. However, once we attempt to go beyond this low-energy window—probing high-dimension operators or bulk curvature invariants near the UV cutoff—the differences between distinct UV completions become significant. In such cases, one must specify a particular exact CFT to ensure consistency.

\subsection{Summary}

In this paper, we proposed a \emph{conformal Cauchy--slice formulation} of gravity in asymptotically Anti--de Sitter spacetimes and proved its classical equivalence to the familiar ADM description.  
Whenever a \emph{unique} maximal-volume slice $\Sigma_{K=0} \subset \Omega$ exists for every classical saddle, and the matter sector satisfies certain conditions, the condition $K=0$ serves as a valid gauge-fixing condition for the Hamiltonian constraint. Solving the Lichnerowicz equation then eliminates the conformal factor $\phi$, thereby reducing the ADM phase space $\Gamma_{\text{ADM}}$, as originally done in \cite{Witten:2022xxp}.

\medskip
\noindent\textbf{Phase--space enlargement and Theorem 1.}  
We then \emph{enlarged} this reduced phase space by reinstating the pure-gauge pair $(\varphi,\mathsf{\Pi})$ and imposing the \emph{real Weyl--anomaly constraint} $\mathcal{W}+\mathcal{A}=0$, where $\mathcal{W}$ is the generator of local Weyl rescalings of all fields and $\mathcal{A}$ is the holographic conformal anomaly of the dual CFT. We proved that $\{\mathcal{W}+\mathcal{A},D_a,G^A\}$ closes under the Poisson bracket, so the enlarged ``alternative'' phase space
\[
\Gamma_{\text{ALT}}=(\mathcal{P}_{\text{ALT}},\omega_{\text{ALT}};\;\mathcal{W}+\mathcal{A},\;D_a,\;G^A)
\]
possesses only first-class constraints. Imposing $\mathcal{W}+\mathcal{A}=0$ along with its gauge-fixing condition $R[\gamma]=2\Lambda$ (equivalently $\varphi=1$) eliminates $(\varphi,\mathsf{\Pi})$ again and yields a reduced phase space $\Gamma^{\text{red}}_{\text{ALT}}$ that is \emph{symplectomorphic} to $\Gamma^{\text{red}}_{\text{ADM}}$. This establishes Theorem~1: the ADM and alternative phase spaces are physically equivalent at the kinematical level.

\medskip
\noindent\textbf{Quantum theory and Theorem 2.}  
Quantizing $\Gamma_{\text{ALT}}$ à la Dirac, we required physical wave-functionals to satisfy
\[
(\widehat{\mathcal{W}}+\mathcal{A})\Psi = \widehat{D}_a\Psi = \widehat{G}^A\Psi = 0.
\]
A key result is that the partition function of a $d$-dimensional \emph{holographic} CFT, analytically continued to \emph{imaginary} central charge $c \to ic$, obeys precisely these constraints when defined on $\Sigma_{K=0}$ with boundary state $\psi_{\text{CFT}}$. Consequently (Theorem~2)
\begin{equation}
     \Psi_{\text{QG}}[\mathfrak{g}_{ab},\Phi^i] = Z^{(ic)}_{\text{CFT}}[\mathfrak{g}_{ab},\Phi^i;\psi_{\text{CFT}}] \nonumber
\end{equation}
furnishes a non-perturbative, background-independent quantum-gravity state in the bulk.

\medskip
\noindent\textbf{Implications.}  
The alternative formulation localises quantum-gravitational data on a single, geometrically distinguished slice~$\Sigma_{K=0}$, sidestepping the branching ambiguities that plague Wheeler–DeWitt states and providing a clearer notion of “bulk time.”  
Because quantum gravity states are CFT partition functions, bulk UV physics is encoded in \emph{finite} CFT correlators on $\Sigma_{K=0}$, offering a new, manifestly background-independent handle on questions such as singularity resolution and bulk locality. We believe that further developing the dynamical dictionary will illuminate these issues and may yield a tractable, non-perturbative definition of quantum gravity in AAdS space-times in the language of bulk physics.

\vspace{4pt}
{\centering
\noindent\rule{8cm}{0.8pt}
\par}
\vspace{-6pt}

{\paragraph{Acknowledgements:} \small We are grateful for helpful conversations with Aron Wall, Ronak Soni, Eva Silverstein, Henry Maxfield, Shoaib Akhtar, Alex Frenkel, Raghu Mahajan, Albert Law, and Xiaoliang Qi. We would especially like to thank Aron Wall, Ronak Soni, and Henry Maxfield for their valuable comments on the draft.}

\begin{appendices}

\section{Lichnerowicz  Equation and Covariant Conformal Decomposition in \texorpdfstring{$d=2$}{}}\label{appendix d=2}

Let us explain the $d=2$ case. Everything proceeds almost identically to the higher-$d$ case, with only minor differences. Here, we assume the reader is familiar with sections~\ref{Section Lichnerowicz Equation} and~\ref{section: Covariant Conformal Decomposition}. We will emphasize only the differences that arise in the $d=2$ case as compared to higher dimensions.

In the conformal decomposition, the metric is expressed as
\begin{equation}\label{d=2 conformal decomosition of metric}
    g_{ab} = e^{2\omega} \gamma_{ab}.
\end{equation}
Here, $\omega$ represents the conformal factor, which is a scalar field (not restricted in sign), and vanishes at the boundary: $\omega|_{\partial\Sigma}=0$. $\gamma_{ab}$ is the conformal part of the metric on $\Sigma$. Under this decomposition, the Ricci scalar transforms as
\begin{align}
    e^{2\omega} R_{[g]}  &= R_{[\gamma]}-2\nabla^2_{[\gamma]} \omega.
\end{align}
Similar to Proposition~\ref{Prop: Yamabe}, the uniformization theorem tells us that, in this case, the conformal part $\gamma_{ab}$ of any AH metric $g_{ab}$ can be uniquely chosen such that $R_{[\gamma]} = 2\Lambda$.

Then, on the maximal volume slice ($\Pi = 0$), the Hamiltonian constraint becomes
\begin{align}
  2\nabla^2_{[\gamma]} \omega
+|\pi|^2 e^{-2\omega} 
- 2\Lambda
+2\Lambda e^{2\omega}
+ \frac{2\kappa}{\sqrt{g}}\mathcal{H}_{\text{matter}}e^{2\omega}&=0.    
\end{align}
Here, $|\pi|^2:=\left(\frac{2\kappa}{\sqrt{\gamma}}\right)^2\gamma_{ac}\gamma_{bd}\pi^{ab}\pi^{cd} \geq 0$, and  $\pi^{ab} = e^{2\omega} (\Pi^{ab}-\frac{1}{2}\Pi g^{ab})$.

Let us similarly assume
\begin{align}
 \frac{2\kappa}{\sqrt{g}}\mathcal{H}_{\text{matter}} &= A_i e^{2n_i\omega} + B e^{-2\omega} + C_j e^{2m_j\omega},    
\end{align}
where $n_i<-1<m_j$, with $A_i \geq 0$, $B > 2\Lambda$, and $C_j \leq 0$, allowing the weaker bound $C_j < -2\Lambda$ when $m_j = 0$. The Hamiltonian constraint then becomes
\begin{align}\label{d=2 lichnerowicz equation}
  2\nabla^2_{[\gamma]} \omega - F(\omega,x) &=0,
\end{align}
where 
\begin{equation}
    F(\omega,x) = 
    -|\pi|^2 e^{-2\omega} 
- A_i e^{2(n_i+1)\omega}
-(B-2\Lambda)
- C_j e^{2(m_j+1)\omega}
-2\Lambda e^{2\omega}.
\end{equation}

Now the proof of existence is almost identical to the one in section \ref{Section Lichnerowicz Equation}, with only minor changes. The method of sub- and supersolutions, as stated in proposition \ref{Prop: supersubsolution}, does not require the sub- and supersolutions to be positive functions; it remains valid even when they are sign-changing functions. A large enough positive constant $c_+ \gg 1$ serves as a constant supersolution, and a large enough negative constant $c_- \ll -1$ serves as a constant subsolution. However, we now want our solution to vanish at the boundary.

In the FG gauge, express the conformal metric for some neighbourhood near the boundary $\partial\Sigma$ as
\begin{equation}
    ds^2 = \gamma_{ab}\, dx^a dx^b =\frac{L^2}{z^2} (dz^2 + \sigma(\theta,z)d\theta^2), \quad \text{and} \quad \sigma(\theta,z) = \sigma^{(0)}(\theta)+z^2 \sigma^{(1)} (\theta)+\cdots,
\end{equation}
where $z$ is again the radial coordinate increasing inwards, and $L = -1/\Lambda$. But the Laplacian in this case is
\begin{align}
\nabla^2_{[\gamma]} \omega
&= \frac{z^2}{L^2}\,\partial_z^2 \omega
\;+\frac{z^2}{L^2}\;\Bigl[\frac{1}{2\sigma}\partial_z\sigma
         \Bigr]\,\partial_z \omega
\;+\;\frac{z^2}{L^2}\,\nabla^2_{[\sigma]} \omega.
\end{align}

Now choose a small $\epsilon>0$ and two smooth monotonic interpolating functions $u_{+}(z)$ and $u_{-}(z)$ such that $u_{+}(\epsilon)=c_{+}$, $u_{+}(0)=0$, $u_{-}(\epsilon)=c_{-}$, and $u_{-}(0)=0$. Then define the functions $\omega_{+}$ and $\omega_{-}$ on $\Sigma$ as
\begin{equation}
\omega_{+}(z) =
\begin{cases}
c_{+} & \text{for } z \geq \epsilon, \\
u_{+}(z) & \text{for } 0 < z < \epsilon,
\end{cases}
\qquad
\omega_{-}(z) =
\begin{cases}
c_{-} & \text{for } z \leq \epsilon, \\
u_{-}(z) & \text{for } 0 < z < \epsilon.
\end{cases}
\end{equation}

By the same arguments used before in the previous proof in section \ref{Section Lichnerowicz Equation}, here too we have $F(\omega_+,x)>0$ and $F(\omega_-,x)<0$ everywhere on $\Sigma$. In fact, from the asymptotic conditions, for some small $\delta > 0$, we can make $F(\omega_+,x) > \delta$ and $F(\omega_-,x) < \delta$ everywhere on $\Sigma$. $\nabla^2_{[\sigma]} \omega_{\pm} = 0$, and $\partial^2 \omega_{\pm}$ can be made arbitrarily close to zero by choosing the interpolating functions arbitrarily close to a linear function. But now the difference comes in the following point. The leading contribution in $\nabla^2_{[\gamma]} \omega_{\pm}$ is $z^3 \frac{\sigma^{(1)}}{L^2 \sigma^{(0)}} \, \partial_z \omega_{\pm}$, which is sign-indefinite, as opposed to the situation in section \ref{Section Lichnerowicz Equation}. However, for a linear interpolating function, $\partial_z \omega_{\pm} \sim c_{\pm} / \epsilon$, and so although sign-indefinite, the linear term in this $d=2$ case is of order $O(z^2)$ (even when $z \sim \epsilon$).\footnote{In the previous proof for $d > 2$, we had $\partial_z \phi_{\pm} \sim c_{\pm} \epsilon$ and the dominating linear term was $|z \partial_z \phi_{\pm}| \sim 1$ when $z \sim \epsilon$, and so this term is not necessarily small; its sign definiteness enabled the proof. In this $d=2$ case, the linear term is small and does not matter, although it is sign-indefinite.} So we have $2\nabla^2_{[\gamma]} \omega_{+} - F(\omega_{+},x) \leq 0$ and $2\nabla^2_{[\gamma]} \omega_{-} - F(\omega_{-},x) \geq 0$ everywhere on $\Sigma$. Also, $\omega_{+}|_{\partial\Sigma} = \omega_{-}|_{\partial\Sigma} = 1$. Thus, $\omega_{+}$ and $\omega_{-}$ are a supersolution and a subsolution respectively, and both approach $0$ at the boundary with $\omega_{-} < \omega_{+}$. So, from proposition \ref{Prop: supersubsolution}, there exists a solution to equation \eqref{d=2 lichnerowicz equation}.

Next, the proof for uniqueness is also almost identical to the one in section \ref{Section Lichnerowicz Equation}. The analogue of equation \eqref{adding together to prove uniqueness} for the $d=2$ case is
\begin{align}
 0 &= 2\nabla^2_{[h]} \omega_0 - F_{h}(\omega_0,x) + F_h(0,x),
\end{align}
but notice that the difference here is that $F_h(0,x)$ is not multiplied by anything. $\omega_0$ is analogous to $\phi_0$, and $h$ is defined similarly to the previous proof. Then the same arguments lead to uniqueness.

Next, for the covariant conformal decomposition in $d=2$, the metric $g_{ab}$\footnote{We use the same notation for the variables in the ADM phase space and alternative phase space to explain the $d=2$ case, and the context will make it clear which variables are being used.} is decomposed as in \ref{d=2 conformal decomosition of metric}, with $R_{[\gamma]}=2\Lambda$, and the momenta $\mathsf{\Pi}^{ab}$ are decomposed as
\begin{equation}
     \Pi^{ab} = e^{-2\omega}\, \Sigma^{ab} + \sqrt{\gamma}\, e^{-2\omega}\, Y^{ab}_{\gamma} f,
\end{equation}
where $Y^{ab}_\gamma$ and $f$ are defined exactly as before in section \ref{section: Covariant Conformal Decomposition}.
\begin{align}
    \Pi^{ab} \delta g_{ab} &= ( \Sigma^{ab} + \sqrt{\gamma}\,  Y^{ab}_{\gamma} f) ( \delta \gamma_{ab} + 2 \gamma_{ab} \delta\omega)= \Sigma^{ab} \delta \gamma_{ab} 
    + 2\, \Pi \, \delta\omega 
    +  (\sqrt{\gamma}\,  Y^{ab}_{\gamma} f) \delta \gamma_{ab},
\end{align}
so the symplectic potential is
\begin{equation}
    \theta =\int_\Sigma d^2x (\Pi^{ab}\delta g_{ab} + \Pi_{\Phi^i}\delta\Phi^i) =  \int_\Sigma d^2x (\Sigma^{ab} \delta \gamma_{ab} 
    +  \Pi_\omega \, \delta\omega 
    + \Pi_{\Phi^i}\delta\Phi^i),
\end{equation}
where $\Pi_\omega := 2\Pi$. The term $\int d^2x \sqrt{\gamma}\,  (Y^{ab}_{\gamma} f) \delta \gamma_{ab}$ drops out after integration by parts and using $Y^{ab}_\gamma \delta\gamma_{ab}=0$, and the boundary terms arising here vanish similarly due to Dirichlet boundary conditions on $\gamma_{ab}$ and $f$. The covariant conformal decomposition in $d=2$ is the change of variables from $(g_{ab},\Pi^{ab},\Phi^i,\Pi_{\Phi^i})$ to $(\omega,\Pi_\omega,\gamma_{ab},\Sigma^{ab},\Phi^i,\Pi_{\Phi^i})$.

Thus, the discussion of the Lichnerowicz equation and covariant conformal decomposition for the $d=2$ case is almost identical to the higher $d$ case, with these minor differences, and everything else in the body of this paper holds for the $d=2$ case as well.

\section{\texorpdfstring{$\mathcal{W}+\mathcal{A}$}{} for Einstein Gravity \texorpdfstring{$+$}{} Massive Scalar in \texorpdfstring{$d=4$}{}}\label{appendix d=4 massive scalar}

Consider a massive scalar field $\Phi$ minimally coupled to Einstein gravity in $(4+1)$-dimensional spacetime. Although this model violates the strong energy condition—so uniqueness of the maximal volume slice can fail—we do not study it here in the context of phase space reduction. Instead, we present it to illustrate again the derivation of the Weyl-anomaly constraint from the Hamiltonian constraint, while still exhibiting a more complicated anomaly. The Hamiltonian and diffeomorphism constraints are
\begin{align}
    \mathcal{H}&=\frac{2\kappa}{\sqrt{g}}\Big(\Pi_{ab}\Pi^{ab}-\frac{1}{d-1}\Pi^2\Big) -\frac{\sqrt{g}}{2\kappa}(R-2\Lambda)+\frac{1}{2\sqrt{g}}
\Pi_\Phi^2
+\frac{1}{2}\sqrt{g}\left(g^{ab}\nabla_a\Phi\nabla_b\Phi+m^2\Phi^2\right),\\
D_a &= -2\nabla_b \Pi^b_a + \partial_a \Phi \, \Pi_{\Phi}.
\end{align}
Rescale the variables as
\begin{equation}
 \quad  g_{ab} \rightarrow \epsilon^{-2} g_{ab},  \quad \Pi^{ab} \rightarrow \epsilon^{2} \Pi^{ab}, \quad \Phi \rightarrow \epsilon^{\Delta_{\Phi}} \Phi, \quad \Pi_{\Phi^i} \rightarrow \epsilon^{-\Delta_{\Phi}} \Pi_{\Phi}.
\end{equation}
Under this rescaling, notice that $D_a$ does not change, and the Hamiltonian constraint $\mathcal{H}$ changes as follows:
\begin{align}
    \mathcal{H}^{\text{rescaled}} &= \epsilon^{4}\frac{2\kappa}{\sqrt{g}}\Big(\Pi_{ab}\Pi^{ab}-\frac{1}{d-1}\Pi^2\Big) 
    -\epsilon^{-2}\frac{\sqrt{g}}{2\kappa} R
    +\epsilon^{-4}\frac{\sqrt{g}}{2\kappa}2\Lambda \nonumber\\
    &+\epsilon^{4-2\Delta_\Phi}\frac{1}{2\sqrt{g}}
\Pi_\Phi^2
+\epsilon^{-2+2\Delta_\Phi}\frac{1}{2}\sqrt{g} g^{ab}\nabla_a\Phi\nabla_b\Phi 
+\epsilon^{-4+2\Delta_\Phi}\frac{1}{2}\sqrt{g}m^2\Phi^2 .
\end{align}
The $\sqrt{g}\Lambda$ and $\sqrt{g}R$ terms are relevant, and whether a term in the matter part is relevant or not depends on $\Delta_\Phi$: the $\frac{1}{\sqrt{g}}\Pi_\Phi^2$ term is relevant when $\Delta_\Phi > 2$, the $\sqrt{g}\Phi^2$ term is relevant when $\Delta_\Phi < 2$, and the $\sqrt{g} g^{ab}\nabla_a\Phi\nabla_b\Phi$ term is relevant when $\Delta_\Phi < 1$. Let us consider the case when
\begin{equation}
    \Delta_\Phi \in [1,2).
\end{equation}
Then all of the relevant terms are the $\sqrt{g}\Lambda$, $\sqrt{g}R$, and $\sqrt{g}\Phi^2$ terms. So, choose the holographic counterterms as
\begin{equation}
    CT = \int d^3x\, \sqrt{g} \left(a_\Lambda + a_R R + a_\Phi \Phi^2  \right).
\end{equation}
The transformation $e^{(\{\cdot,CT\})}$ leaves $g_{ab}$ and $\Phi$ invariant and changes $\Pi^{ab}$ and $\Pi_\Phi$ as:
\begin{align}
    \Pi^{ab} &\longrightarrow e^{(\{\cdot ,CT\})} \Pi^{ab} = \Pi^{ab} + \{\Pi^{ab},CT\} = \Pi^{ab} - \frac{1}{2} \sqrt{g} \left(a_\Lambda  + a_\Phi \Phi^2  \right) g^{ab} + a_R  \sqrt{g} G^{ab},\\
     \Pi_\Phi &\longrightarrow e^{(\{\cdot ,CT\})} \Pi_\Phi = \Pi_\Phi + \{\Pi_\Phi,CT\} = \Pi_\Phi  - 2\, a_\Phi \, \sqrt{g}  \, \Phi .
\end{align}
So we get
\begin{align}
  e^{(\{\cdot ,CT\})}   \mathcal{H}^{\text{rescaled}} &= \frac{1}{3} a_\Lambda\kappa \epsilon^{4} \left(       2\Pi - \frac{6 a_\Phi }{\kappa a_\Lambda}  \epsilon^{-2\Delta_\Phi}
    \, \Phi \Pi_\Phi \right) + \epsilon^{4}2\kappa a_R^2  \sqrt{g} \Big(    G^{ab}    G_{ab} - \frac{1}{3} G^2\Big) \label{firstlinereferingd=4appendixeqn}\\
 & +\epsilon^{-2+2\Delta_\Phi}\frac{1}{2}\sqrt{g} g^{ab}\nabla_a\Phi\nabla_b\Phi  - \frac{2}{3} \epsilon^{4}\kappa    a_\Phi    a_R  \sqrt{g}\, \Phi^2  R 
 - \frac{2\kappa  a_\Phi^2}{3}\epsilon^{4} \sqrt{g}    \Phi^4 \nonumber\\
 &+ \epsilon^{4}\frac{2\kappa}{\sqrt{g}}\Big(\Pi_{ab}\Pi^{ab}-\frac{1}{3}\Pi^2\Big) + 4\kappa \epsilon^{4}  a_R   \Big(G^{ab} \Pi_{ab} - \frac{1}{3} G\, \Pi\Big)  \nonumber\\
    &+\epsilon^{4-2\Delta_\Phi}\frac{1}{2\sqrt{g}}
\Pi_\Phi^2 +  a_\Phi\epsilon^{4}\frac{2\kappa}{3}    \Phi^2   \Pi 
 \nonumber\\
&+ \left(2 \epsilon^{4-2\Delta_\Phi}
 a_\Phi^2   +\epsilon^{-4+2\Delta_\Phi}\frac{m^2}{2} - \frac{4\kappa  a_\Lambda   a_\Phi}{3}\epsilon^{4} \right) \sqrt{g}  \, \Phi^2  
 \nonumber\\
 &-\left(  \frac{2}{3} \epsilon^{4}\kappa     a_\Lambda     a_R  + \epsilon^{-2}\frac{1}{2\kappa} \right)\sqrt{g}\, R \nonumber\\
 &+\left(- \frac{2\kappa a_\Lambda^2}{3}\epsilon^{4}   +\epsilon^{-4}\frac{1}{2\kappa}2\Lambda    \right)  \sqrt{g}.
\end{align}
To eliminate the relevant terms in the above equation, we need to set their coefficients to zero, which leads to
\begin{align}
    a_\Lambda^2     &= \epsilon^{-8}\frac{3 \Lambda}{2\kappa^2} , \quad a_\Lambda     a_R   = - \epsilon^{-6}\frac{3}{4\kappa^2}, \quad m^2 = - 4 \epsilon^{8-4\Delta_\Phi}
 a_\Phi^2  + \frac{8\kappa  a_\Lambda   a_\Phi}{3}\epsilon^{8-2\Delta_\Phi} .
\end{align}
This gives:
\begin{equation}
    a_\Lambda    = \pm  
 i \epsilon^{-4}\sqrt{\frac{3 |\Lambda|}{2\kappa^2}}.
\end{equation}
We must now choose the sign that leads to the Weyl-anomaly equation with the correct sign in the anomaly. It turns out that the $(+)$ sign is the correct one, so this gives
\begin{align}
    a_\Lambda    =   
 i \epsilon^{-4}\sqrt{\frac{3 |\Lambda|}{2\kappa^2}}, \quad     a_R   = i \epsilon^{-2}\frac{1}{4\kappa}  \sqrt{\frac{6}{ |\Lambda|}}, 
 \quad m^2 = - 4 \epsilon^{8-4\Delta_\Phi}
 a_\Phi^2  + i 8     a_\Phi 
 \epsilon^{4-2\Delta_\Phi}   
  \sqrt{\frac{|\Lambda|}{6}}.\label{thirdeqnreferenced=4appendix}
\end{align}
Also, note that the term in the first bracket in \eqref{firstlinereferingd=4appendixeqn} is the Weyl generator, by identifying the coefficient of $\Phi\Pi_\Phi$ in it to be the negative of the conformal dimension $\Delta_\Phi$:
\begin{equation}
    -\Delta_\Phi =  -\frac{6 a_\Phi }{\kappa a_\Lambda}  \epsilon^{-2\Delta_\Phi},
\end{equation}
which gives
\begin{equation}
     a_\Phi  =  i \frac{\Delta_\Phi}{6} \epsilon^{2\Delta_\Phi-4}     \sqrt{\frac{3 |\Lambda|}{2}},
\end{equation}
and from the third equation in \eqref{thirdeqnreferenced=4appendix}, this gives the usual holographic relation between the mass $m$ and the conformal dimension $\Delta_\Phi$ of the scalar field $\Phi$:
\begin{equation}
     \quad m^2 =  \Delta_\Phi \left( \Delta_\Phi
  -  4  \right) \frac{|\Lambda|}{6}.
\end{equation}
So now we have eliminated the relevant $\sqrt{g}\Lambda$, $\sqrt{g}R$, and $\sqrt{g}\Phi^2$ terms:
\begin{align}
  e^{(\{\cdot ,CT\})}   \mathcal{H}^{\text{rescaled}} &=    
 i \sqrt{\frac{|\Lambda|}{6}}   \left(       2\Pi -\Delta_\Phi \Phi \Pi_\Phi \right) -  
   \frac{3}{ 4\kappa |\Lambda|}
 \sqrt{g} \Big(    G^{ab}    G_{ab} - \frac{1}{3} G^2\Big)\nonumber\\
 & +\epsilon^{2\Delta_\Phi-2}\frac{1}{2}\sqrt{g} g^{ab}\nabla_a\Phi\nabla_b\Phi  
 +  \epsilon^{2\Delta_\Phi-2}      \frac{\Delta_\Phi}{12}       \sqrt{g}\, \Phi^2  R 
 + \epsilon^{4\Delta_\Phi-4} \frac{\kappa |\Lambda| \Delta_\Phi^2}{36}      \sqrt{g}    \Phi^4 \nonumber\\
 &+ \epsilon^{4}\frac{2\kappa}{\sqrt{g}}\Big(\Pi_{ab}\Pi^{ab}-\frac{1}{3}\Pi^2\Big) 
 + i \epsilon^{2}     \sqrt{\frac{6}{ |\Lambda|}}   \Big(G^{ab} \Pi_{ab} - \frac{1}{3} G\, \Pi\Big)  \nonumber\\
    &+\epsilon^{4-2\Delta_\Phi}\frac{1}{2\sqrt{g}}
\Pi_\Phi^2 + i\epsilon^{2\Delta_\Phi}   \frac{\kappa\Delta_\Phi}{6}      \sqrt{\frac{2|\Lambda|}{3}}    \Phi^2   \Pi.
\end{align}
No newly generated terms here are relevant, and so we have eliminated all of the relevant terms. The terms in the first line are all marginal, the terms in the third and fourth lines are irrelevant, and the terms in the second line are irrelevant for $\Delta_\Phi \in (1,2)$ and marginal for $\Delta_\Phi = 1$. Using equation \eqref{imaginary weyl anomaly function as a limit} for our case of $d=4$, we have
\begin{equation}
         \left(\mathcal{W} - i \mathcal{A}\right) = - i \sqrt{\frac{6}{|\Lambda|}} \ \lim_{\epsilon \to 0}  \ e^{(\{\cdot,CT\})} \mathcal{H}^{\text{rescaled}}.
\end{equation}
So when $\Delta_\Phi \in (1,2)$, the imaginary Weyl-anomaly constraint is
\begin{align}
     \left(\mathcal{W} - i \mathcal{A}\right) &=   \left(       2\Pi -\Delta_\Phi \Phi \Pi_\Phi \right) 
 + i c \,\sqrt{g}\, \frac{1}{8\pi^2 } 
  \Big(    G^{ab}    G_{ab} - \frac{1}{3} G^2\Big),     
\end{align}
where the overall coefficient in the anomaly is the central charge   
\begin{equation}
 c= \frac{\pi L_{AdS}^3}{8 G_N }   = \frac{\pi^2 }{\kappa} \left(\frac{6}{|\Lambda|} \right)^{3/2},
\end{equation}
and when $\Delta_\Phi=1$, the imaginary Weyl-anomaly constraint is
\begin{align}
     \left(\mathcal{W} - i \mathcal{A}\right) &=   \left(       2\Pi - \Phi \Pi_\Phi \right) 
 + i \frac{c}{8\pi^2} \sqrt{g}\Bigg(  
  \Big(    G^{ab}    G_{ab} - \frac{1}{3} G^2\Big) -  \frac{2\kappa|\Lambda|}{3} \left( g^{ab}\nabla_a\Phi\nabla_b\Phi  
 +  \frac{1}{6}        \Phi^2  R \right)
 -     \frac{\kappa^2 |\Lambda|^2 }{27}     \Phi^4  \Bigg).     
\end{align}
Notice there are factors of $\kappa$ in the conformal anomaly, but that is because the bulk scalar field was normalized canonically in the action (i.e., the factor in front of the kinetic term in the action is $\frac{1}{2}$). If we normalize it instead with the factor $\frac{1}{\kappa}$ in front of the matter action (as is usually done in string theory), then all factors of $\sqrt{\kappa}$ in the above anomaly would be absorbed into the scalar field. Then the rescaled bulk fields, with this particular normalization convention and also after absorbing the factors of $\sqrt{|\Lambda|}$ into them, serve as the sources of the dual CFT, and the CFT satisfies this imaginary Weyl-anomaly constraint with these sources.

Then the real Weyl-anomaly constraint is obtained simply by replacing the factor $``(-i)"$ in the imaginary Weyl-anomaly constraint with $1$.

\end{appendices}

\vspace{1cm}
\color{black}
\noindent\rule[0.25\baselineskip]{\textwidth}{1pt}

\bibliographystyle{ieeetr}
\bibliography{references}

\begin{thebibliography}{10}

\bibitem{Maldacena:1997re}
J.~M. Maldacena, ``{The Large $N$ limit of superconformal field theories and supergravity},'' {\em Adv. Theor. Math. Phys.}, vol.~2, pp.~231--252, 1998.

\bibitem{Witten:1998qj}
E.~Witten, ``{Anti de Sitter space and holography},'' {\em Adv. Theor. Math. Phys.}, vol.~2, pp.~253--291, 1998.

\bibitem{kaplanAdSCFT}
J.~Kaplan, ``Lectures on ads/cft from the bottom up,'' 2016.
\newblock Lecture notes.

\bibitem{Hubeny:2014bla}
V.~E. Hubeny, ``{The AdS/CFT Correspondence},'' {\em Class. Quant. Grav.}, vol.~32, no.~12, p.~124010, 2015.

\bibitem{Wald:1993nt}
R.~M. Wald, ``{Black hole entropy is the Noether charge},'' {\em Phys. Rev. D}, vol.~48, no.~8, pp.~R3427--R3431, 1993.

\bibitem{Harlow:2019yfa}
D.~Harlow and J.-Q. Wu, ``{Covariant phase space with boundaries},'' {\em JHEP}, vol.~10, p.~146, 2020.

\bibitem{Arnowitt:1962hi}
R.~L. Arnowitt, S.~Deser, and C.~W. Misner, ``{The Dynamics of general relativity},'' {\em Gen. Rel. Grav.}, vol.~40, pp.~1997--2027, 2008.

\bibitem{DeWitt:1967yk}
B.~S. DeWitt, ``{Quantum Theory of Gravity. 1. The Canonical Theory},'' {\em Phys. Rev.}, vol.~160, pp.~1113--1148, 1967.

\bibitem{DeWitt:1967ub}
B.~S. DeWitt, ``{Quantum Theory of Gravity. 2. The Manifestly Covariant Theory},'' {\em Phys. Rev.}, vol.~162, pp.~1195--1239, 1967.

\bibitem{DeWitt:1967uc}
B.~S. DeWitt, ``{Quantum Theory of Gravity. 3. Applications of the Covariant Theory},'' {\em Phys. Rev.}, vol.~162, pp.~1239--1256, 1967.

\bibitem{Freidel:2008sh}
L.~{Freidel}, ``{Reconstructing AdS/CFT},'' {\em arXiv e-prints}, p.~arXiv:0804.0632, Apr. 2008.

\bibitem{McGough:2016lol}
L.~McGough, M.~Mezei, and H.~Verlinde, ``{Moving the CFT into the bulk with $ T\overline{T} $},'' {\em JHEP}, vol.~04, p.~010, 2018.

\bibitem{Hartman:2018tkw}
T.~Hartman, J.~Kruthoff, E.~Shaghoulian, and A.~Tajdini, ``{Holography at finite cutoff with a $T^2$ deformation},'' {\em JHEP}, vol.~03, p.~004, 2019.

\bibitem{Zamolodchikov:2004ce}
A.~B. Zamolodchikov, ``{Expectation value of composite field T anti-T in two-dimensional quantum field theory},'' {\em arXiv e-prints}, 1 2004.

\bibitem{Taylor:2018xcy}
M.~Taylor, ``{$T \bar{T}$ deformations in general dimensions},'' {\em Adv. Theor. Math. Phys.}, vol.~27, no.~1, pp.~37--63, 2023.

\bibitem{Lewkowycz:2019xse}
A.~Lewkowycz, J.~Liu, E.~Silverstein, and G.~Torroba, ``{$ T\overline{T} $ and EE, with implications for (A)dS subregion encodings},'' {\em JHEP}, vol.~04, p.~152, 2020.

\bibitem{CSH}
G.~Araujo-Regado, R.~Khan, and A.~C. Wall, ``{Cauchy slice holography: a new AdS/CFT dictionary},'' {\em JHEP}, vol.~03, p.~026, 2023.

\bibitem{Isenberg_1995}
J.~Isenberg, ``Constant mean curvature solutions of the einstein constraint equations on closed manifolds,'' {\em Classical and Quantum Gravity}, vol.~12, p.~2249, sep 1995.

\bibitem{Andersson_Chrusciel}
L.~Andersson and P.~Chrusciel, ``Solutions of the constraint equations in general relativity satisfying hyperboloidal boundary conditions,'' {\em Dissertationes Mathematicae}, vol.~355, pp.~1--100, 1996.

\bibitem{Sakovich:2009nb}
A.~Sakovich, ``{Constant mean curvature solutions of the Einstein-scalar field constraint equations on asymptotically hyperbolic manifolds},'' {\em Class. Quant. Grav.}, vol.~27, p.~245019, 2010.

\bibitem{Witten:2022xxp}
E.~Witten, ``{A note on the canonical formalism for gravity},'' {\em Adv. Theor. Math. Phys.}, vol.~27, no.~1, pp.~311--380, 2023.

\bibitem{Isenberg:1996kz}
J.~Isenberg and J.~Park, ``{Asymptotically hyperbolic nonconstant mean curvature solutions of the Einstein constraint equations},'' {\em Class. Quant. Grav.}, vol.~14, pp.~A189--A202, 1997.

\bibitem{Couch:2018phr}
J.~Couch, S.~Eccles, T.~Jacobson, and P.~Nguyen, ``{Holographic Complexity and Volume},'' {\em JHEP}, vol.~11, p.~044, 2018.

\bibitem{Chrusciel:2022cjz}
P.~T. Chru{\'s}ciel and G.~J. Galloway, ``{Maximal hypersurfaces in asymptotically Anti-de Sitter spacetime},'' {\em arXiv e-prints}, 8 2022.

\bibitem{Andersson:1992yk}
L.~Andersson, P.~Chrusciel, and H.~Friedrich, ``{On the Regularity of solutions to the Yamabe equation and the existence of smooth hyperboloidal initial data for Einsteins field equations},'' {\em Commun. Math. Phys.}, vol.~149, pp.~587--612, 1992.

\bibitem{Henneaux:1992ig}
M.~Henneaux and C.~Teitelboim, {\em {Quantization of Gauge Systems}}.
\newblock Princeton University Press, 1992.

\bibitem{Hayward:1992ix}
G.~Hayward and K.~Wong, ``{Boundary Schrodinger equation in quantum geometrodynamics},'' {\em Phys. Rev. D}, vol.~46, pp.~620--626, 1992.
\newblock [Addendum: Phys.Rev.D 47, 4778--4779 (1993)].

\bibitem{Shyam:2017qlr}
V.~Shyam, ``{Connecting holographic Wess-Zumino consistency condition to the holographic anomaly},'' {\em JHEP}, vol.~03, p.~171, 2018.

\bibitem{Shyam:2016zuk}
V.~Shyam, ``{General Covariance from the Quantum Renormalization Group},'' {\em Phys. Rev. D}, vol.~95, no.~6, p.~066003, 2017.

\bibitem{Khan:Semiclassical}
R.~Khan, ``{The Semiclassical Approximation: Its Application to Holography and the Information Paradox},'' {\em arXiv e-prints}, 9 2023.

\end{thebibliography}

\end{document}